\documentclass[sn-mathphys]{sn-jnl}
\jyear{2022}

\theoremstyle{thmstyleone}

\theoremstyle{thmstyletwo}

\theoremstyle{thmstylethree}

\usepackage{mymath,journals}
\usepackage[varg]{txfonts}

\usepackage{ifthen}
\newboolean{@comments}
\setboolean{@comments}{true}
\newcommand{\comment}[3]
 {\ifthenelse{\boolean{@comments}}
  {\textcolor{#2}{\textbf{[#1:}~#3\textbf{]}}}
  {}}
\newcommand{\sk}[1]{\comment{SK}{orange}{#1}}

\begin{document}

\title[KFT for cosmic structure formation]{Kinetic Field Theory for Cosmic Structure Formation}

\author{\fnm{Sara} \sur{Konrad}}\email{fz002@uni-heidelberg.de}

\author{\fnm{Matthias} \sur{Bartelmann}}\email{bartelmann@uni-heidelberg.de}

\affil
  {\orgdiv{Institute for Theoretical Physics},
   \orgname{Heidelberg University},
   \orgaddress
     {\street{Philosophenweg 16}, \postcode{69120} \city{Heidelberg},
      \country{Germany}}}

\abstract
  {We apply kinetic field theory to non-linear cosmic structure formation. Kinetic field theory decomposes the cosmic density field into particles and follows their trajectories through phase space. We assume that initial particle momenta are drawn from a Gaussian random field. We place particular emphasis on the late-time, asymptotic behaviour on small spatial scales of low-order statistical measures for the distribution of particles in configuration and velocity space. Our main result is that the power spectra for density and velocity fluctuations in ensembles of particles freely streaming along Zel'dovich trajectories asymptotically fall off with wave number $k$ like $k^{-3}$ for $k\to\infty$, irrespective of the cosmological model and the type of dark matter assumed, with the exponent set only by the number of spatial dimensions. This conclusion remains valid for density-fluctuation power spectra if particle interactions are taken into account in a mean-field approximation. We also show that the bispectrum of freely-streaming particles falls off asymptotically like $k^{-11/2}$ under the same general conditions.}


\maketitle

\section{Introduction}
\label{sec:1}

Our observable universe is permeated by structures on all scales. The Earth is part of the Solar System, located in one spiral arm of the Milky Way galaxy, which is a member of the Local Group of galaxies, which is part of the Virgo Supercluster. Galaxies identified in the Two-Micron All-Sky Survey mark large-scale filamentary structures, galaxy clusters and voids in the nearby universe \cite{2006AJ....131.1163S}. These are structures characterizing the cosmic matter distribution at present, almost 14 billion years after the Big Bang. Temperature fluctuations in the cosmic microwave background on the other hand represent density fluctuations in the early universe, about 400000 years after the Big Bang. At that time, density fluctuations had an amplitude of $\approx10^{-5}$ relative to the mean \cite{2020A&A...641A...6P}. By now, the matter density in the central regions of galaxy clusters exceeds the mean cosmic density by factors of $\approx5\ldots10$ \cite{2012ApJ...757...22C, 2013SSRv..177....3B}. Within 14 billion years, the amplitude of density fluctuations has grown by a factor of $\gtrsim10^6$ on such scales. The cosmological standard model requires dark matter, i.e.\ a dominant form of matter incapable of interacting electromagnetically, to explain this amount of growth \cite{1982ApJ...263L...1P}. With known forms of matter, the density-fluctuation amplitude could not have grown by more than a factor of $\approx10^3$ in the same time (see eg. \cite{dodelson2020modern}).

\begin{figure}[t]
  \includegraphics[width=0.49\hsize]{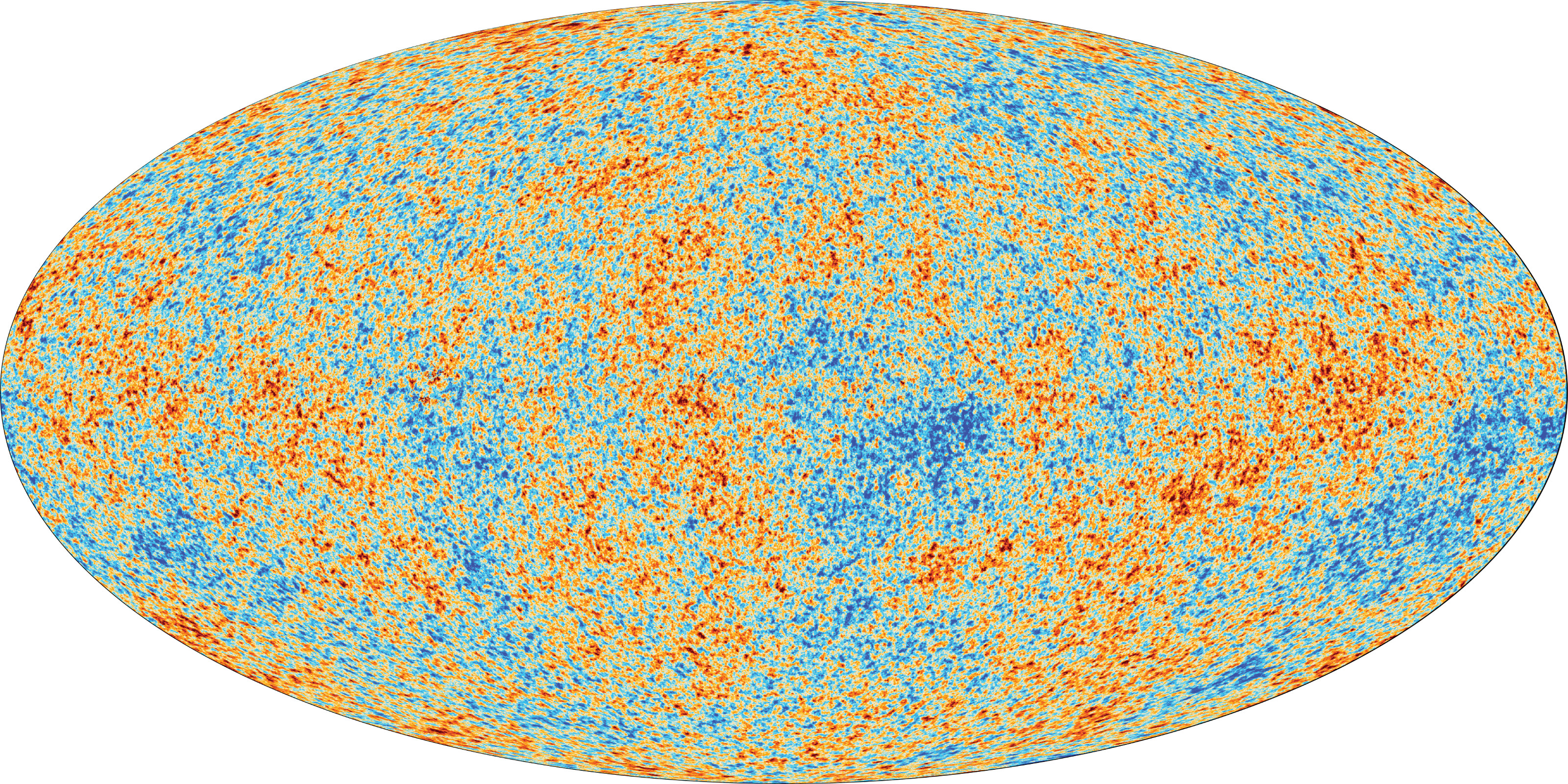}\hfill
  \includegraphics[width=0.49\hsize]{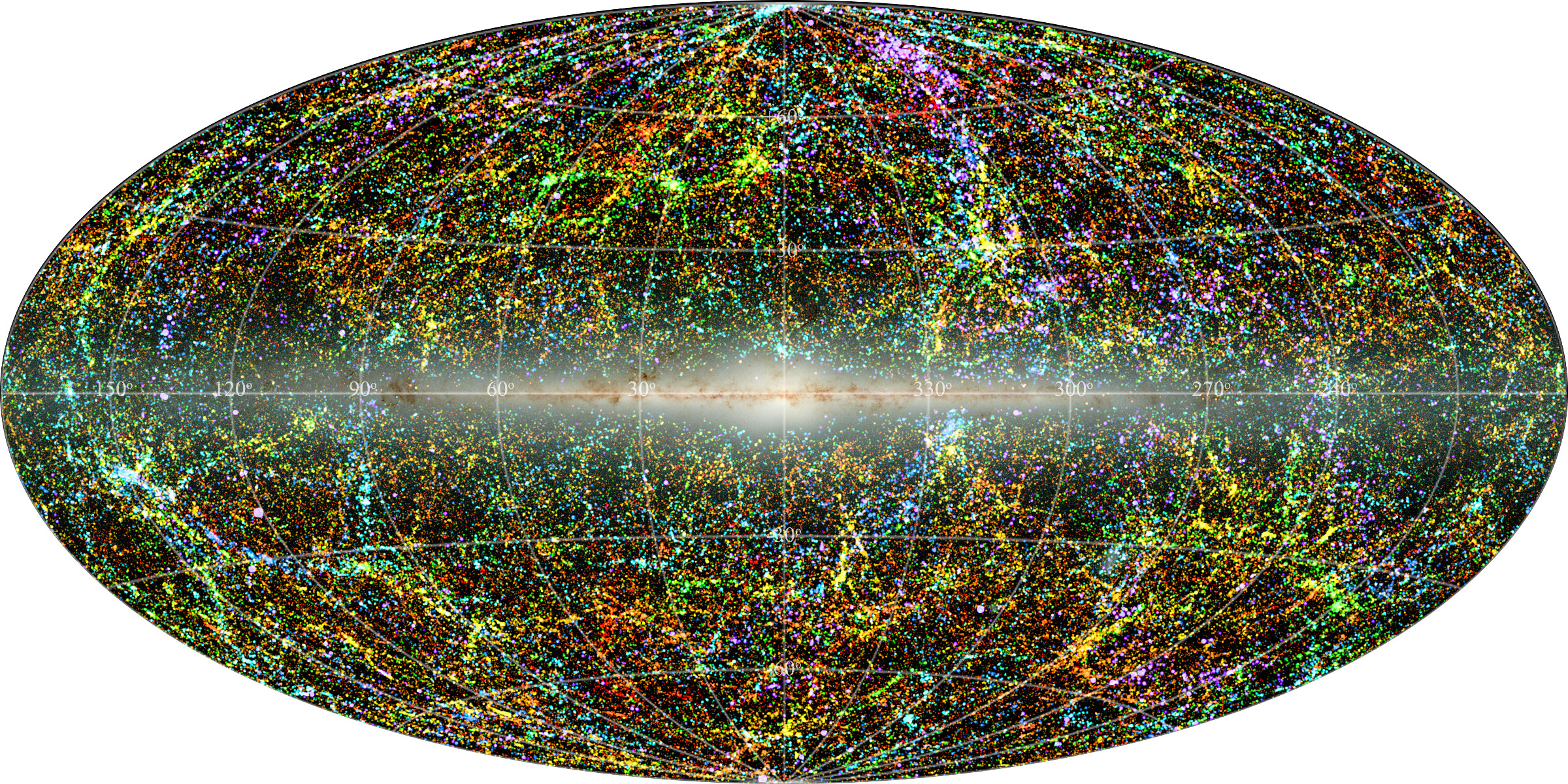}
\caption{Left: Cosmic structures in early cosmic history as revealed by the temperature fluctuations in the cosmic microwave background (Planck 2018). Right: Cosmic structures at the cosmic present as traced by the local galaxy distribution observed by the Two-Micron All-Sky Survey (courtesy of Dr.\ T.\ H.\ Jarrett, IPAC/Caltech).}
\label{fig:1}
\end{figure}

Based on the assumption of dominating dark matter, of cold dark matter in particular, more and more refined and extended numerical simulations have revealed over decades that the morphology and low-order statistical measures of the observed cosmic matter distribution can be well reproduced in detail \cite{1993ApJ...416....1K, 2005Natur.435..629S, 2018MNRAS.475..676S}. Impressively highly resolved simulations have shown furthermore that the radial profiles of the dark-matter density in gravitationally bound objects from a wide range of masses have a self-similar, universal form, and measurements in galaxy clusters have confirmed this profile shape \cite{2005Natur.435..629S, 2017MNRAS.469.1824X, 2018MNRAS.475..676S, 2019MNRAS.484..476C, 2020NatRP...2...42V, 2020Natur.585...39W}.

Cosmic structures are most commonly quantified by their power spectrum, which is the variance of the Fourier modes of their matter-density distribution as a function of the wave number $k$. As long as the relative amplitude of density fluctuations is smaller than unity, their evolution can be well described by the linearized system of the continuity, Euler, and Poisson equations on the expanding cosmic space-time (see e.g.\ \cite{2002PhR...367....1B} for a review). In this linear theory, the Fourier modes of the density field evolve independently and at a rate independent of wave number. The linearly evolved density-fluctuation power spectrum thus has the same shape as it had initially, but an amplitude expected to be approximately $10^6$ times larger than right after the cosmic microwave background was released (see eg. \cite{dodelson2020modern}).

Reconstructions of the linearly evolved power spectrum from many different cosmological measurements confirm that its shape cannot be observationally distinguished from the simplest shape it could be expected to have, growing almost linearly with $k$ at large scales (small $k$), reaching a broad maximum near $k\approx10^{-2}\,h\,\mathrm{Mpc}^{-1}$ and turning over to a decrease approaching an asymptotic fall-off proportional to $k^{-3}$ on small scales (large $k$) \cite{1970PhRvD...1.2726H, 1970ApJ...162..815P, 1972MNRAS.160P...1Z, 1982ApJ...263L...1P}, \cite{2020A&A...641A...1P}. Numerical simulations show that this power spectrum is characteristically deformed by non-linear evolution on small scales: at wave numbers $\gtrsim1\,h\,\mathrm{Mpc}^{-1}$, its amplitude is enhanced by way more than an order of magnitude, and it seems to approach an asymptotic fall-off proportional to $k^{-3}$ on small scales \cite{2016MNRAS.459.1468M, 2021MNRAS.506.2871S}. Measurements of the non-linearly evolved power spectrum via the weak gravitational lensing effect confirm these numerical results on the scales accessible to observation \cite{2020A&A...634A.127A, 2017MNRAS.471.4412K}. Structures with power spectra approaching an asymptotic, logarithmic slope of $-3$ are distinguished because their power, defined as the number of density fluctuations times their variance in Fourier space, becomes scale-independent towards small scales: every decade in scale then adds an equal amount of fluctuation power.

\begin{figure}[t]
  \centering
  \includegraphics[width=0.6\hsize]{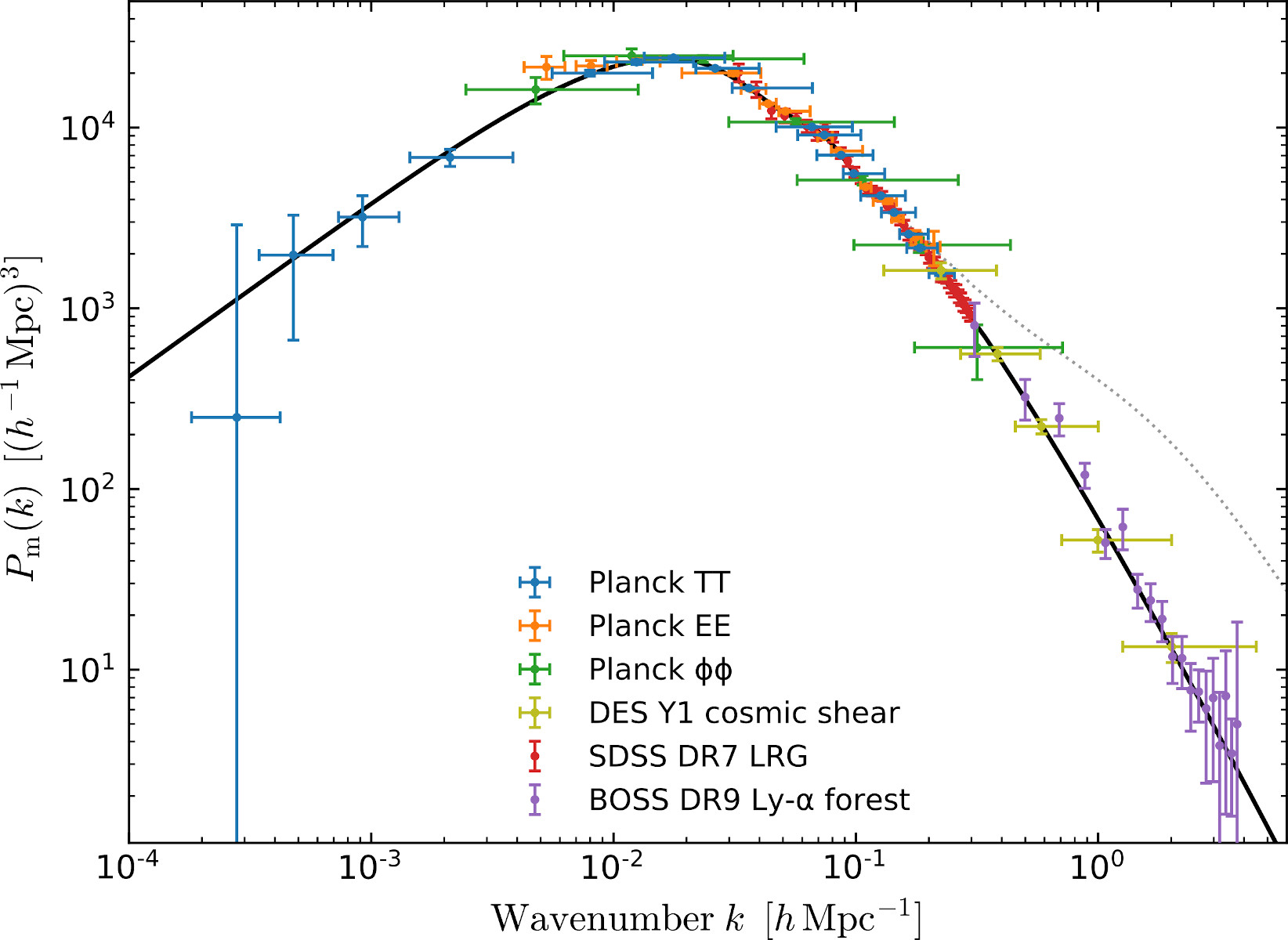}
\caption{Linearly evolved, density-fluctuation power spectrum reconstructed from different classes of cosmological measurements. The dotted line shows the non-linear evolution of the power spectrum modelled by numerical simulations (Planck 2018 I).}
\label{fig:2}
\end{figure}

While numerical simulations reproduce the statistical properties of the cosmic matter distribution very well and thus strongly support the cosmological standard model as well as the hypothesis of cold dark matter, they cannot identify any fundamental reasons for the self-similar density profiles of gravitationally-bound cosmic structures or the scale-independence of fluctuation power on small scales. Understanding the fundamental origin of this kind of universality of cosmic structures is necessary to decide whether it is caused by any specialities of the cosmological model, the assumed properties of dark matter, the functional shape of the gravitational law, or due to any other reason. Testing wide classes of theoretical possibilities with sufficiently detailed numerical simulations seems forbiddingly costly. The inevitable shot noise in numerically simulated matter distributions on small scales caused by the necessarily finite number of simulation particles adds another motivation to search for rigorous statements on the statistics of cosmic structures on small scales (see e.g.\ \cite{2020NatRP...2...42V} for a recent review on cosmic $N$-body simulations).

Analytic approaches to cosmic structure formation exist. They fall into the two main classes of Eulerian \cite{1995ApJ...455....7M, 2001A&A...379....8V, 2004ApJ...612...28M, 2006PhRvD..73f3519C, 2006PhRvD..73f3520C, 2008JCAP...10..036P, 2011JCAP...06..015A, 2012JCAP...12..013A, 2012JCAP...01..019P} and Lagrangian perturbation theory  \cite{1992MNRAS.254..729B, 1993MNRAS.264..375B, 1994MNRAS.267..811B, 1995A&A...296..575B, 1997GReGr..29..733E, 2008PhRvD..77f3530M, 2008PhRvD..78h3503B, 2012JCAP...06..021R, 2013PhRvD..87h3522V}, and enter into the non-linear regime of cosmic of structure formation by perturbative or effective methods \cite{2012JCAP...07..051B, 2014JCAP...05..022P, 2014JCAP...07..057C, 2014PhRvD..89d3521H, 2015JCAP...05..007B, 2018PhRvD..97f3526L, 2019JCAP...11..027K, 2020JCAP...07..011F}. These different approaches suffer from one essential problem: they describe the evolution of cosmic structures in terms of dynamical equations for the cosmic density and velocity fields, assumed to be smooth and differentiable. In this sense, dark matter is modelled as a fluid, based on the ideal or viscous hydrodynamical equations. Once convergent streams of dark-matter paricles cross, however, the velocity field is no longer uniquely valued, and the fluid description becomes inadequate. This is the origin of the notorious shell- or stream-crossing problem.

For this main reason, we choose a different approach here, based on kinetic field theory \cite{2016NJPh...18d3020B, 2017NJPh...19h3001B, 2018JSMTE..04.3214F, 2019AnP...53100446B, 2021JCAP...06..035K, 2021ScPP...10..153B, 2021arXiv211007427K}. This theory is kinetic in the sense of describing the statistical properties of a large number of microsopic particles. These particles are classical, subject to Hamiltonian dynamics, and as an ensemble need not be in any kind of equilibrium. The field that the theory is acting on is the bundle of particle trajectories. The theory differs from conventional approaches to kinetic theory in that it does not assume a smooth phase-space density function of the particles to exist. Its central mathematical object is thus not a dynamical equation for a phase-space density, but a generating functional characterizing the initial statistical properties of the particle ensemble and the time evolution of the particle trajectories. Statistical properties of the ensemble are extracted from the generating functional by functional derivation. Since phase-space trajectories subject to Hamiltonian dynamics cannot cross, the theory avoids the shell-crossing problem by construction. By design, it is formally identical to a statistical quantum-field theory.

We shall focus in this paper on rigorous statements that can be derived on small-scale cosmic structures within the framework of kinetic field theory. Other aspects of the theory have been worked out elsewhere, most noticeable on its relation to conventional kinetic theory or cosmological perturbation theory \cite{2021JCAP...06..035K}, macroscopic reformulations including resummation \cite{2019JCAP...04..001L}, applications to mixtures of dark matter and gas \cite{2019JCAP...05..017G, 2021JCAP...01..046G}, and others. We develop kinetic field theory in Sect.~\ref{sec:2} where we describe in detail how the theory can be adapted to the expanding cosmological background space-time. In Sect.~\ref{sec:3}, we first characterize the statistical properties of the initial state of the particle ensemble and then describe how low-order statistical measures for the evolved particle distribution in configuration and velocity space can be derived. The asymptotic, small-scale behaviour of the density-fluctuation power spectrum, the density-fluctuation bispectrum, and the velocity power spectrum is derived for freely-streaming particles in Sect.~\ref{sec:4}. In this section, we also show how particle interactions can be included in a mean-field approximation, and demonstrate that the asymptotic behaviour of the density-fluctuation power spectrum is unchanged by such interactions. In Sect.~\ref{sec:5}, we summarize our results and present our conclusions.

\section{Kinetic field theory in the cosmological context}
\label{sec:2}

Kinetic field theory studies the evolution of classical particle ensembles in and out of equilibrium. We are studying canonical ensembles of particles here whose motion is described by Hamiltonian mechanics. Their phase-space trajectories $(q_i(t), p_i(t)) =: x_i(t)$, with the index $i = 1,\ldots,N$ enumerating the particles, are completely determined once their initial values $(q^\mathrm{(i)}, p^\mathrm{(i)})$ are given at some point in time. Phase-space trajectories cannot cross: if they did, they would have one point in common at a certain time, which would force them to have the same past and to continue identically since the solutions of the Hamiltonian equations are unique. This is one of the major advantages of kinetic field theory compared to other approaches: it avoids by construction any problems with matter streams crossing in configuration space.

\subsection{Notation}

We shall introduce two essential pieces of notation here: bundles of particle trajectories and low-order statistical measures for density fluctuations.

\subsubsection{Bundles of particle trajectories}

The phase-space trajectories for all $N$ particles of the ensemble together are the fundamental field that kinetic field theory is concerned with. For notational convenience, we denote this field as
\begin{equation}
  \tens x(t) = x_i(t)\otimes\hat e_i\;,
\label{eq:1}
\end{equation} 
where the $\hat e_i$ are the Cartesian unit vectors of $\mathbb{R}^N$. Let now $H(x, t)$ be the Hamiltonian on the single-particle phase space. With the symplectic matrix
\begin{equation}
  M = \matrix{cc}{0 & \id 3 \\ -\id 3 & 0}\;,
\label{eq:2}
\end{equation} 
the Hamiltonian equation for a single trajectory can be written as
\begin{equation}
  \dot x_i(t) = M\partial_{x_i}H(x_i, t)\;.
\label{eq:3}
\end{equation}
Introducing the matrix
\begin{equation}
  \mathcal{M} = M\otimes\id N
\label{eq:4}
\end{equation}
for the entire particle ensemble, further the $N$-particle phase-space gradient
\begin{equation}
  \partial_{\tens x} = \partial_{x_i}\otimes\hat e_i
\label{eq:5}
\end{equation}
and the $N$-particle Hamiltonian $H(x_1,\ldots,x_N, t)$, the equations of motion for the entire trajectory bundle can be compactly written as
\begin{equation}
  \dot{\tens x}(t) = \mathcal{M}\partial_{\tens x}H\;.
\label{eq:6}
\end{equation}

For brevity and convenience, we introduce the short-hand notations
\begin{equation}
  \int_q = \int\D^3q\;,\quad \int_k = \int\frac{\D^3k}{(2\pi)^3}\;,\quad
  \int_{\tens q} = \int\prod_{i=1}^N\D^3q_i\;,\quad
  \int_{\tens k} = \int\prod_{i=1}^N\frac{\D^3k_i}{(2\pi)^3}
\label{eq:7}
\end{equation} 
and fix the Fourier convention by
\begin{equation}
  \tilde f(k) = \int_qf(q)\,\E^{-\I k\cdot q}\;,\quad
  f(q) = \int_k\tilde f(k)\,\E^{\I k\cdot q}\;.
\label{eq:8}
\end{equation}
We assume three spatial dimensions unless explicitly stated otherwise.

\subsubsection{Correlation functions and power spectra}

The cosmic matter density $\rho$ is often written as a mean density $\bar\rho$ times a fluctuation,
\begin{equation}
  \rho(q) = \bar\rho\left(1+\delta(q)\right)\;,
\label{eq:9}
\end{equation} 
where $\delta(q)$ is called the relative density contrast at position $q$. The probability for finding one matter particle in a small volume $\D V$ around position $q_1$ and another one within $\D V$ around position $q_2$ is then given by
\begin{equation}
  P(q_2\vert q_1)P(q_1) = \frac{(\bar\rho\D V)^2}{m^2}\left\langle
    \left(1+\delta(q_1)\right)\left(1+\delta(q_2)\right)
  \right\rangle =
  \frac{(\bar\rho\D V)^2}{m^2}\left(
    1+\xi_\delta(q_1,q_2)
  \right)\;,
\label{eq:10}
\end{equation} 
where $\xi_\delta(q_1,q_2)$ is the correlation function of the density fluctuations $\delta$ between positions $q_1$ and $q_2$. Thus, the correlation function quantifies the conditional probability for finding a particle at $q_2$ given another particle at $q_1$. Due to statistical homogeneity, $\xi_\delta$ may only depend on the separation vector $q_2-q_1$ between the two points, and due to statistical istropy, it may only depend on the absolute value $q=\vert q_2-q_1\vert$,
\begin{equation}
  \xi_\delta(q_1,q_2) = \xi_\delta(q_2-q_1) = \xi_\delta(q) =
  \left\langle
    \delta(q_1)\delta(q_2)
  \right\rangle\;.
\label{eq:11}
\end{equation}
The Fourier transform of the correlation function is the power spectrum,
\begin{equation}
  P_\delta(k) = \int_q\xi_\delta(q)\,\E^{-\I k\cdot q}\;.
\label{eq:12}
\end{equation}

Since the Fourier transform of the density is
\begin{equation}
  \tilde\rho(k) = \bar\rho\int_q\left(1+\delta(q)\right)\,\E^{-\I k\cdot q} =
  \bar\rho\left[
    (2\pi)^3\delta_\mathrm{D}(k)+\tilde\delta(k)
  \right]\;,
\label{eq:13}
\end{equation}
the two-point function of the Fourier-transformed density is related by
\begin{equation}
  \left\langle
    \tilde\rho(k_1)\tilde\rho(k_2)
  \right\rangle = \bar\rho^2\left[
    (2\pi)^6\delta_\mathrm{D}(k_1)\delta_\mathrm{D}(k_2)+\left\langle
      \tilde\delta(k_1)\tilde\delta(k_2)
    \right\rangle
  \right]
\label{eq:14}
\end{equation}
to the two-point function of the Fourier-transformed density contrast. The latter is related to the power spectrum by
\begin{equation}
  \left\langle
    \tilde\delta(k_1)\tilde\delta(k_2)
  \right\rangle = (2\pi)^3\delta_\mathrm{D}(k_1+k_2)P_\delta(k_1)\;,
\label{eq:15}
\end{equation}
we can write (\ref{eq:14}) as
\begin{equation}
  \frac{1}{\bar\rho^2}\left\langle
    \tilde\rho(k_1)\tilde\rho(k_2)
  \right\rangle = (2\pi)^3\delta_\mathrm{D}(k_1+k_2)\left[
    (2\pi)^3\delta_\mathrm{D}(k_1)+P_\delta(k_1)
  \right]\;.
\label{eq:16}
\end{equation}
We shall return to this equation later in (\ref{eq:110}). On the other hand, the connected part of the two-point function of the Fourier-transformed density, i.e.\ the two-point density cumulant in Fourier space, is
\begin{align}
  \left\langle
    \tilde\rho(k_1)\tilde\rho(k_2)
  \right\rangle_\mathrm{c} &= \left\langle
    \tilde\rho(k_1)\tilde\rho(k_2)
  \right\rangle-
  \left\langle\tilde\rho(k_1)\right\rangle
  \left\langle\tilde\rho(k_2)\right\rangle \nonumber\\ &=
  (2\pi)^3\delta_\mathrm{D}(k_1+k_2)P_\delta(k_1)
\label{eq:17}
\end{align}
by combining (\ref{eq:14}) with (\ref{eq:13}) and (\ref{eq:15}).

\subsection{Generating functional for a classical particle ensemble}

\begin{figure}[t]
  \centering
  \includegraphics[width=0.6\hsize]{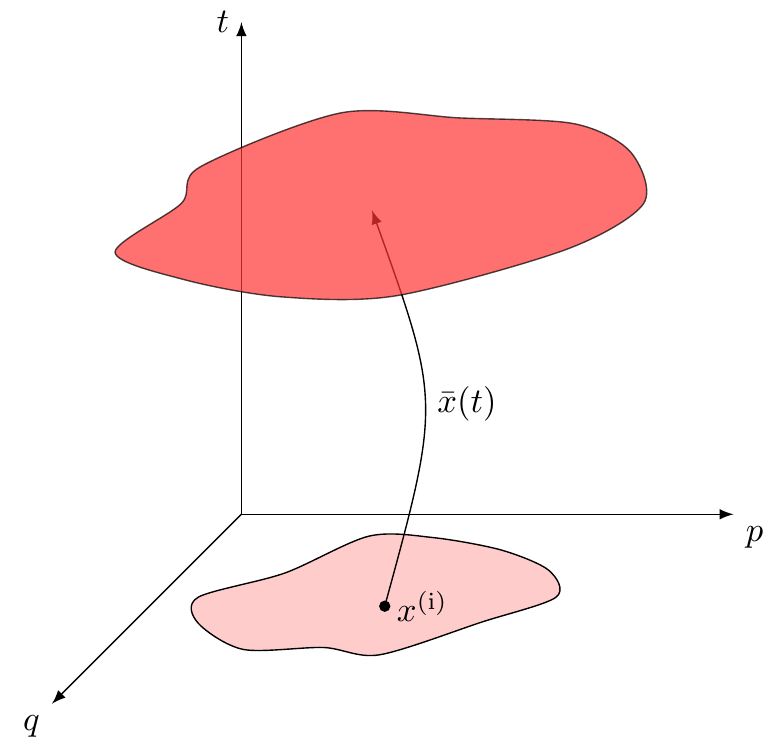}
\caption{Phase space is shown schematically as a function of time. Particle trajectories originating at an initial position transport the initial probability forward in time.}
\label{fig:3}
\end{figure}

A bundle of phase-space trajectories $\tens x(t)$ beginning at the initial phase-space points $\tens x^\mathrm{(i)}$ follows the classical Hamiltonian flow $\phi_\mathrm{cl}(\tens x^\mathrm{(i)}, t)$. The probability to find the particles at the phase-space points $\tens x(t)$ at time $t$ is
\begin{equation}
  P(\tens x(t)) = \int\D\tens x^\mathrm{(i)}\,
  P\left(\tens x(t)\vert\tens x^\mathrm{(i)}\right)
  P\left(\tens x^\mathrm{(i)}\right)\;,
\label{eq:18}
\end{equation} 
where $P(\tens x^\mathrm{(i)})$ is the probability for the initial particle positions $\tens x^\mathrm{(i)}$ to be occupied, and $P(\tens x(t)\vert\tens x^\mathrm{(i)})$ is the transition probability for the particle ensemble from there to $\tens x(t)$. Since the particles are classical and follow deterministic trajectories, this transition probability is a functional Dirac delta distribution,
\begin{equation}
  P\left(\tens x(t)\vert\tens x^\mathrm{(i)}\right) = \delta_\mathrm{D}\left[
    \tens x(t)-\phi_\mathrm{cl}\left(\tens x^\mathrm{(i)}, t\right)
  \right]\;.
\label{eq:19}
\end{equation}
Integrating $P(\tens x(t))$ over the trajectory bundle $\tens x(t)$ and introducing a generator field $\tens J(t)$, we arrive at the generating functional
\begin{equation}
  Z[\tens J] = \int\D\tens x^\mathrm{(i)}P\left(\tens x^\mathrm{(i)}\right)\,\int\mathcal{D}\tens x\,\delta_\mathrm{D}\left[
    \tens x(t)-\phi_\mathrm{cl}\left(\tens x^\mathrm{(i)}, t\right)
  \right]\E^{\I(\tens J,\tens x)}\;,
\label{eq:20}
\end{equation} 
where the parentheses in the exponent denote a suitably defined scalar product between $\tens J$ and $\tens x$ including a time integral,
\begin{equation}
  \left(\tens J,\tens x\right) = \left(
    J_i\otimes\hat e_i,x_j\otimes\hat e_j
  \right) = \delta_{ij}\int_0^t\D t'\,J_i(t')\cdot x_j(t') =
  \int_0^t\D t'\,J_i(t')\cdot x_i(t')\;.
\label{eq:21}
\end{equation} 
Evaluating the path integral in (\ref{eq:20}) over all possible trajectory bundles $\tens x(t)$, the delta distribution selects the classical solution
\begin{equation}
  \bar{\tens x}(t) = \phi_\mathrm{cl}\left(\tens x^\mathrm{(i)}, t\right)
\label{eq:22}
\end{equation} 
of the Hamiltonian equations (\ref{eq:6}). The generating functional for the entire particle ensemble thus becomes
\begin{equation}
  Z[\tens J] = \int\D\Gamma\,\E^{\I(\tens J, \bar{\tens x})}\;,
\label{eq:23}
\end{equation} 
with the integral measure $\D\Gamma$ on the initial phase space,
\begin{equation}
  \D\Gamma = P\left(\tens x^\mathrm{(i)}\right)\D\tens x^\mathrm{(i)} =
  P\left(\tens x^\mathrm{(i)}\right)
  \prod_{i=1}^N\D x_i^\mathrm{(i)}\;.
\label{eq:24}
\end{equation} 
The generating functional $Z[\tens J]$ in (\ref{eq:23}) is already the central mathematical object of kinetic field theory. By functional derivatives with respect to the generator field $\tens J$, then setting $\tens J = 0$, statistical information on the particle ensemble at any time $t$ can be extracted from $Z[\tens J]$ \cite{2016NJPh...18d3020B, 2019AnP...53100446B}. Aiming at statistical information on the particle ensemble, we are not solving dynamical equations for density and velocity fields assumed to be sufficiently smooth. Instead, the time evolution of the particle trajectories implies the time evolution of the generating functional, from which statistical information on the particle distribution in phase-space can be extracted at any required time. Studying particle trajectories instead of smooth fields avoids the notorious shell-crossing problem, which arises in other approaches when a velocity field assumed to be smooth ceases to be uniquely valued after matter streams cross.

It should be noted that the generating functional $Z[\tens J]$ contains the exact information on the time evolution of the entire particle ensemble if we insert the exact particle trajectories $\bar{\tens x}(t)$ into (\ref{eq:23}). Approximation schemes for the trajectories will be introduced as we go along to arrive at tractable expressions. We will now turn to describing particle trajectories with suitably chosen Green's functions.

\subsection{Particle trajectories}

Beginning with the Lagrange function of point particles of equal mass in the expanding cosmic background, we shall derive Green's functions here solving the equations of motion and thus characterizing the particle trajectories through phase space. We shall pay particular attention to choosing appropriate reference trajectories and deriving the force acting between two particles relative to these trajectories (see eg. \cite{2015PhRvD..91h3524B, 2019AnP...53100446B}).

\subsubsection{Lagrange function, transformation to comoving coordinates}

The trajectories of particles with mass $\tilde m$ on the expanding cosmic space-time are determined by their Lagrange function
\begin{equation}
  \tilde L = \frac{\tilde m}{2}\vec r^{\,2}-\tilde m\Phi\;,
\label{eq:25}
\end{equation} 
where the gravitational potential $\Phi$ satisfies the Poisson equation
\begin{equation}
  \nabla_r^2\Phi = 4\pi G\rho-\Lambda
\label{eq:26}
\end{equation}
with the matter density $\rho$ and the cosmological constant $\Lambda$ \cite{peebles1980large}. In terms of the comoving spatial coordinate $\vec q$, the physical spatial coordinate $\vec r$ is $\vec r = a\vec q$. During the matter-dominated epoch, the scale factor $a$ obeys Friedmann's equation
\begin{equation}
  \frac{\ddot a}{a} = -\frac{4\pi G}{3}\bar\rho+\frac{\Lambda}{3}
\label{eq:27}
\end{equation}
containing the mean matter density $\bar\rho$, which is a function of time only.

Transforming to comoving coordinates, subjecting the Lagrangian to the gauge transformation
\begin{equation}
  \tilde L \mapsto \tilde L-\frac{\D f}{\D t}\quad\mbox{with}\quad
  f = \frac{\tilde m}{2}a\dot a\vec q^{\,2}
\label{eq:28}
\end{equation} 
and dropping the particle mass $\tilde m$ leads to the Lagrangian
\begin{equation}
  \tilde L = \frac{a^2}{2}\dot{\vec q}^{\,2}-\phi\;,
\label{eq:29}
\end{equation}
with the gravitational potential $\phi$ now satisfying the comoving Poisson equation
\begin{equation}
  \nabla_q^2\phi = 4\pi Ga^2\bar\rho\delta
\label{eq:30}
\end{equation}
containing the density contrast
\begin{equation}
  \delta = \frac{\rho}{\bar\rho}-1\;.
\label{eq:31}
\end{equation}

We now introduce the convenient time coordinate
\begin{equation}
  \tau = H_\mathrm{i}^{-1}\left[D_+(t)-1\right]
\label{eq:32}
\end{equation}
instead of the cosmic time $t$, and $H_\mathrm{i}$ is the Hubble constant at some initial time $t_\mathrm{i}$. The growth factor $D_+(t)$ for cosmic density fluctuations is the growing solution of the linearized growth equation (\ref{eq:40}) below.

We set $t_\mathrm{i}$ early in the matter-dominated era, e.g.\ right after the cosmic microwave background has decoupled. We normalize the growth factor to unity initially such that $\tau = 0$ at $t=t_\mathrm{i}$.

The inverse Hubble constant at the initial time, $H_\mathrm{i}^{-1}$, sets an appropriate time scale. We express the mean density $\bar\rho$ by the critical density and the density parameter $\Omega_\mathrm{i}$ at the initial time,
\begin{equation}
  \bar\rho = \frac{3H_\mathrm{i}^2}{8\pi G}\,\Omega_\mathrm{i}\,a^{-3}\;,
\label{eq:33}
\end{equation} 
where the scale factor is also normalized to unity at the initial time $t_\mathrm{i}$. Early in the matter-dominated phase, we can safely replace the density parameter by unity, $\Omega_\mathrm{i}=1$. If we finally drop the common factor $H_\mathrm{i}^2$ from both terms of the Lagrange function $\tilde L$, we find the equivalent Lagrange function
\begin{equation}
  L = \frac{m}{2}\dot{\vec q}^{\,2}-m\varphi\;.
\label{eq:34}
\end{equation} 
The dot now and in the following represents the derivative with respect to the time $\tau$, the potential $\varphi$ satisfies the Poisson equation
\begin{equation}
  \nabla^2\varphi = A_\varphi\delta \quad\mbox{with}\quad
  A_\varphi = \frac{3a}{2m^2}\;,
\label{eq:35}
\end{equation}
and $m$ is the effective, dimension-less, but time-dependent particle mass
\begin{equation}
  m = a^2\frac{\D\tau}{\D t}\;.
\label{eq:36}
\end{equation}
Note that the time-dependence of the effective particle mass $m$ is a mathematically exact and physically equivalent consequence of introducing comoving coordinates. In physical coordinates, particles are diluted by cosmic expansion such that the gravitational acceleration between them decreases over time. In comoving coordinates, their mean separation remains constant, and the decreasing gravitational acceleration is mapped to their increasing mass.

\begin{figure}[t]
  \centering
  \includegraphics[width=\hsize]{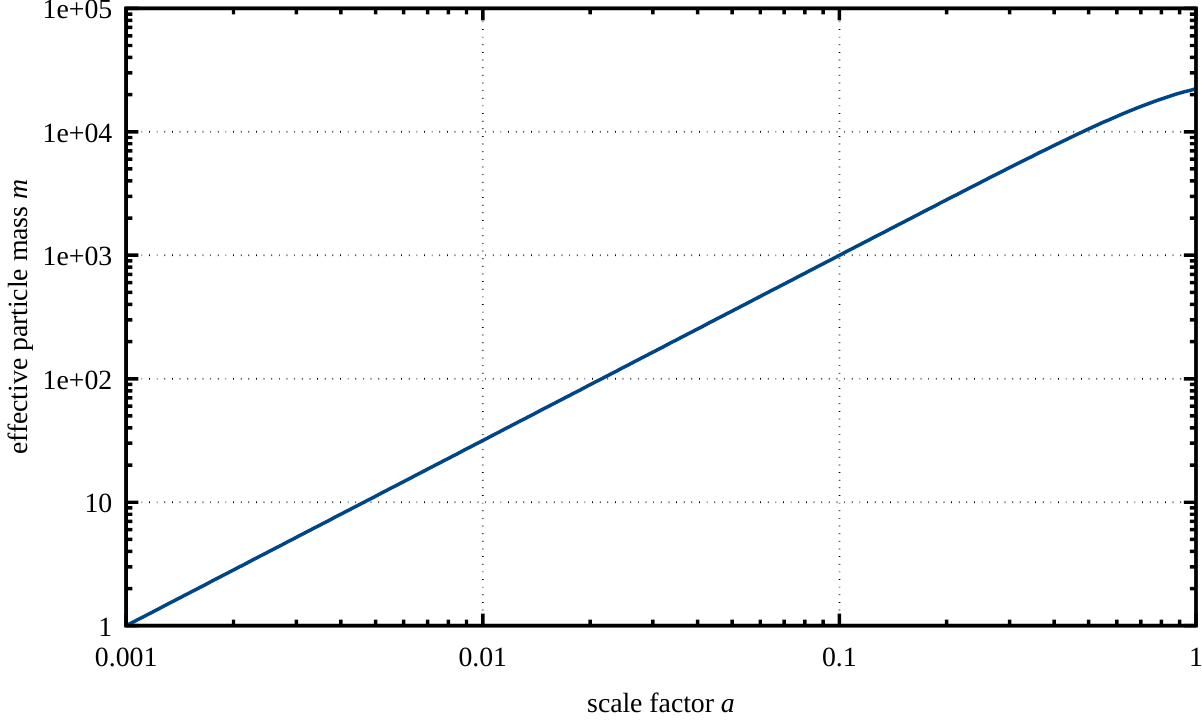}
\caption{Effective particle mass $m$ as a function of the cosmic scale factor $a$ (see \ref{eq:36}).}
\label{fig:4}
\end{figure}

We will need the time derivative of the mass later, i.e.\ the derivative of $m$ with respect to the growth factor,
\begin{equation}
  \dot m = \frac{\D m}{\D D_+} = \frac{m'}{D_+'}\;,
\label{eq:37}
\end{equation} 
where the prime indicates the derivative with respect to the scale factor $a$. We insert (\ref{eq:36}) into (\ref{eq:37}) and use
\begin{equation}
  \frac{\D\tau}{\D t} = \frac{\D D_+}{\D a}\dot a = D_+'aH\;,
\label{eq:38}
\end{equation} 
where $H=\dot a/a$ is the Hubble function. Measuring time in units of the initial Hubble time $H_\mathrm{i}^{-1}$, we replace $H$ by the expansion function $E = H/H_\mathrm{i}$. Then,
\begin{equation}
  m' = \frac{\D}{\D a}\left(a^3D_+'E\right) =
  a^3E\left[D_+''+\left(\frac{3}{a}+\frac{E'}{E}\right)D_+'\right]\;,
\label{eq:39}
\end{equation} 
(see eg. \cite{2021ScPP...10..153B}).
The growth factor itself satisfies the linear growth equation,
\begin{equation}
  D_+''+\left(\frac{3}{a}+\frac{E'}{E}\right)D_+' =
  \frac{3}{2}\frac{\Omega_\mathrm{i}}{a^5E^2}D_+\;,
\label{eq:40}
\end{equation}
which simplifies the right-hand side of (\ref{eq:39}). Early in the matter-dominated era, $\Omega_\mathrm{i}\approx1$ to excellent approximation. Combining this with (\ref{eq:35})--(\ref{eq:40}), we can bring $\dot m$ into the simple form
\begin{equation}
  \dot m = \frac{3}{2}\frac{D_+}{D_+'a^2E} = \frac{3}{2}\frac{aD_+}{m} =
  mA_\varphi D_+\;,
\label{eq:41}
\end{equation}
with $A_\varphi$ specified in (\ref{eq:35}). From now on, we shall write $t$ instead of $\tau$, expressing all times by the linear growth factor as defined in (\ref{eq:32}).

\subsubsection{Hamiltonian Green's function}

The Lagrange function $L$ from (\ref{eq:34}) implies the canonical momentum $\vec p = \dot{\vec q}/m$ with the dimension-less, time-dependent mass $m$. Due to our normalizing both the scale factor $a$ and the growth factor $D_+$ to unity at the initial time, the effective particle mass $m$ is also unity initially and increases from there. The Hamilton function
\begin{equation}
  H = \frac{\vec p^{\,2}}{2m}+m\varphi = H_0+H_\mathrm{I}
\label{eq:42}
\end{equation} 
splits into a free, kinetic part $H_0 = \vec p^{\,2}/(2m)$ and an interacting part $H_\mathrm{I} = m\varphi$ (see eg. \cite{2015PhRvD..91h3524B, 2016NJPh...18d3020B, 2019AnP...53100446B, 2021ScPP...10..153B}). The free part implies the equation of motion
\begin{equation}
  \dot x = M\partial_xH_0 = A(t)x \quad\mbox{with}\quad
  A(t) = \matrix{cc}{0 & m^{-1} \\ 0 & 0}\;,
\label{eq:43}
\end{equation}
which is solved by the exponential
\begin{equation}
  x(t) = \E^{\bar A(t,0)} x_0 \quad\mbox{with}\quad
  \bar A(t,t') = \int_{t'}^t\D\bar t\,A(\bar t)\;.
\label{eq:44}
\end{equation}
Since $\bar A(t,t')$ is nilpotent, $\bar A^2 = 0$, the solution (\ref{eq:44}) simplifies to
\begin{equation}
  x(t) = \left[\id 3+\bar A(t,0)\right]x_0\;.
\label{eq:45}
\end{equation}
Varying the constant $x_0$ leads to the solution
\begin{equation}
  x(t) = \left[\id 3+\bar A(t,0)\right]x^\mathrm{(i)}-
  \int_0^t\D t'\,\left[\id 3+\bar A(t,t')\right]
  \cvector{0 \\ m\nabla\varphi}
\label{eq:46}
\end{equation}
for the phase-space trajectory or
\begin{equation}
  q(t) = q^\mathrm{(i)}+g_\mathrm{H}(t,0)p^\mathrm{(i)}-
  \int_0^t\D t'\,g_\mathrm{H}(t,t')m\nabla\varphi
\label{eq:47}
\end{equation} 
for the trajectory in configuration space, where
\begin{equation}
  g_\mathrm{H}(t,t') = \int_{t'}^t\frac{\D\bar t}{m}
\label{eq:48}
\end{equation}
is what we call the Hamiltonian propagator \cite{2015PhRvD..91h3524B}.

\begin{figure}[t]
  \centering
  \includegraphics[width=\hsize]{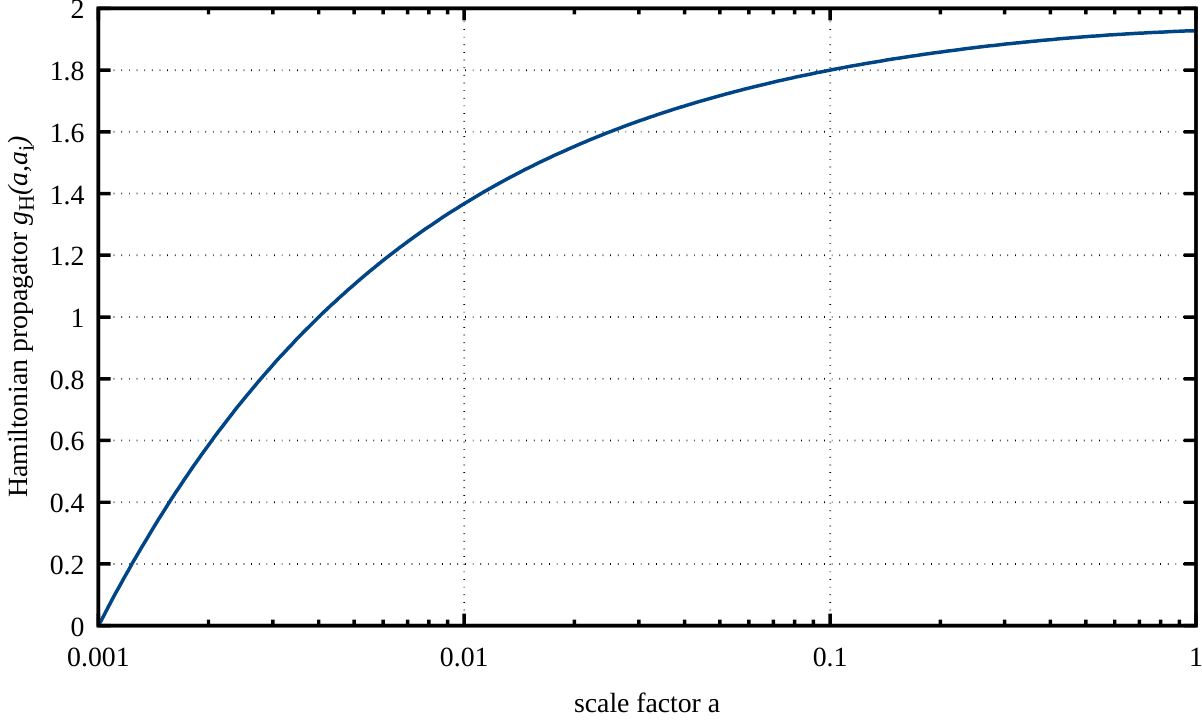}
\caption{Hamiltonian propagator $g_\mathrm{H}$ as defined in (\ref{eq:48}), shown here as a function of the cosmic scale factor with $a_i = 0.001$. The propagator is bounded from above because of cosmic expansion.}
\label{fig:5}
\end{figure}

\subsubsection{Motion relative to Zel'dovich trajectories}

It is convenient in cosmology to replace this Hamiltonian propagator $g_\mathrm{H}(t,t')$ by the time difference $t-t'$, which we have chosen to be the difference between linear growth factors (see eg. \cite{2015PhRvD..91h3524B}). To achieve this, we re-write the trajectory (\ref{eq:47}) as
\begin{equation}
  q(t) = q^\mathrm{(i)}+tp^\mathrm{(i)}-
  \int_0^t\D t'\left\{
    g_\mathrm{H}(t,t')m\nabla\varphi+\left[
      1-\dot g_\mathrm{H}(t',0)
    \right]
  \right\}p^\mathrm{(i)}\;,
\label{eq:49}
\end{equation} 
where the dot denotes the time derivative with respect to the first time argument of the Hamiltonian propagator, and define a function $A_\mathrm{p}(t)$ implicitly by
\begin{equation}
  \int_0^t\D t'\,\left[
    1-\dot g_\mathrm{H}(t',0)
  \right] \stackrel{!}{=}
  \int_0^t\D t'\,g_\mathrm{H}(t,t')A_\mathrm{p}(t')\;.
\label{eq:50}
\end{equation}
Differentiating (\ref{eq:50}) twice with respect to $t$ gives $A_\mathrm{p} = \dot m = mA_\varphi D_+$; cf.\ (\ref{eq:41}) and the definition of $g_\mathrm{H}$ in (\ref{eq:48}).

We replace the potential $\varphi$ by the shifted potential
\begin{equation}
  \phi = \varphi+A_\varphi D_+\psi\;,
\label{eq:51}
\end{equation}
where $\psi$ is an initial velocity potential defined to satisfy
\begin{equation}
  \nabla\psi = p^\mathrm{(i)}\;.
\label{eq:52}
\end{equation}
By assuming a scalar potential for the initial particle velocities, we neglect any initial vortical flows, which does however not mean that the flow remains non-vortical in the course of the further evolution. Since initial velocities and density fluctuations have to satisfy the continuity equation, the velocity potential must be related to the initial density fluctuations $\delta^\mathrm{(i)}$ by the Poisson equation
\begin{equation}
  \nabla^2\psi = -\delta^\mathrm{(i)}\;.
\label{eq:53}
\end{equation} 

Considering (\ref{eq:50}) and replacing $\varphi$ by $\phi$, we can bring the trajectory (\ref{eq:49}) into the form
\begin{equation}
  q(t) = q^\mathrm{(i)}+tp^\mathrm{(i)}-
  \int_0^t\D t'\,g_\mathrm{H}(t,t')m\nabla\phi\;,
\label{eq:54}
\end{equation} 
with the potential $\phi$ satisfying the Poisson equation
\begin{equation}
  \nabla^2\phi = \nabla^2\varphi+A_\varphi D_+\nabla^2\psi =
  A_\varphi\left(\delta-\delta^\mathrm{(lin)}\right)\;,
\label{eq:55}
\end{equation} 
where $\delta^\mathrm{(lin)} = D_+\delta^\mathrm{(i)}$ is the linearly evolving density contrast.

This is an important result in our context. The potential $\phi$ is sourced by the difference between the actual density contrast $\delta$ and its linearly evolving representative $\delta^\mathrm{(lin)}$, i.e.\ $\phi$ is the potential created by the non-linear part of the density contrast only. Initially, therefore, $\phi = 0$ and the particles follow the inertial trajectories
\begin{equation}
  q(t) \approx q^\mathrm{(i)}+tp^\mathrm{(i)}\;,
\label{eq:56}
\end{equation}
representing the Zel'dovich approximation \cite{1970A&A.....5...84Z}. As the density contrast develops a non-linear contribution at later times and on small scales, particles are deflected from the Zel'dovich trajectories. This is the essential reason for introducing the Zel'dovich trajectories (\ref{eq:56}) as reference trajectories here: the force with respect to these trajectories initially vanishes, and it only builds up as density fluctuations become non-linear.

\begin{figure}[t]
  \includegraphics[width=0.49\hsize]{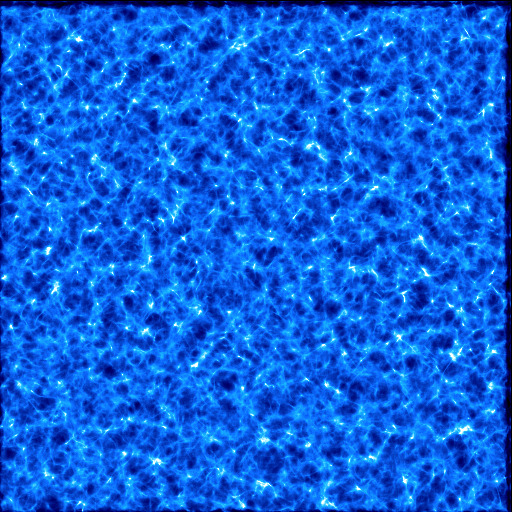}
  \includegraphics[width=0.49\hsize]{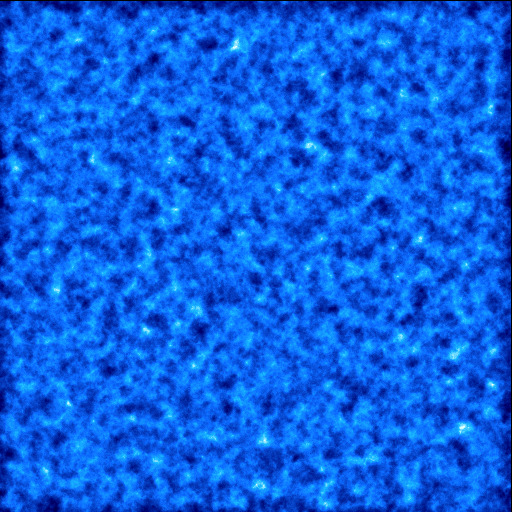}
\caption{Illustration of the Zel'dovich approximation, e.g.\ simulated density field at an early stage of the evolution (left) and a later stage (right).}
\label{fig:6}
\end{figure}

We can finally replace the Hamiltonian propagator $g_\mathrm{H}$ under the integral in (\ref{eq:54}) by implicitly defining an effective force $f(t)$ acting relative to the fiducial, Zel'dovich trajectories (\ref{eq:56}) (see eg. \cite{2021ScPP...10..153B}). Then,
\begin{equation}
  \int_0^t\D t'\,g_\mathrm{H}(t,t')m\nabla\phi \stackrel{!}{=}
  -\int_0^t\D t'\,\left(t-t'\right)f(t')\;.
\label{eq:57}
\end{equation} 
Differentiating this equation twice with respect to $t$ results in the effective force
\begin{equation}
  f(t) = -m\nabla\phi+\frac{\dot m}{m^2}\int_0^t\D t'\,m\nabla\phi\;.
\label{eq:58}
\end{equation}
With the Green's function
\begin{equation}
  G(t,t') = \matrix{cc}{\id 3 & (t-t')\id 3 \\ 0 & \id 3}\;,
\label{eq:59}
\end{equation}
the full phase-space trajectory can be written as
\begin{equation}
  x(t) = G(t,0)x^\mathrm{(i)}+
  \int_0^t\D t'\,G(t,t')\cvector{0 \\ f(t')}\;.
\label{eq:60}
\end{equation}

\subsubsection{Effective interaction potential}

The Poisson equation (\ref{eq:55}) for the potential is interesting in its own right. In Fourier space, it reads
\begin{equation}
  \tilde\phi = -\frac{A_\varphi}{k^2}
  \left(\tilde\delta-\tilde\delta^\mathrm{(lin)}\right)\;.
\label{eq:61}
\end{equation}
At the same time, $\phi$ is the convolution $\delta\bar n\ast v$ of the fluctuation $\delta\bar n$ of the mean particle-number density $\bar n$ and the one-particle potential $v$, thus
\begin{equation}
  \tilde\phi = \bar n\tilde\delta\tilde v
\label{eq:62}
\end{equation} 
according to the convolution theorem. The effective one-particle potential is then determined by
\begin{equation}
  \tilde v\tilde\delta = -\frac{A_\varphi}{\bar nk^2}\left(
    \tilde\delta-\tilde\delta^\mathrm{(lin)}
  \right)\;.
\label{eq:63}
\end{equation} 
Multiplying this equation once with $\tilde\delta$ and once with $\tilde\delta^\mathrm{(lin)}$, taking the ensemble average and eliminating $\langle\tilde\delta\tilde\delta^\mathrm{(lin)}\rangle$ between the resulting two equations gives
\begin{equation}
  \tilde v = -\frac{A_\varphi}{\bar nk^2}\left(
    1-\sqrt{\frac{P_\delta^\mathrm{(lin)}}{P_\delta}}
  \right)\;.
\label{eq:64}
\end{equation}
At large wave numbers, $P_\delta^\mathrm{(lin)}\ll P_\delta$, and the one-particle potential approaches the expected Newtonian form $\propto k^{-2}$. At small wave numbers, $P_\delta^\mathrm{(lin)} = P_\delta$ and the potential drops to zero. Relative to the Zel'dovich trajectories (\ref{eq:56}), the potential thus acquires a large-scale cutoff. In good approximation, its shape resembles the Yukawa form
\begin{equation}
  \tilde v = -\frac{A_\varphi}{\bar n(k_0^2+k^2)}
\label{eq:65}
\end{equation} 
with a time-dependent cutoff scale $k_0$ delineating the regimes of linear and non-linear structure formation \cite{2021ScPP...10..153B}.

\begin{figure}[t]
  \centering
  \includegraphics[width=\hsize]{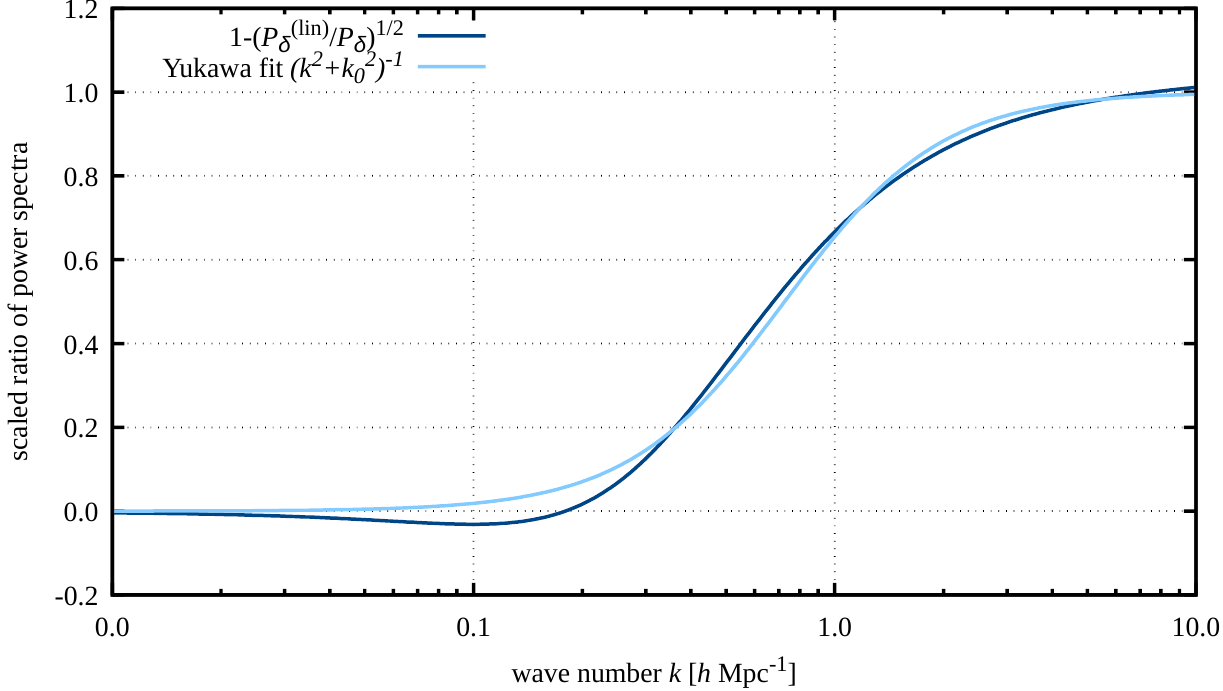}
\caption{Yukawa fit function (light blue line) from (\ref{eq:65}) compared to the scaled ratio of the power spectra (dark blue line), (\ref{eq:64}).}
\label{fig:7}
\end{figure}

\subsection{Remarks concluding this section}

We finish this subsection by five remarks which appear important here also in view of wrong statements frequently repeated about the kinetic field theory of cosmic structure formation.

\begin{enumerate}
  \item For completeness, we repeat that the time-dependent, effective particle mass $m$ is due to the exact transformation from physical to comoving coordinates. It is not at all an approximation.
  \item The trajectories (\ref{eq:60}) are not approximate either, but exact. By a sequence of transformations, we have introduced inertial trajectories with respect to a time coordinate $t$ given by the linear growth factor as introduced in (\ref{eq:32}). This time coordinate is non-uniform in cosmic time. These inertial, or Zel'dovich, trajectories are chosen to incorporate part of the gravitational interaction between the particles, as quantified by the Poisson equation (\ref{eq:56}) and the definition (\ref{eq:58}) of the effective force. Despite their simple form, the Zel'dovich trajectories are subject to gravity by linear density fluctuations because the initial particle momenta are correlated with the initial density fluctuations via the velocity potential $\psi$: where particles are overdense, their flow converges.
  \item The trajectories (\ref{eq:60}) show that the kinetic field theory of cosmic structure formation goes beyond the Zel'dovich approximation inasmuch as the effective force is taken into account. We shall show later how the gravitational interaction relative to Zel'dovich trajectories can be incorporated in a mean-field approach.
  \item Introducing reference trajectories of the Zel'dovich form and an effective force relative to them is nothing mysterious or unusual, and not at all any limitation of the kinetic field theory of structure formation. Newton's axioms introduce reference trajectories defined by inertial motion in coordinate time, and forces as reasons for deviations from inertial motion. The equivalence principle introduces the trajectories of freely-falling particles as a reference, in consequence of which gravitational force is transformed away altogether, and gravitational interaction is reduced to the gravitational tidal field. This illustrates that reference trajectories, and thus Green's functions, can be chosen at will if forces acting relative to them are suitably adapted. The discussion of the effective gravitational potential above shows that the interaction relative to the Zel'dovich trajectories has not an infinite range any more.
  \item Deviations from the Zel'dovich trajectories are exactly quantified by the time integral over the force term in (\ref{eq:61}). Since this force is sourced by non-linear density fluctuations only, it is initially small even at small scales, and remains so at large scales. The integral in (\ref{eq:61}) thus precisely determines a quantity suitable for a perturbative approach, if perturbation theory is what is wanted. Other approaches, such as the mean-field approximation described below, are possible, less tedious, and more efficient.
\end{enumerate}

\section{Statistics of the particle ensemble}
\label{sec:3}

Having set up the generating functional of kinetic field theory including the particle trajectories, we shall proceed in this section by characterizing the initial state of the particle ensemble in the cosmological situation, and by describing generally how statistical information on the evolved particle ensemble can be extracted from the generating functional.

\subsection{The initial state}

Since this section is about the initial particle configuration in phase space only, we drop here the superscript $(\mathrm{i})$ on the initial density contrast $\delta^\mathrm{(i)}$ and the initial phase-space positions $\tens x^\mathrm{(i)} =(\tens q^\mathrm{(i)},\tens p^\mathrm{(i)})$, understanding that $\tens x = (\tens q,\tens p)$ are initial positions. The set of the initial phase-space coordinates for all particles is what we call the initial state of the ensemble. Two assumptions are crucial for this initial state: initial velocities are non-vortical, and the initial velocity potential is a Gaussian random field \cite{2020A&A...641A...9P}.

\subsubsection{Probability distribution of initial phase-space positions}

As discussed before, we assume that an initial velocity field exists which is the gradient $\nabla\psi$ of a velocity potential $\psi$. By continuity, the initial density contrast $\delta$ must then be the negative Laplacian of this potential, $\delta = -\nabla^2\psi$; see (\ref{eq:52}) and (\ref{eq:53}). This initial density field is now sampled by $N$ point particles, placed by a Markov process such that the probability $P(\tens q\vert\delta)$ of finding particles at positions $\tens q=q_i\otimes\hat e_i$ in the density field characterized by the density contrast $\delta$ is
\begin{equation}
  P(\tens q\vert\delta) = \frac{1}{V^N}\prod_{i=1}^N\left[
    1+\delta(q_i)
  \right]\;,
\label{eq:66}
\end{equation}
(see eg. \cite{2016NJPh...18d3020B}).
Given the joint distribution $P(\tens\delta,\tens p)$ of the density-contrast values $\tens\delta=\delta(q_i)\otimes\hat e_i$ at the particle positions and the momenta $\tens p$, the probability $P(\tens q,\tens p)$ of the initial phase-space positions for the particle ensemble is
\begin{equation}
  P(\tens q,\tens p) = \int\D\tens\delta\,
  P(\tens q\vert\tens\delta)\,P(\tens\delta,\tens p)\;.
\label{eq:67}
\end{equation}

Assuming that $\psi$ is a Gaussian random field, the joint probability $P(\tens\delta,\tens p)$ is a multi-variate Gaussian. Its characteristic function
\begin{equation}
  \Phi(\tens r,\tens s) = \left\langle
    \E^{-\I(\tens r\cdot\tens\delta+\tens s\cdot\tens p)}
  \right\rangle
\label{eq:68}
\end{equation}
is then given by
\begin{equation}
  \Phi(\tens r,\tens s) = \exp\left(
    -\frac{1}{2}\tens k^\top C\tens k
  \right) \quad\mbox{with}\quad
  \tens k = \cvector{r_i \\ s_i}\otimes\hat e_i\;,
\label{eq:69}
\end{equation}
characterized by the covariance matrix
\begin{equation}
  C = \left\langle\tens d\otimes\tens d\right\rangle \quad\mbox{with}\quad
  \tens d = \cvector{\delta_i \\ p_i}\otimes\hat e_i\;.
\label{eq:70}
\end{equation}
The joint probability $P(\tens q,\tens p)$ is then
\begin{equation}
  P(\tens\delta,\tens p) = \int_{\tens r}\int_{\tens s}
  \Phi(\tens r,\tens s)\,
  \E^{\I(\tens r\cdot\tens\delta+\tens s\cdot\tens p)}\;.
\label{eq:71}
\end{equation}

With this, we return to (\ref{eq:67}), where we introduce for later convenience a source field $\tens R$ associated with $\tens\delta$,
\begin{equation}
  P(\tens q,\tens p) = \left. \int\D\tens\delta\,
  P(\tens q\vert\tens\delta)\,P(\tens\delta,\tens p)\,
  \E^{-\I\tens R\cdot\tens\delta}\,\right\vert_{\tens R=0}\;.
  \label{eq:72}
\end{equation} 
This allows us to elevate the conditional probability $P(\tens q\vert\tens\delta)$ from (\ref{eq:67}) to an operator acting on the source field $\tens R$,
\begin{equation}
  P(\tens q\vert\tens\delta) \mapsto
  \hat P\left(\tens q\vert\I\partial_{\tens R}\right)\;,
\label{eq:73}
\end{equation}
which we can pull in front of the integral in (\ref{eq:72}).
This, together with (\ref{eq:71}), turns (\ref{eq:67}) into
\begin{equation}
  P(\tens q,\tens p) = \left.
    \hat P(\tens q\vert\I\partial_{\tens R}) 
    \int\D\tens\delta\,
    \E^{-\I\tens R\cdot\tens\delta}
    \int_{\tens r}\int_{\tens s}
    \Phi(\tens r,\tens s)\,
    \E^{\I(\tens r\cdot\tens\delta+\tens s\cdot\tens p)}\,
  \right\vert_{\tens R=0}\;.
\label{eq:74}
\end{equation}
Performing the integral over $\tens\delta$ results in the Dirac delta distribution $\delta_\mathrm{D}(\tens r-\tens R)$ such that the ensuing integration over $\tens r$ replaces $\tens r$ by $\tens R$,
\begin{equation}
  P(\tens q,\tens p) = \hat D\int_{\tens s}
  \Phi(\tens R,\tens s)\,\E^{\I\tens s\cdot\tens p}
\label{eq:75}
\end{equation}
with the differential operator
\begin{equation}
  \hat D = \hat P(\tens q\vert\I\partial_{\tens R})
  \Big\vert_{\tens R=0}\;.
\label{eq:76}
\end{equation}
As we shall discuss in detail later, this differential operator can in cosmological applications at late times safely be approximated by
\begin{equation}
  \hat D \approx V^{-N}\,\hat 1\,\Big\vert_{\tens R=0}\;,
\label{eq:77}
\end{equation}
such that
\begin{align}
  P(\tens q,\tens p) &= \frac{1}{V^N}
  \int_{\tens s}\Phi(0,\tens s)\E^{\I\tens s\cdot\tens p} \nonumber\\ &=
  \frac{1}{V^N\sqrt{(2\pi)^{3N}\det C_{pp}(\tens q)}}
  \exp\left(-\frac{1}{2}\tens p^\top C_{pp}^{-1}(\tens q) \tens p\right)
\label{eq:78}
\end{align}
where
\begin{equation}
  C_{pp} = \left\langle\tens p\otimes\tens p\right\rangle =
  C_{p_ip_j}\otimes E_{ij}\quad\mbox{with}\quad
  E_{ij} = \hat e_i\otimes\hat e_j
\label{eq:79}
\end{equation} 
is the correlation matrix for the entire set of particle momenta, and $C_{p_ip_j}$ is the correlation matrix of the momenta of particles $i$ and $j$ \cite{2016NJPh...18d3020B}.

\subsubsection{Momentum correlations}

Due to the definition $p=\nabla\psi$ of the initial momentum field in terms of the velocity potential $\psi$, we have
\begin{equation}
  C_{p_ip_j}(q) = \left(\nabla_i\otimes\nabla_j\right)\xi_\psi(q)\;,
\label{eq:80}
\end{equation}
where $\xi_\psi$ is the auto-correlation function
\begin{equation}
  \xi_\psi(q) = \int_kP_\psi(k)\E^{\I k\cdot q}
\label{eq:81}
\end{equation}
of the initial velocity potential taken at the separation $q = \vert q_i-q_j\vert$ of the particles $i$ and $j$. Thus,
\begin{equation}
  C_{p_ip_j}(q) = \int_k\left(k\otimes k\right)
  P_\psi(k)\,\E^{\I k\cdot q}\;.
\label{eq:82}
\end{equation}
Notice that $C_{p_ip_j}=C_{p_jp_i}$ because a sign change in $q$ can be cancelled by an irrelevant sign change in $k$ in (\ref{eq:82}).

\begin{figure}[t]
  \centering
  \includegraphics[width=\hsize]{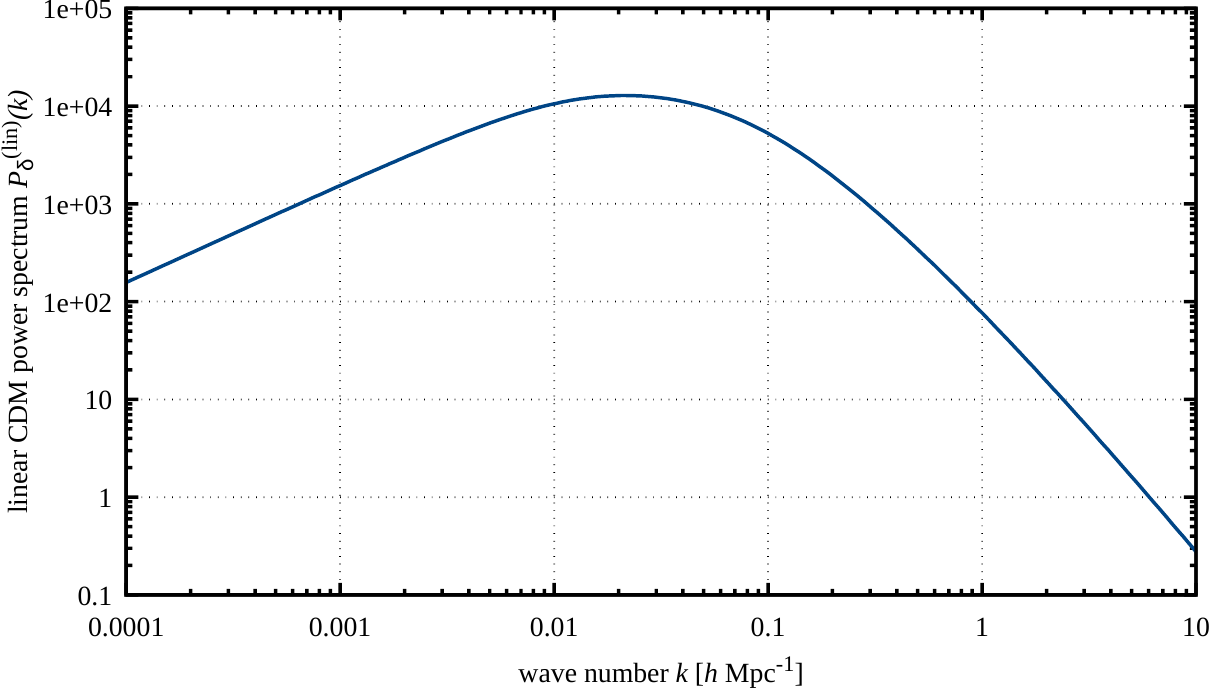}
\caption{Linearly evolved CDM power spectrum at $z=0$.}
\label{fig:8}
\end{figure}

Carrying out the integral over the directions of the wave vector $k$, we obtain
\begin{equation}
  C_{p_ip_j}(q) = -\id 3\,a_1(q)-\pi_\parallel\,a_2(q)
\label{eq:83}
\end{equation}
with the correlation functions
\begin{align}
  a_1(q) &= -\frac{1}{2\pi^2}
  \int_0^\infty\D k\,P_\delta^\mathrm{(i)}(k)\,\frac{j_1(kq)}{kq}\;,\nonumber\\
  a_2(q) &= \frac{1}{2\pi^2}
  \int_0^\infty\D k\,P_\delta^\mathrm{(i)}(k)\,j_2(kq)
\label{eq:84}
\end{align}
and the projector
\begin{equation}
  \pi_\parallel = \hat q\otimes\hat q
\label{eq:85}
\end{equation}
parallel to the line connecting the two points which are being correlated \cite{2017NJPh...19h3001B}. The vector $\hat q$ is the unit vector in $q$ direction, $\hat q = q/\vert q\vert$. We have used in (\ref{eq:84}) that $P_\delta^\mathrm{(i)} = k^4P_\psi$ due to the Poisson equation (\ref{eq:53}) between the velocity potential $\psi$ and the initial density contrast $\delta^\mathrm{(i)}$. The functions $a_{1,2}(q)$ have the limits
\begin{equation}
  \lim_{q\to0}a_1(q) = -\frac{\sigma_1^2}{3}\;,\quad
  \lim_{q\to0}a_2(q) = 0\;,
\label{eq:86}
\end{equation}
where $\sigma_1^2$ is one of the moments
\begin{equation}
  \sigma_n^2 = \frac{1}{2\pi^2}\int_0^\infty\D k\,
  k^{2n-2}\,P_\delta^\mathrm{(i)}(k)
\label{eq:87}
\end{equation} 
of the initial density-fluctuation power spectrum; see also the asymptotic expansions (\ref{eq:156}) below. Of course, we implicitly need to assume here that these moments exist up to the order $n$ needed in later expressions.

\begin{figure}[t]
\includegraphics[width=\hsize]{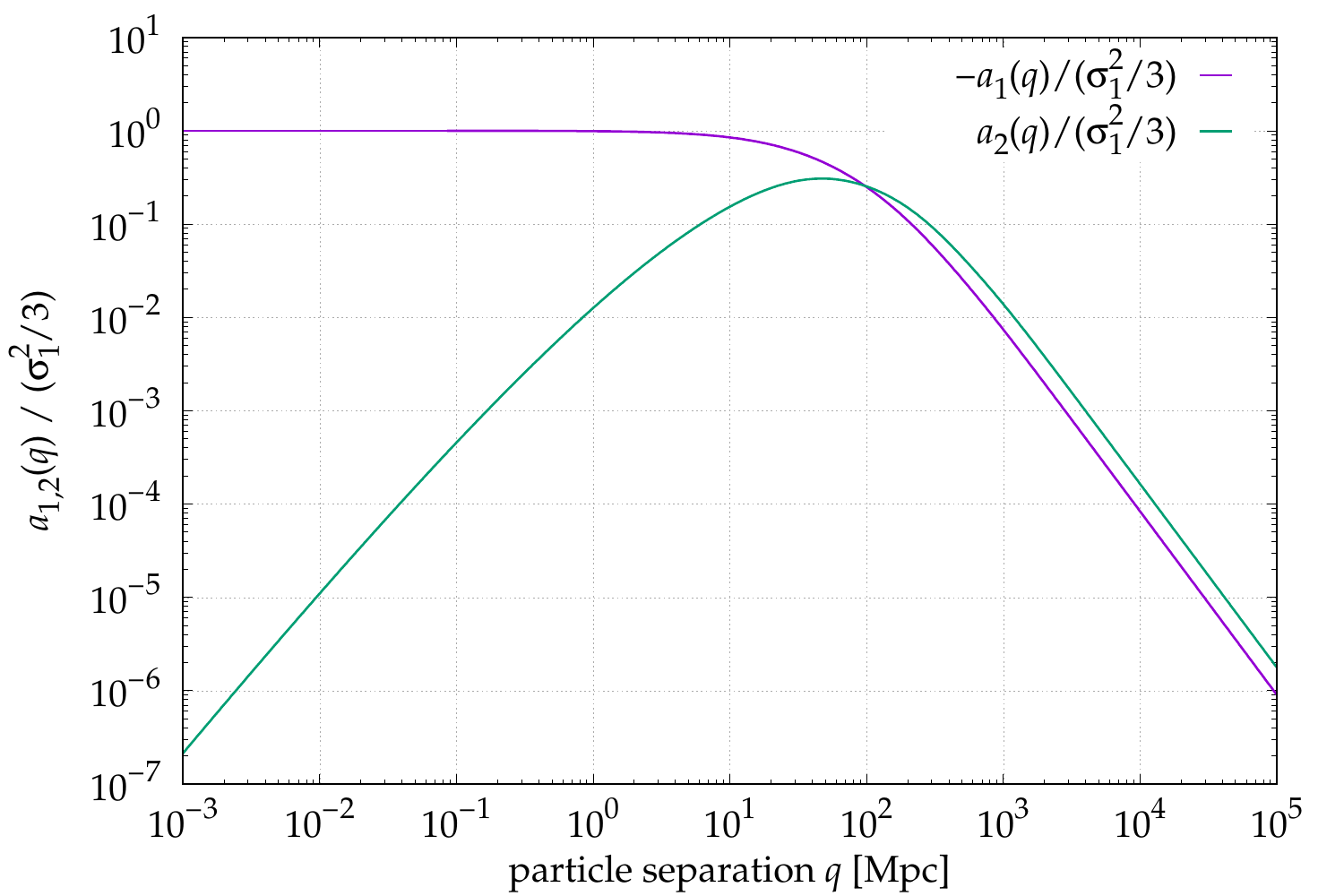}
\caption{The initial momentum correlation functions $a_{1,2}(q)$ as defined in (\ref{eq:86}) normalized to $\frac{\sigma_1^2}{3}$ (\ref{eq:87}) are shown as a function of particle separation.}
\label{fig:9}
\end{figure}

\subsection{Statistics of the evolved particle distribution}

Statistical information can be extracted from the time-evolved generating functional by applying suitable operators (see eg. \cite{2016NJPh...18d3020B, 2019AnP...53100446B}). We shall first introduce density operators here and discuss density correlation functions before we proceed to power spectra for density and velocity fluctuations.

\subsubsection{Density operators}

The number density of our particle ensemble is a sum over delta distributions,
\begin{equation}
  \rho(q,t) = \sum_{i=1}^N\delta_\mathrm{D}\left(q-q_i(t)\right)\;.
\label{eq:88}
\end{equation}
We Fourier transform this expression,
\begin{equation}
  \rho(q,t) \mapsto \tilde\rho(k,t) =
  \sum_{i=1}^N\exp\left(-\I k\cdot q_i(t)\right)
\label{eq:89}
\end{equation}
and elevate it to an operator by replacing the particle positions $q_i(t)$ by the respective functional derivative with respect to $J_{q_i}(t)$,
\begin{equation}
  q_i(t) \mapsto -\I\,\frac{\delta}{\delta J_{q_i}(t)}\;.
\label{eq:90}
\end{equation}
We thus obtain the density operator
\begin{equation}
  \hat\rho(k,t) = \sum_{i=1}^N\hat\rho_i(k,t) = \sum_{i=1}^N
  \exp\left(-k\cdot\frac{\delta}{\delta J_{q_i}(t)}\right)\;,
\label{eq:91}
\end{equation} 
which is a sum of one-particle density operators $\hat\rho_i(k,t)$.

\subsubsection{Density correlation functions}

Since the operators $\hat\rho_i(k)$ contain a functional derivative in the exponential, they create a translation. Thus, any sequence of $n$ one-particle density operators applied to the generating functional gives, setting $\tens J = 0$ at the end,
\begin{equation}
  \hat\rho_1(1)\cdots\hat\rho_n(n)\,Z[\tens J]\,
  \Big\vert_{\tens J = 0} = Z[\tens L]\;,
\label{eq:92}
\end{equation}
where the short-hand notation $\hat\rho(j) = \hat\rho(k_j,t_j)$ was introduced. The shift tensor is
\begin{equation}
  \tens L = -\sum_{j=1}^nk_j\cdot\frac{\delta\tens J}{\delta J_{q_j}(t_j)}
\label{eq:93}
\end{equation}
Applying finally $n$ density operators to the generating functional results in
\begin{align}
  \left\langle\rho(1)\cdots\rho(n)\right\rangle &=
  \sum_{j_1,\ldots,j_n=1}^N\,\left.
    \hat\rho_{j_1}(1)\cdots\hat\rho_{j_n}(n)\,
    Z[\tens J]
  \right\vert_{\tens J=0} \nonumber\\ &=
  \prod_{r=0}^{n-1}(N-r)\,Z[\tens L] \approx N^n\,Z[\tens L]\;,
\label{eq:94}
\end{align} 
where the first equality in the second line holds because the particles of the ensemble are indistinguishable, so each of the $N(N-1)\cdots(N-n+1)$ particle tuples $(j_1,\ldots,j_n)$ must contribute the same statistical result. The final approximate equality holds if $N\gg n$, as will usually be most safely the case. If the thermodynamic limit $N\to\infty$ could be taken or would be inappropriate, shot-noise terms could occur \cite{2016NJPh...18d3020B}. In cosmology, however, we can assume that any reasonably large volume will be filled with an extremely large number of particles. We shall therefore replace (\ref{eq:94}) by an equality from here on.

\subsection{Low-order statistics of the free particle distribution}

After this general discussion, we shall now specify the results obtained so far to low-order statistical measures for the free particle distribution, i.e.\ for the distribution of the particle ensemble flowing along force-free trajectories.

\subsubsection{Free power spectrum}

For power spectra, we have with $n = 2$ in (\ref{eq:94})
\begin{equation}
  \left\langle\rho(1)\rho(2)\right\rangle = N^2Z[\tens L]\;.
\label{eq:95}
\end{equation}
For synchronous power spectra, $t_2 = t_1$, and the tensor $\tens L$ in (\ref{eq:95}) is
\begin{equation}
  \tens L = -\cvector{1\\ 0}\delta_\mathrm{D}\left(t-t_1\right)
  \left(
    k_1\otimes\hat e_1+k_2\otimes\hat e_2
  \right)
\label{eq:96}
\end{equation} 
such that the scalar product $(\tens L,\bar{\tens x}(t))$ turns out to be
\begin{equation}
  \left(\tens L,\bar{\tens x}(t)\right) =
  -k_1\cdot\bar q_1(t_1)-k_2\cdot\bar q_2(t_1)\;.
\label{eq:97}
\end{equation}

If we neglect the contribution to the particle trajectories (\ref{eq:54}) due to the particle interactions and insert the Zel'dovich trajectories (\ref{eq:56}) into (\ref{eq:97}), we obtain from (\ref{eq:23}) what we call the free generating functional,
\begin{equation}
  Z_0[\tens L] = \int\D\Gamma\,
  \E^{-\I k_1\left(q_1+tp_1\right)-\I k_2\left(q_2+tp_2\right)}\;,
\label{eq:98}
\end{equation}
where $(q_i,p_i)$ are now meant to be initial particle positions and momenta.

Due to spatial homogeneity, we can refer the positions of all particles to the position of any particular particle, say particle $1$, replacing $q_i\mapsto q_i-q_1$, and integrate over $q_1$. Since the momentum-correlation matrix depends only on particle separations, but not on absolute particle positions, this results in a delta distribution, leaving the free generating functional in the form
\begin{equation}
  Z_0[\tens L] = (2\pi)^3\delta_\mathrm{D}\left(k_1+k_2\right)
  \int\D\Gamma_{\hat 1}\,\E^{-\I k_1t(p_1-p_2)+\I k_1q_2}\;,
\label{eq:99}
\end{equation} 
where the Gaussian integral measure
\begin{equation}
  \D\Gamma_{\hat 1} = \D\tens q_{\hat 1}\D\tens p\,
  \frac{1}{V^N\sqrt{(2\pi)^{3N}\det C_{pp}}}\,
  \exp\left(-\frac{1}{2}\tens p^\top C^{-1}_{pp}\tens p\right)
\label{eq:100}
\end{equation}
now does not contain $\D q_1$ any more,
\begin{equation}
  \D\tens q_{\hat 1} = \prod_{i=2}^N\D q_i\;.
\label{eq:101}
\end{equation}
Writing
\begin{equation}
  -k_1t\left(p_1-p_2\right) = \tens L_p\cdot\tens p\quad\mbox{with}\quad
  \tens L_p = -k_1t\otimes\left(\hat e_1-\hat e_2\right)\;,
\label{eq:102}
\end{equation} 
the momentum integration in (\ref{eq:99}) can be carried out directly, leading to
\begin{equation}
  Z_0[\tens L] = (2\pi)^3\delta_\mathrm{D}(k_1+k_2)V^{-N}
  \int\D\tens q_{\hat 1}\,\exp\left(
    -\frac{1}{2}\tens L_p^\top C_{pp}\tens L_p
  \right)\E^{\I k_1q_2}\;.
\label{eq:103}
\end{equation}
Combining (\ref{eq:102}) with (\ref{eq:79}), the quadratic form in the exponential of (\ref{eq:103}) is
\begin{equation}
  \tens L_p^\top C_{pp}\tens L_p = 2t^2k_1^\top\left(
    C_{p_1p_1}-C_{p_1p_2}
  \right)k_1 = 2t^2k_1^2\left(
    \frac{\sigma_1^2}{3}+a_\parallel(q_2,\mu)
  \right)
\label{eq:104}
\end{equation} 
with the definition
\begin{equation}
  a_\parallel(q_2,\mu) = a_1(q_2)+\mu^2a_2(q_2)
\label{eq:105}
\end{equation}
containing the cosine $\mu = \hat k\cdot\hat q_2$ of the angle between $k_1$ and $q_2$ (see eg. \cite{2017NJPh...19h3001B, 2019AnP...53100446B}). Notice the limit
\begin{equation}
  \lim_{q\to0}a_\parallel(q,\mu) = -\frac{\sigma_1^2}{3}\;.
\label{eq:106}
\end{equation} 
of the momentum-correlation function $a_\parallel(q,\mu)$.

Since the integrand in (\ref{eq:103}) depends only on the (relative) position $q_2$, we can integrate over all initial particle positions $q_j$ with $j>2$, which results in a factor $V^{N-2}$. We can further pull part of (\ref{eq:107}) in front of the integral, arriving at
\begin{equation}
  Z_0[\tens L] = (2\pi)^3\delta_\mathrm{D}\left(k_1+k_2\right)V^{-2}
  \E^{-Q_\mathrm{D}}\int_q\exp\left(
    -t^2k_1^2a_\parallel(q,\mu)
  \right)\E^{\I k_1\cdot q}\;,
\label{eq:107}
\end{equation}
where $q$ now abbreviates $q_2$ for simplicity, and $Q_\mathrm{D}$ is
\begin{equation}
  Q_\mathrm{D} = \frac{\sigma_1^2}{3}t^2k_1^2\;.
\label{eq:108}
\end{equation} 

Returning to (\ref{eq:105}), the free two-point density correlator is now
\begin{equation}
  \left\langle\rho(1)\rho(2)\right\rangle = N^2\,Z_0[\tens L]\;.
\label{eq:109}
\end{equation} 
As shown in (\ref{eq:16}), the free power spectrum $\mathcal{P}(k)$ is defined in terms of this expression, except for the preceding delta distribution and the prefactor $\bar\rho^2 = (N/V)^2$,
\begin{equation}
  (2\pi)^3\delta_\mathrm{D}\left(k_1+k_2\right)\,\left[
    (2\pi)^3\delta_\mathrm{D}(k)+\mathcal{P}(k)
  \right] =
  \frac{1}{\bar\rho^2}\left\langle\rho(1)\rho(2)\right\rangle\;.
\label{eq:110}
\end{equation} 
Combining (\ref{eq:110}), (\ref{eq:109}) and (\ref{eq:107}), we thus find
\begin{align}
  \mathcal{P}(k) &= \E^{-Q_\mathrm{D}}
  \int_q\left[
    \E^{-t^2k^2a_\parallel(q,\mu)}-1
  \right]\E^{\I k\cdot q} \nonumber\\ &=
  \E^{-Q_\mathrm{D}}
  \int_q\E^{-t^2k^2a_\parallel(q,\mu)}\E^{\I k\cdot q}
  \quad\mbox{for}\quad k\ne0\;.
\label{eq:111}
\end{align}
It is evident from the argument of the first exponential in the integrand of (\ref{eq:111}) that multiplying the amplitude of the initial power spectrum, and thus the correlation function $a_\parallel$, by a certain factor is equivalent to leaving this amplitude unchanged but multiplying the time coordinate by the root of the same factor. This will later be reflected by our results containing the time coordinate only in combination with $\sigma_2$, suggesting to introduce $\tau_2 = t\sigma_2$ as the time coordinate relevant for structure formation.

\begin{figure}[t]
\includegraphics[width=\hsize]{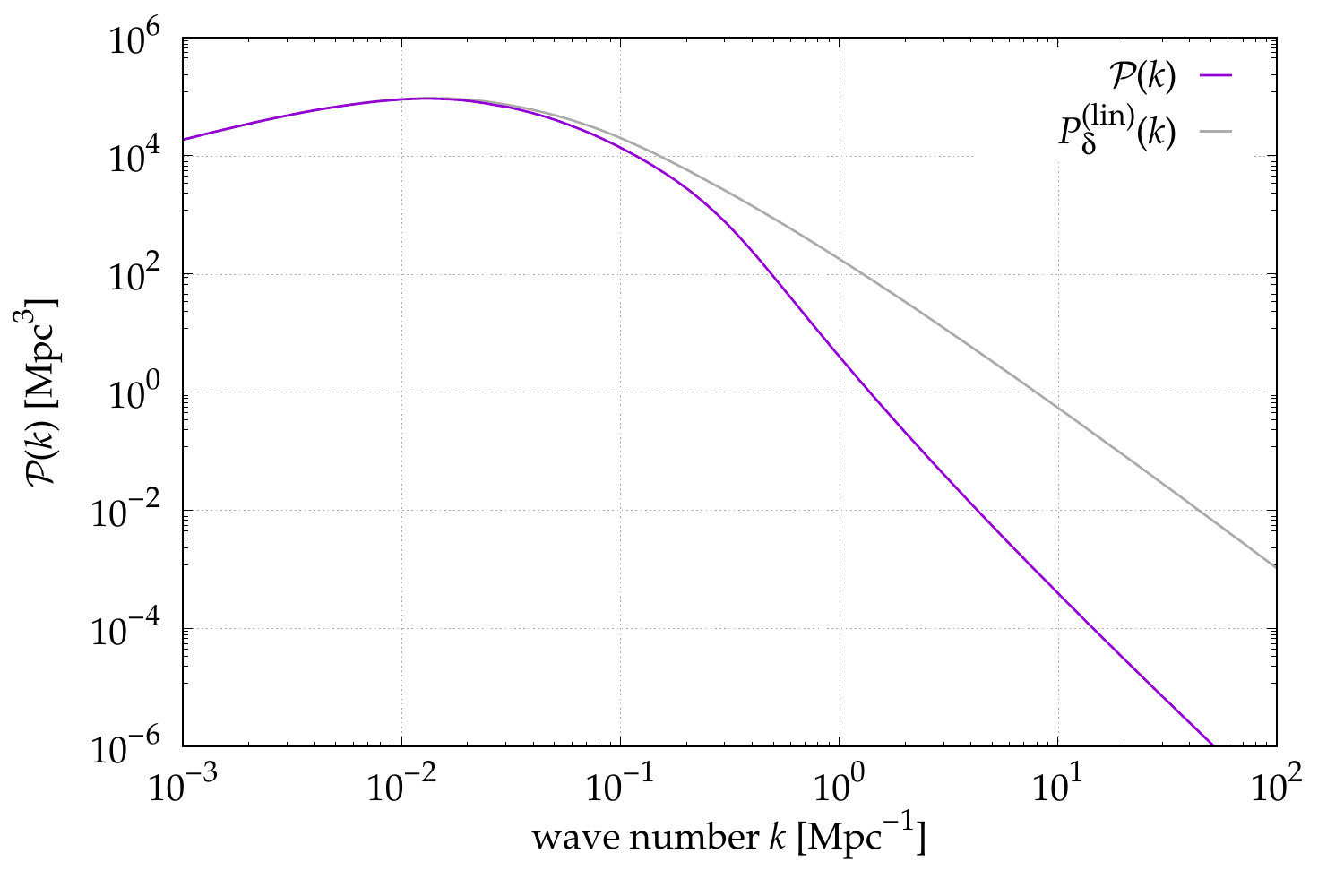}
\caption{The non-linear power spectrum $\mathcal{P}(k)$, (\ref{eq:111}) at redshift $z=0$ (purple line), together with the linearly evolved power spectrum (gray line) is shown for WIMP dark matter, where the initial power spectrum is cut-off with a Gaussian kernel at a wave number $k_s = 10^6$ Mpc$^{-1}$.}
\label{fig:10}
\end{figure}

The exponential prefactor $\exp(-Q_\mathrm{D})$ seems to indicate that the spectrum will be damped exponentially on small scales, but we will show later that it will be compensated exactly on small scales. If the argument of the exponential in the integrand of (\ref{eq:111}) is small enough, we can approximate
\begin{align}
  \int\D q\left[
    \E^{-t^2k^2a_\parallel(q,\mu)}-1
  \right]\E^{\I k\cdot q} &\approx
  -t^2k^2\int_qa_\parallel(q,\mu)\,\E^{\I k\cdot q} \nonumber\\ &=
  t^2P_\delta^\mathrm{(i)}(k) =
  P_\delta^\mathrm{(lin)}(k)\;,
\label{eq:112}
\end{align} 
where we have used in the last step that our time coordinate $t$ is the linear growth factor. Since $a_\parallel$ is bounded, this approximation holds for sufficiently small $k^2t^2$, i.e.\ for any given time on sufficently large scales, as expected \cite{2017NJPh...19h3001B}.

\subsubsection{Free bispectrum}

For the bispectrum, we proceed in an exactly analogous way (see eg. \cite{2016NJPh...18d3020B}). The shift tensor $\tens L$ contains three wave vectors $(k_1,k_2,k_3)$ now, thus (\ref{eq:97}) is replaced by
\begin{equation}
  \left(\tens L,\bar{\tens x}(t)\right) = -\sum_{i=1}^3k_i\cdot\bar q_i(t)\;.
\label{eq:113}
\end{equation}
Consequently, the free generating functional, evaluated at $\tens L$, is extended to
\begin{equation}
  Z_0[\tens L] = (2\pi)^3\delta_\mathrm{D}\left(\sum_{i=1}^3k_i\right)
  V^{-3}\E^{-Q_\mathrm{D}^{(3)}}\int_q\int_{q'}
  \exp\left(-t^2Q_\mathrm{C}^{(3)}\right)\E^{\I(k_2\cdot q+k_3\cdot q')}
\label{eq:114}
\end{equation}
where
\begin{equation}
  Q_\mathrm{D}^{(3)} = \frac{\sigma_1^2}{6}t^2\left(\sum_{i=1}^3k_i^2\right)
\label{eq:115}
\end{equation}
and
\begin{equation}
  Q_\mathrm{C}^{(3)} =
  k_1^\top C_{p_1p_2}\left(q\right)k_2+
  k_1^\top C_{p_1p_3}\left(q'\right)k_3+
  k_2^\top C_{p_2p_3}\left(\left\vert q-q'\right\vert\right)k_3\;.
\label{eq:116}
\end{equation} 

Similar to (\ref{eq:110}), the free bispectrum $\mathcal{B}$ is defined by
\begin{equation}
  (2\pi)^3\delta_\mathrm{D}\left(\sum_{i=1}^3k_i\right)\mathcal{B}(k_2,k_3) =
  \frac{1}{\bar\rho^3}\left\langle
    \rho(1)\rho(2)\rho(3)
  \right\rangle
\label{eq:117}
\end{equation}
for non-degenerate configurations of the three wave vectors $k_{1,2,3}$, i.e.\ if none of them vanishes. The free bispectrum depends on only two wave vectors because the delta distribution in (\ref{eq:114}) ensures that $k_1=-(k_2+k_3)$. Taking account of (\ref{eq:94}) with $n = 3$, then comparing (\ref{eq:117}) to (\ref{eq:114}), we find the expression
\begin{equation}
  \mathcal{B}(k_2,k_3) = \E^{-Q_\mathrm{D}^3}\int_q\int_{q'}
  \exp\left(-t^2Q_\mathrm{C}^{(3)}\right)\E^{\I(k_2\cdot q+k_3\cdot q')}\;,
\label{eq:118}
\end{equation}
understanding again that the wave vector $k_1$ in $Q_\mathrm{C}^{(3)}$ and $Q_\mathrm{D}^{(3)}$ is fixed by the condition $k_1+k_2+k_3 = 0$.

\begin{figure}[t]
\includegraphics[width=\hsize]{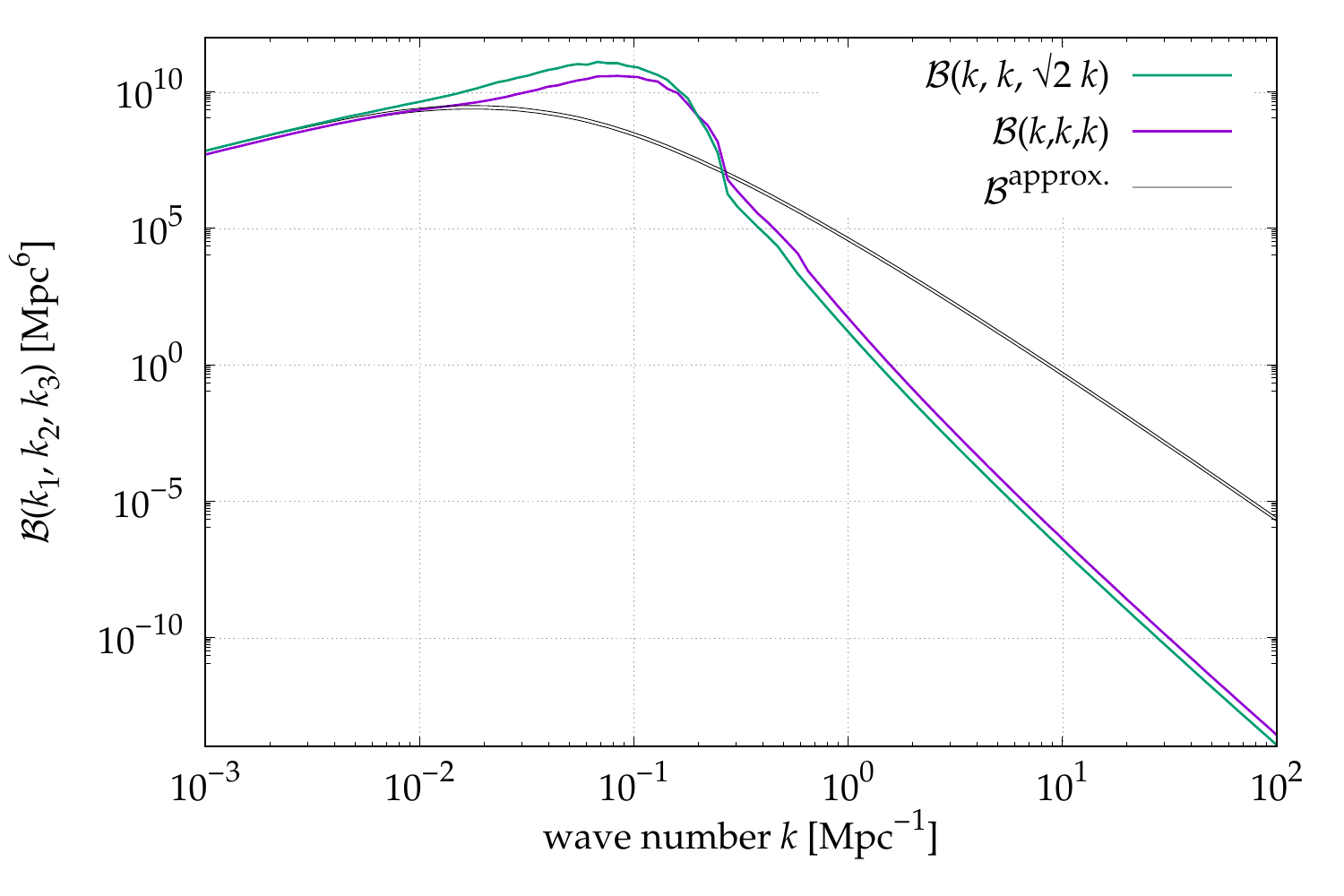}
\caption{The non-linear bispectrum $\mathcal{B}(k)$, (\ref{eq:118}) at redshift $z=0$ for the isosceles (green line) and the equilateral (purple line) configuration of $k$ vectors, together with the bispectrum approximation (thin black lines) as in (\ref{eq:119}) for light WIMP dark matter ($k_s = 10^5$ Mpc$^{-1}$ as in Fig. (\ref{fig:9})).}
\label{fig:11}
\end{figure}

In the large-scale limit, when the norm of all wave vectors involved is getting small, $k_1,k_2,k_3\to 0$, the expression resulting from (\ref{eq:114}) can be brought into the familiar form
\begin{equation}
  \mathcal{B}(k_2,k_3) \approx 
  F(k_2,k_3)P_\delta^\mathrm{(lin)}(k_2)P_\delta^\mathrm{(lin)}(k_3)+
  \mbox{cyc.}
\label{eq:119}
\end{equation} 
with the kernel function
\begin{equation}
  F(k_2,k_3) = \left(1+\frac{k_2\cdot k_3}{k_2^2}\right)
  \left(1+\frac{k_2\cdot k_3}{k_3^2}\right)\;.
\label{eq:120}
\end{equation}

\subsubsection{Free velocity power spectrum}

The generating functional of kinetic field theory contains the complete statistical information on the phase-space trajectories of the particle ensemble and thus also allows calculation velocity power spectra \cite{2019AnP...53100446B}. The momentum $p_j$ of a particle $j$ needs to be localized at the spatial position $q_j$ by the expression
\begin{equation}
  p_j\delta_\mathrm{D}(q-q_j)\;,
\label{eq:121}
\end{equation}
which we elevate to the velocity operator
\begin{equation}
  \hat\Pi_j = \hat p_j\otimes\hat\rho_j =
  -\I\frac{\partial}{\partial J_{p_j}}\otimes\hat\rho_j\;.
\label{eq:122}
\end{equation}
The density operator locates the particle $j$ and the derivative with respect to the source field $J_{p_j}$ extracts this particle's momentum.

Since the derivative with respect to $J_{q_j}$ contained in the density operator and the derivative with respect to $J_{p_j}$ commute, we can apply multiple operators $\hat\Pi_1\ldots\hat\Pi_n$ by shifting all involved density operators $\hat\rho_1\ldots\hat\rho_n$ to the right to apply them first to the generating functional, which they will translate by an amount $\tens L$,
\begin{equation}
  \hat\Pi_1(1)\cdots\hat\Pi_n(n)\,Z_0[\tens J]\Big\vert_{\tens J=0} =
  \left(-\I\frac{\partial}{\partial J_{p_1}(t_1)}\right)\otimes\cdots\otimes
  \left(-\I\frac{\partial}{\partial J_{p_n}(t_n)}\right)\,
  Z_0[\tens J+\tens L]\Big\vert_{\tens J=0}\;.
\label{eq:123}
\end{equation}
If we specialize to synchronous spectra, all operators are applied at the same time $t_j=t$. The free generating functional $Z_0[\tens J+\tens L]$ can then be brought into a form similar to (\ref{eq:98}),
\begin{equation}
  Z_0[\tens L+\tens J] = \int\D\Gamma\,
  \E^{\I(\tens L_q+\tens J_q,\tens q)+\I(\tens L_p+\tens J_p,\tens p)}\;,
\label{eq:124}
\end{equation}
with the components
\begin{equation}
  \tens L_q = -k_j\otimes\hat e_j\;,\quad
  \tens L_p = -k_jt\otimes\hat e_j
\label{eq:125}
\end{equation}
of the shift tensor $\tens L$, generalizing the definition of $\tens L_p$ in (\ref{eq:102}). Expression (\ref{eq:124}) shows that we can extract information on particle momenta $p_j$ by taking derivatives with respect to $L_{p_j}$ rather than $J_{p_j}$. Therefore, for a velocity power spectrum with $n=2$,
\begin{equation}
  \left\langle\Pi(1)\Pi(2)\right\rangle = -N^2
  \left(
    \frac{\partial}{\partial L_{p_1}}\otimes
    \frac{\partial}{\partial L_{p_2}}
  \right)Z_0[\tens L]\;,
\label{eq:126}
\end{equation}
analogous to (\ref{eq:109}). Taking the free generating functional from (\ref{eq:103}) and defining the velocity power spectrum $\mathcal{P}_\Pi(k)$ in analogy to (\ref{eq:110}) results in
\begin{equation}
  \mathcal{P}_\Pi(k) = -D_1\otimes D_2\int\D q\,\E^{-Q}\,\E^{\I k\cdot q}\;,
\label{eq:127}
\end{equation}
which is a second-rank tensor because it correlates all momentum components at one position with all momentum components at another position. Here, $D_j = \partial/\partial L_{p_j}$ abbreviates the derivative with respect to $L_{p_j}$ and $Q$ is the quadratic form
\begin{equation}
  Q = \frac{1}{2}\tens L_p^\top C_{pp}\tens L_p\;.
\label{eq:128}
\end{equation}
The derivatives of $Q$ with respect to $L_{p_{1,2}}$ are
\begin{align}
  D_1Q &= \tens L^\top C_{pp_1} =
  \left(C_{p_1p_2}-C_{p_1p_1}\right)kt =
  -\left(\frac{\sigma_1^2}{3}+a_1\right)kt-a_2\pi_\parallel kt\;,
  \nonumber\\
  D_2Q &= \tens L^\top C_{pp_2} = -D_1Q\;,
\label{eq:129}
\end{align}
while the second derivative is
\begin{equation}
  D_1\otimes D_2Q = C_{p_1p_2}\;.
\label{eq:130}
\end{equation}
Thus, the velocity power spectrum reads
\begin{equation}
  \mathcal{P}_\Pi(k) = \int\D q\,\mathcal{Q}\,\E^{-Q}\,\E^{\I k\cdot q}
\label{eq:131}
\end{equation}
with the matrix
\begin{equation}
  \mathcal{Q} = D_1Q\otimes D_2Q-C_{p_1p_2}\;.
\label{eq:132}
\end{equation}

\begin{figure}[t]
\includegraphics[width=\hsize]{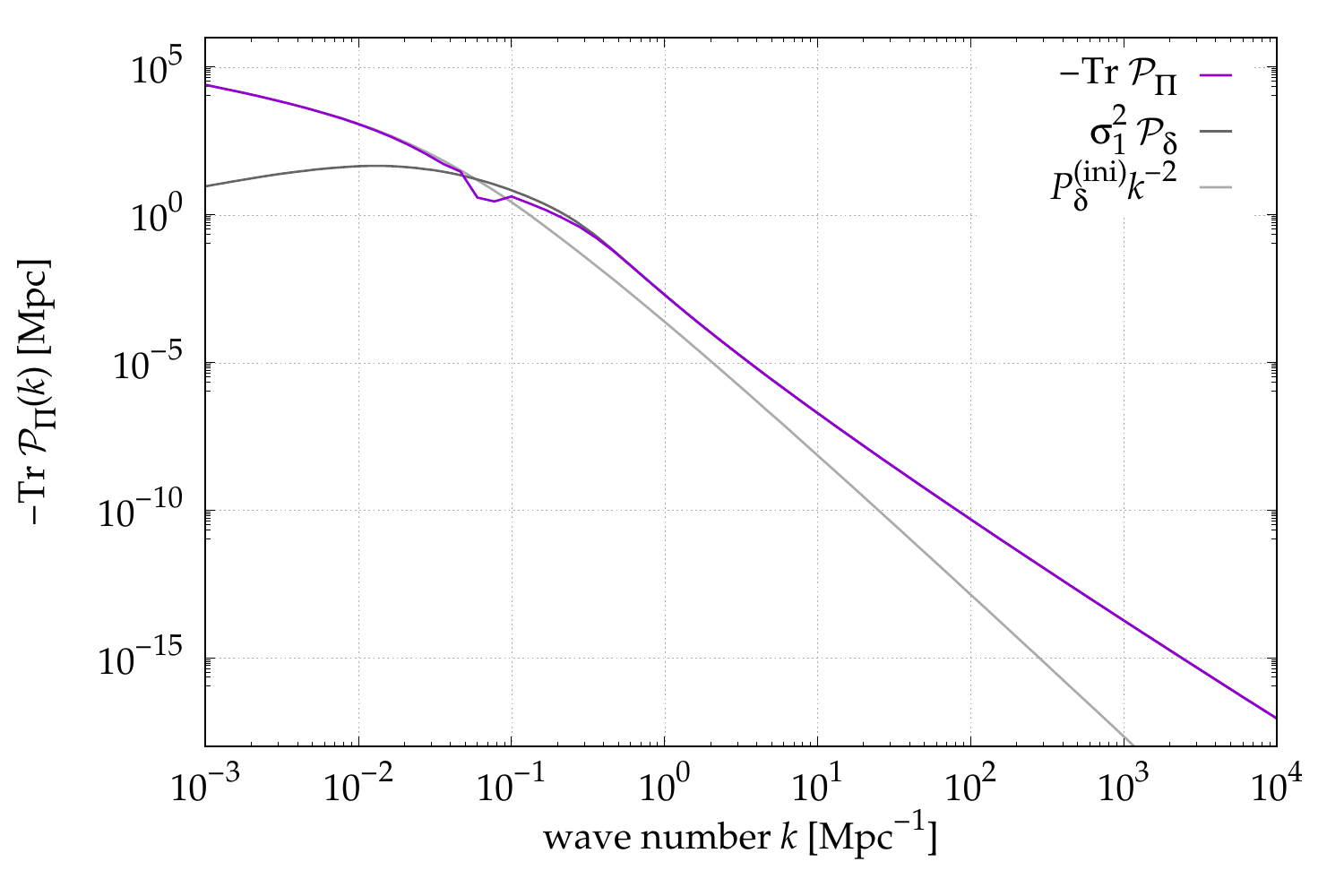}
\caption{The absolute value of the trace of the free velocity power spectrum Tr $\mathcal{P}_{\Pi}$ (purple line) from (\ref{eq:131}) at redshift $z=0$ is shown together with $\frac{P_{\delta}^{(ini)}}{k^2}$ (light gray line), which matches the trace at large scales, and the free power spectrum, multiplied by $\sigma_1^2$ (dark gray line), matching the trace at small scales. The initial power spectrum is for WIMP dark matter with a Gaussian cut-off at $k_s = 10^{6}$ Mpc$^{-1}$.}
\label{fig:12}
\end{figure}

\section{Asymptotic small-scale behaviour of power spectra}
\label{sec:4}

We shall now turn to deriving rigorous statements on the small-scale behaviour of the main quantities derived above, in particular the free density-fluctuation power spectrum, the free density-fluctuation bispectrum, and the free velocity power spectrum. We shall then generalize some of our results by including the complete set of initial correlations and by including interactions in the mean-field approximation.

\subsection{Free density-fluctuation power spectrum}

We shall focus on density-fluctuation power spectra first, deriving their asymptotic small-scale behaviour in two different ways \cite{2021arXiv211007427K}.

\subsubsection{Leading-order asymptotics from Morse's lemma}

We are interested in general statements on the power spectrum on small scales. For this reason, we first study the asymptotic behaviour of the free power spectrum $\mathcal{P}(k)$ as derived in (\ref{eq:111}).

We begin with Laplace's method for $d$-dimensional integrals of the form
\begin{equation}
  J(\lambda) = \int_\Omega\E^{-\lambda f(x)}\,g(x)\,\D^dx
\label{eq:133}
\end{equation}
over functions $f,g\in C^\infty$ on a domain $\Omega\subset\mathbb{R}^d$. The integral is supposed to converge absolutely for sufficiently large $\lambda\in\mathbb{R}$, the function $f$ is assumed to have a minimum at $x_0\in\Omega$ and only there, and the Hessian $A$ of $f$ in $x_0$ is supposed to be positive definite. Then, $J(\lambda)$ has the asymptotic expansion
\begin{equation}
  J(\lambda) \sim \E^{-\lambda f(x_0)}
  \sum_{n=0}^\infty\frac{c_n}{\lambda^{d/2+n}}
\label{eq:134}
\end{equation}
for $\lambda\to\infty$ (see eg. \cite{bleistein1975asymptotic, wong2001asymptotic}).

For specifying the coefficients $c_n$, we introduce some elements of notation. Under the given conditions, Morse's lemma ensures that neighbourhoods $U,V$ of $y=0$ and $x_0$ and a diffeomorphism $h:U\to V$ exist such that
\begin{equation}
  (f\circ h)(y) = f(x_0)+\frac{1}{2}y^\top Qy\;,\quad
  Q = \mathrm{diag}(\mu_1,\ldots,\mu_d)\;.
\label{eq:135}
\end{equation}
With the Jacobian determinant $\det H$ of $h$, we define the function $G:U\to\mathbb{R}$ by
\begin{equation}
  G(y) = (g\circ h)(y)\det H(y)\;.
\label{eq:136}
\end{equation}
We further introduce the multi-index $\alpha = (\alpha_1,\ldots,\alpha_d)$ and agree on the notation
\begin{align}
  \vert\alpha\vert = \sum_{k=1}^d\alpha_k\;&,\quad
  \alpha! = \prod_{k=1}^d\alpha_k!\;,\nonumber\\
  \Gamma(\alpha) = \prod_{k=1}^d\Gamma(\alpha_k)\;&,\quad
  \mu^\alpha = \prod_{k=1}^d\mu_1^{\alpha_1}\cdots\mu_d^{\alpha_d}\;,
\label{eq:137}
\end{align}
where $\mu$ in the last equation is a $d$-dimensional vector. Moreover, we define the symbol
\begin{equation}
  \delta(\alpha) =
  \begin{cases}
    1 & \mbox{all $\alpha_j$ even} \\ 0 & \mbox{else}
  \end{cases}
\label{eq:138}
\end{equation} 
and the derivative operator
\begin{equation}
  D^\alpha G(0) = \left.\frac
   {\partial^{\vert\alpha\vert}}
   {\partial^{\alpha_1}y_1\cdots\partial^{\alpha_d}y_d}\,G(y)
   \right\vert_{y=0}\;.
\label{eq:139}
\end{equation} 
Then, the coefficients $c_n$ are given by
\begin{equation}
  c_n = \sum_{\vert\alpha\vert=2n}\delta(\alpha)
  \left(\frac{2}{\mu}\right)^{(\alpha+1)/2}
  \Gamma\left(\frac{\alpha+1}{2}\right)\frac{D^\alpha G(0)}{\alpha!}\;.
\label{eq:140}
\end{equation} 
Applying the result (\ref{eq:134}) to integrals of the Laplace-Fourier type,
\begin{equation}
  J(\lambda, k) = \int_\Omega\E^{-\lambda f(x)}\E^{\I k\cdot x}\D^dx
\label{eq:141}
\end{equation}
and resumming the coefficients, we find
\begin{equation}
  J(\lambda, k) \sim \E^{-\lambda f(0)}
  \sqrt{\frac{(2\pi)^d}{\lambda^d\det A}}
  \exp\left(-\frac{k^\top A^{-1}k}{2\lambda}\right)
\label{eq:142}
\end{equation}
for $\lambda\to\infty$. Next, we specialize this statement to integrals of the form
\begin{equation}
  P(k) = \int_\Omega\E^{-\vert k\vert^sf(x)}\E^{\I k\cdot x}\D^dx
\label{eq:143}
\end{equation}
with $s\ge2$. We have shown in \cite{2021arXiv211007427K} that although the kernel does not meet the aforementioned conditions, the theorem can still be applied. We thus obtain
\begin{equation}
  P(k) \sim \E^{-\vert k\vert^sf(0)}
  \sqrt{\frac{(2\pi)^d}{\vert k\vert^{sd}\det A}}
  \exp\left(-\frac{k^\top A^{-1}k}{2\vert k\vert^s}\right)
\label{eq:144}
\end{equation}
to leading asymptotic order for $\vert k\vert\to\infty$.

Comparing the integrals in (\ref{eq:143}) and (\ref{eq:111}), we now set $s = 2$, $d = 3$ and $f(q) = t^2a_\parallel(q)$, use the limit (\ref{eq:106}) of $a_\parallel(q,\mu)$ and the Hessian
\begin{equation}
  A = t^2\left.\left(
    \frac{\partial^2a_\parallel(q)}{\partial q_i\partial q_j}
  \right)\right\vert_{q = 0} = \frac{\sigma_2^2t^2}{15}\left(
    \id{3}+2\hat k\otimes\hat k
  \right)\;.
\label{eq:145}
\end{equation}
Since the inverse of the $A$ is
\begin{equation}
  A^{-1} = \frac{15}{\sigma_2^2t^2}\left(
    \id{3}-\frac{2}{3}\hat k\otimes\hat k
  \right)\;,
\label{eq:146}
\end{equation}
and its determinant is
\begin{equation}
  \det A = 3\left(\frac{\sigma_2^2t^2}{15}\right)^3\;,
\label{eq:147}
\end{equation}
we immediately find
\begin{equation}
  \mathcal{P}(k) \sim \frac{3(4\pi)^{3/2}}{k^3}\Sigma^{3/2}(t)\E^{-\Sigma(t)}
\label{eq:148}
\end{equation}
with
\begin{equation}
  \Sigma(t) = \frac{5}{2\tau_2^2}\;,\quad \tau_n^2 = t^2\sigma_n^2\;.
\label{eq:149}
\end{equation} 
Recently, this result has also been derived in the framework of Lagrangian perturbation theory \cite{2020JCAP...06..033C}. This is the leading-order asymptotic term, from which important conclusions can be drawn. Before we get to them, we derive the full asymptotic series for the free power spectrum in a different manner from Erdélyi's theorem.

\begin{figure}[t]
\includegraphics[width=\hsize]{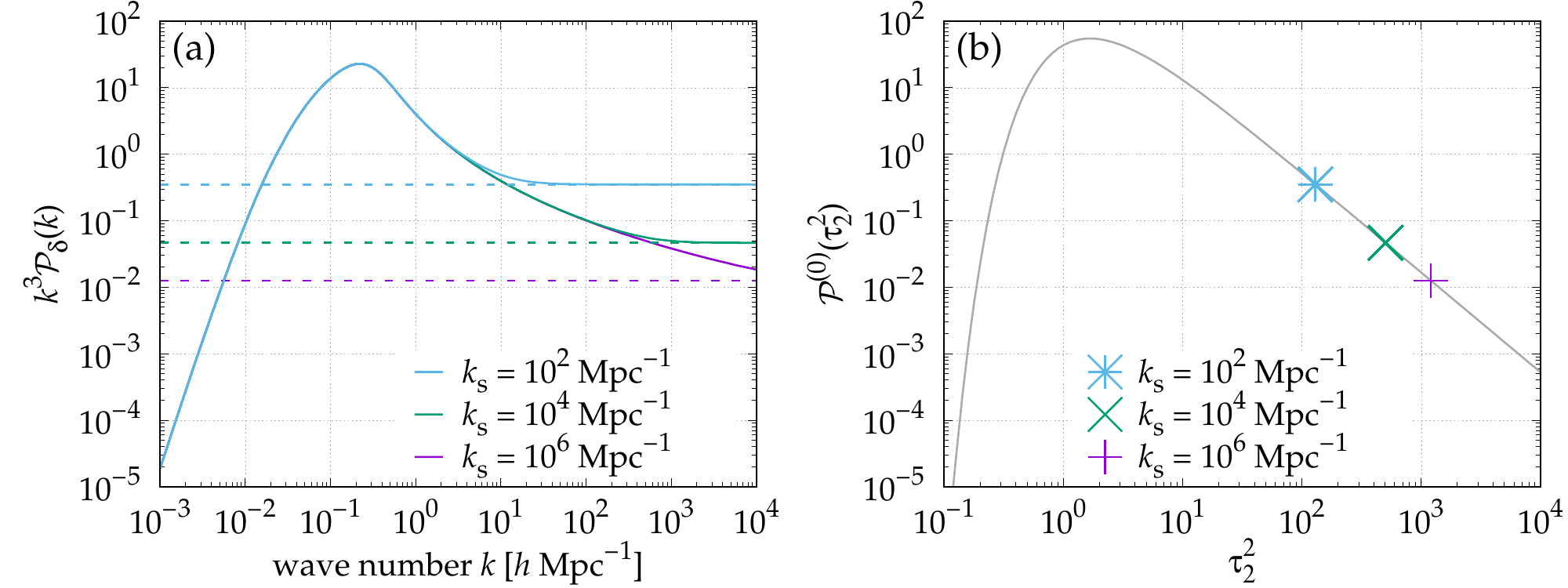}
\caption{Left: The dimensionless free power spectrum $k^3\mathcal{P}$ at redshift $z=0$ (solid colored lines) from (\ref{eq:111}) together with the first order asymptotics (dashed lines) from (\ref{eq:148}) for three different values of the initial small-scale smoothing wave number $k_s$ are shown. Right: The universal time evolution of the asymptotic amplitude $\mathcal{P}^{(0)}$ from (\ref{eq:148}) as a function of the time coordinate $\tau_2^2 = \sigma_2^2 t^2$ (gray line) together with the asymptotic amplitudes the spectra on the left panel.}
\label{fig:13}
\end{figure}

\subsubsection{Asymptotic series from Erdélyi's theorem}

We begin with the integral (\ref{eq:111}), introduce spherical polar coordinates with $\hat k$ as the polar axis, and introduce a finite upper limit $q_\mathrm{max}>0$ for the integration over $q$,
\begin{equation}
  \mathcal{P}(k,t) \sim 2\pi\E^{-Q_\mathrm{D}}\int_{-1}^1\D\mu
  \int_0^{q_\mathrm{max}}q^2\D q\,\E^{-t^2k^2a_\parallel(q)}\E^{\I kq\mu}
\label{eq:150}
\end{equation}
for $k\to\infty$. This is asymptotically correct since the asymptotic behaviour of $\mathcal{P}$ for large wave numbers is determined by the behaviour of the exponent next to its critical point, so the contribution to the integral from $q_\mathrm{max}$ to $\infty$ can be ignored \cite{fulks1961asymptotics, 2021arXiv211007427K}. Then, we can use Erdélyi's theorem (see eg. \cite{doetsch1955anwendungen, erdelyi1956asymptotic, erdelyi1961general})  to derive the asymptotic series for $\mathcal{P}$ \cite{2021arXiv211007427K}.

This theorem states that one-dimensional integrals of the form
\begin{equation}
  I(\lambda) = \int_a^b\E^{-\lambda f(x)}\,g(x)\,\D x
\label{eq:151}
\end{equation}
over functions $f$ and $g$ admitting the asymptotic series
\begin{equation}
  f(x) \sim f(a)+\sum_{k=0}^\infty a_k(x-a)^{\alpha+k}\;,\quad
  g(x) \sim \sum_{k=0}^\infty b_k(x-a)^{k+\beta-1}\;.
\label{eq:152}
\end{equation}
have the asymptotic expansion
\begin{equation}
  I(\lambda) \sim \E^{-\lambda f(a)}
  \sum_{n=0}^\infty\frac{\Gamma(\nu)c_n}{\lambda^\nu}\quad\mbox{with}\quad
  \nu = \frac{n+\beta}{\alpha}
\label{eq:153}
\end{equation}
with the coefficients
\begin{equation}
  c_n = \frac{1}{\alpha a_0^\nu}\sum_{m=0}^n\frac{b_{n-m}}{m!}d_{m,n}\;,\quad
  d_{m,n}=\lim_{x\to0}\frac{\D^m}{\D x^m}\left(
    1+\sum_{j=1}^\infty\frac{a_j}{a_0}x^j
  \right)^{-\nu}\;,
\label{eq:154}
\end{equation}
provided the function $f$ has a global minimum at $a$, the function $f$ can be term-wise differentiated, $f'$ and $g$ are continuous in a neighbourhood of $a$, except possibly at $a$ itself, and $I(\lambda)$ converges absolutely for sufficiently large $\lambda$.

We need to set $f(q) = a_\parallel(q)$ here with $a_\parallel(q)$ from (\ref{eq:105}) and $a_{1,2}(q)$ from (\ref{eq:84}). Using the series expansions
\begin{equation}
  j_\nu(z) = z^\nu\sum_{n=0}^\infty
  \frac{\left(-z^2/2\right)^n}{n!(2\nu+2n+1)!!}
\label{eq:155}
\end{equation} 
for the spherical Bessel functions, we find first
\begin{align}
  a_1(q) &\sim -\frac{\sigma_1^2}{3}-\sum_{n=1}^\infty
  \frac{\left(-q^2\right)^n\sigma_{n+1}^2}{(2n+3)(2n+1)!}\;,\nonumber\\
  a_2(q) &\sim q^2\sum_{n=0}^\infty
  \frac{\left(-q^2\right)^n\sigma_{n+2}^2}{(2n+5)(2n+3)(2n+1)!}
\label{eq:156}
\end{align}
with the moments $\sigma_n^2$ from (\ref{eq:87}). These imply the asymptotic series
\begin{equation}
  f(q) \sim -\frac{\sigma_1^2}{3}+\sum_{m=0}^\infty a_{2m}(\mu)q^{2m+2}
\label{eq:157}
\end{equation}
for $f(q)$ with
\begin{equation}
  a_{2m}(\mu) = \frac{(-1)^{m+2}\sigma_{m+2}^2}{(5+2m)(3+2m)!}\left[
    1+2(m+1)\mu^2
  \right]\;.
\label{eq:158}
\end{equation}
Setting $g(q) = q^2\exp(\I kq\mu)$, we further find
\begin{equation}
  g(q) \sim \sum_{m=0}^\infty b_m(\mu)q^{m+2}\;,\quad
  b_m(\mu) = \frac{(\I k\mu)^m}{m!}\;.
\label{eq:159}
\end{equation}

Comparing (\ref{eq:157}) and (\ref{eq:159}) to (\ref{eq:152}), we can read off $\alpha=2$ and $\beta=3$. Inserting the coefficients $a_{2m}$ and $b_m$ into (\ref{eq:154}), setting $\nu=(n+3)/2$, using (\ref{eq:153}) and integrating the resulting expressions over $\mu$, we can derive the complete asymptotic series
\begin{equation}
  \mathcal{P}(k,t) \sim \sum_{m=0}^\infty\frac{\mathcal{P}^{(m)}(t)}{k^{3+2m}}
  \quad (k\to\infty)\;.
\label{eq:160}
\end{equation}
The two lowest-order terms are
\begin{align}
  \mathcal{P}^{(0)}(t) &= 3(4\pi)^{3/2}\Sigma^{3/2}(t)\E^{-\Sigma(t)}\;,
  \nonumber\\
  \mathcal{P}^{(1)}(t) &= \frac{(4\pi)^{3/2}}{28}
  \frac{\sigma_3^2}{\sigma_2^2}\Sigma^{5/2}(t)\E^{-\Sigma(t)}\left[
    123-132\Sigma(t)+20\Sigma^2(t)
  \right]\;.
\label{eq:161}
\end{align}
The leading-order term in (\ref{eq:160}) reproduces the result (\ref{eq:148}), as it should. Generally, the functions $\mathcal{P}^{(m)}(t)$ are proportional to the moments $\sigma_n^2$ of the initial power spectrum,
\begin{equation}
  \mathcal{P}^{(m)}(t) \propto \sigma_{m+2}^2\;,
\label{eq:162}
\end{equation}
and also depend on lower order moments. Explicit expressions for the coefficient functions $\mathcal{P}^{(m)}(t)$ are given in \cite{2021arXiv211007427K}.

\begin{figure}[t]
\includegraphics[width=\hsize]{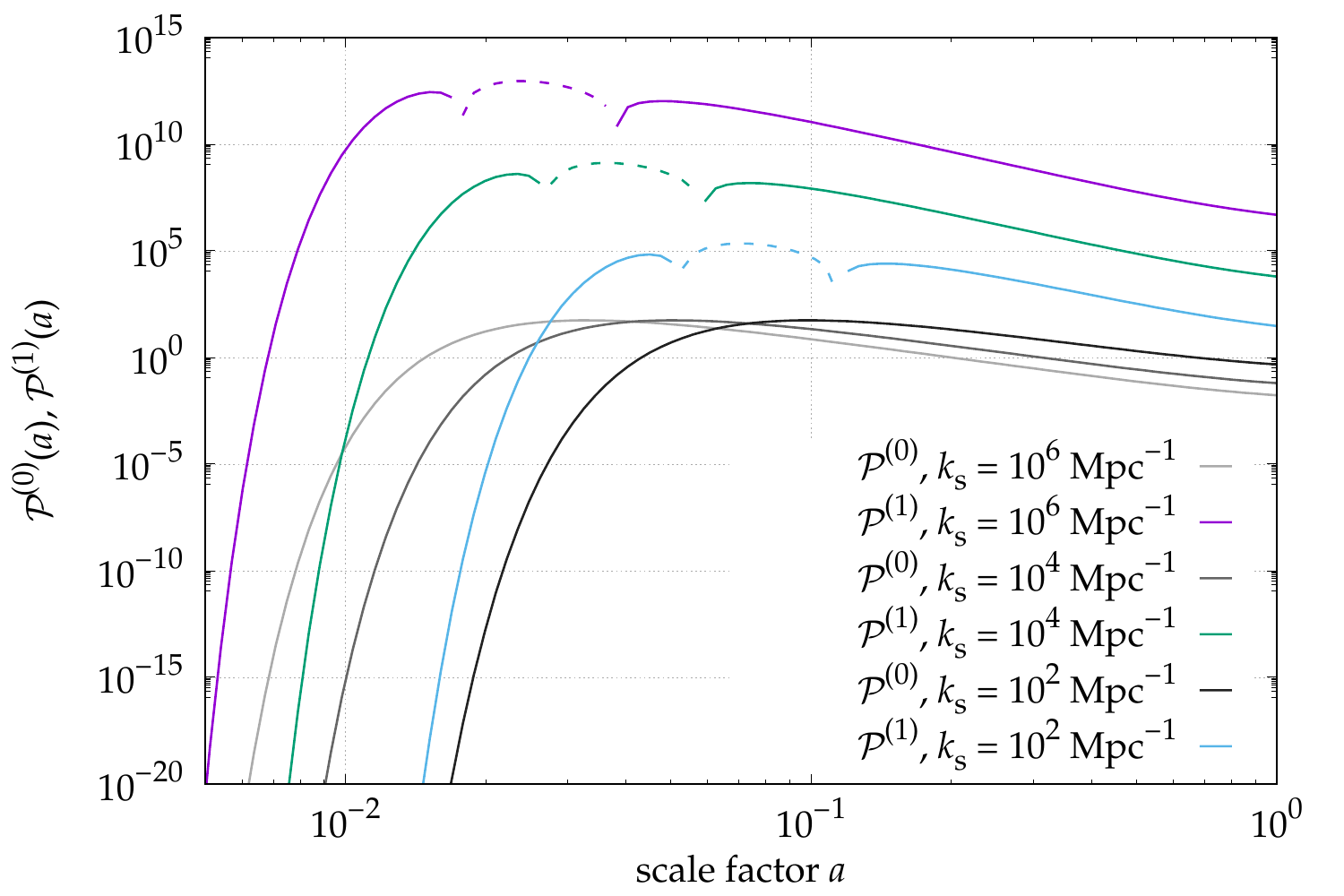}
\caption{Comparison of leading (gray lines) and next-to-leading (colored) order amplitudes of the free power spectrum asymptotics from (\ref{eq:161}) as a function of scale factor $a$ for three different values of the initial small-scale smoothing wave number $k_s$. Dashed lines indicate negative values of the amplitude $\mathcal{P}^{(1)}$.}
\label{fig:14}
\end{figure}

\subsubsection{Conclusions from the asymptotics of the free power spectrum}

Our conclusions on the asymptotic behaviour of the free power spectrum rest upon rather general assumptions. We have assumed that the initial momenta of the particles in our ensemble are drawn from a Gaussian random velocity-potential field and correlated in such a way as to satisfy the continuity equation between the initial density and velocity fields. For deriving the asymptotic series of the free power spectrum, we did not have to specify the shape of the initial power spectrum, but only had to assume that its moments $\sigma_2^2$ for the leading-order and $\sigma_3^2$ for the next-to-leading order terms exist.

Moreover, none of our results obtained so far depends on anything specific for cosmology. We have transformed the Hamiltonian equations of motion to the expanding spatial background, which resulted in a time-dependent particle mass and a specific form of the Poisson equation. Having focussed on the free power spectrum, however, we could describe the particle trajectories purely kinematically, without invoking any specific dynamics. All we have assumed in this regard is that a time coordinate $t$ exists in terms of which particle trajectories take on the inertial form (\ref{eq:56}). The formation of the asymptotic $k^{-3}$ tail of the free power spectrum is thus an effect of collective free streaming of classical particles with phase-space positions drawn from an initially Gaussian random field, and the exponent $-3$ is set solely by the number of spatial dimensions.

Most importantly, the exponential damping factor $\exp(-Q_\mathrm{D})$ appearing in (\ref{eq:107}) is exactly cancelled in the asymptotic terms. Freely-streaming particles with correlated initial momenta thus do not lead to exponential damping of small-scale structures.

The amplitude $\mathcal{P}^{(0)}(t)$ of the leading-order asymptotic term starts at zero for $t = 0$, reaches a maximum value of
\begin{equation}
  \mathcal{P}^{(0)}_\mathrm{max} = 3\left(
    \frac{6\pi}{\E}
  \right)^{3/2} \approx 54.78
\label{eq:163}
\end{equation}
when $\Sigma_\mathrm{max} = 3/2$ and then decreases again. The increase is due to the fact that freely-streaming particles create structures where their flow is locally convergent, while the decrease is due to the fact that they fly past each other and erase these structures again after they pass the point of convergence.

The values of $\Sigma_\mathrm{max}$ and $\mathcal{P}^{(0)}_\mathrm{max}$ have an absolute meaning, irrespective of the cosmological background and the shape of the initial density-fluctuation power spectrum. The definition of $\Sigma$ in (\ref{eq:149}) shows that $\tau_2 = t\sigma_2$ is the relevant time coordinate for structure formation by collective streaming. The lower the moment $\sigma_2$ of the initial power spectrum is, the more time $t$ it will take the asymptotic $k^{-3}$ tail to reach its maximum amplitude, with
\begin{equation}
  t_\mathrm{max} = \sqrt{\frac{5}{3}}\sigma_2^{-1}\;.
\label{eq:164}
\end{equation}
This time scale $t_\mathrm{max}$ is expected to set an important scale for structure formation.

The ratio of the asymptotic terms of the next-to-leading and the leading order is
\begin{equation}
  k^{-2}\frac{\mathcal{P}^{(1)}(t)}{\mathcal{P}^{(0)}(t)} =
  \frac{k^{-2}}{84}\frac{\sigma_3^2}{\sigma_2^2}\,
  \Sigma\left(123-132\Sigma+20\Sigma^2\right)\;.
\label{eq:165}
\end{equation}
When this value drops below unity above a certain wave number $k_0$, the free power spectrum attains its asymptotic behaviour $\propto k^{-3}$ for scales smaller than $k_0$.

\begin{figure}[t]
\includegraphics[width=\hsize]{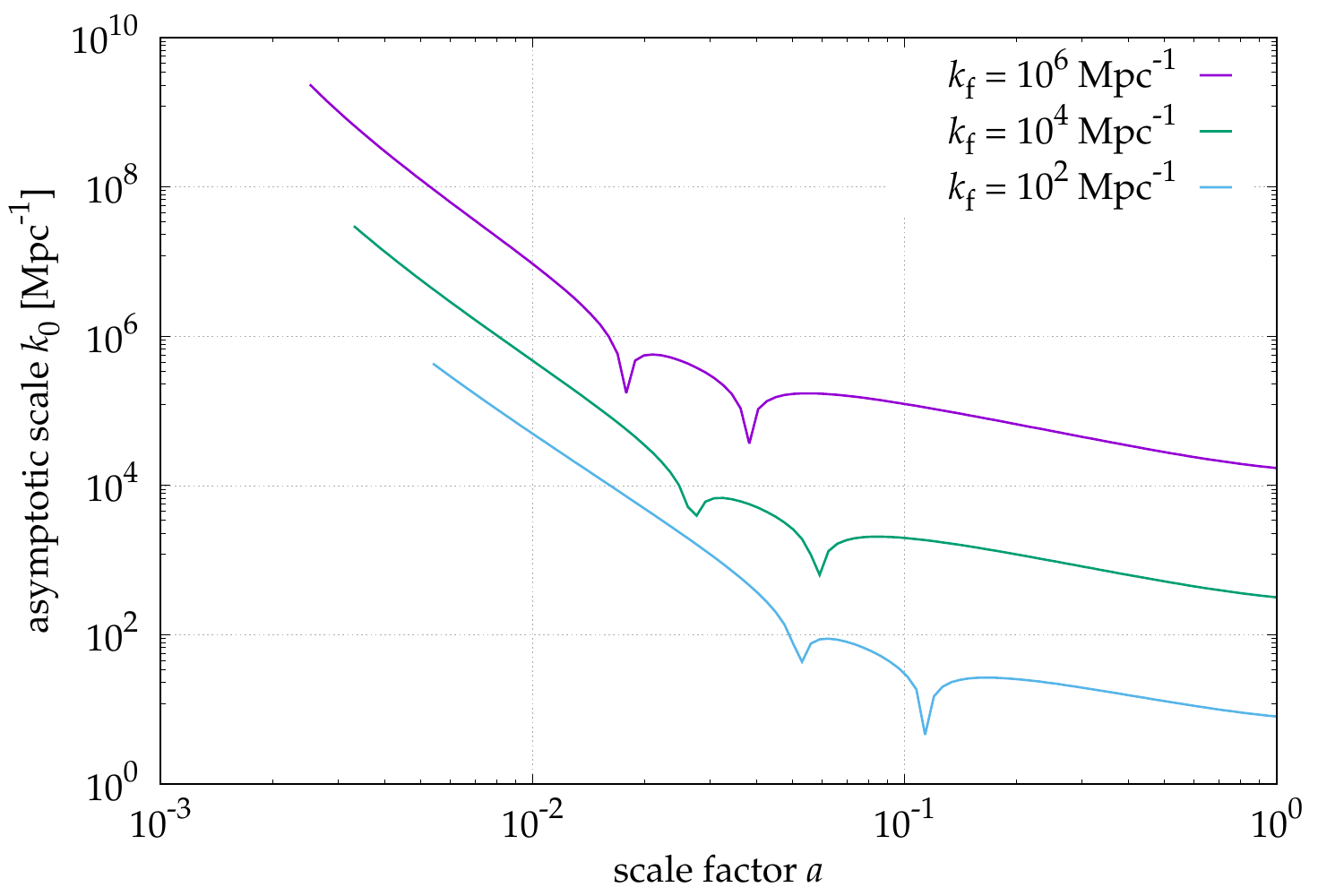}
\caption{The wave number $k_0$ that indicates the transition to the small-scale $k^{-3}$ asymptotics of the free power spectrum, as defined in (\ref{eq:165}) as a function of scale factor $a$ is shown for three different values of the initial small-scale smoothing wave number $k_s$. When the smoothing of initial scales is stronger (smaller values of $k_s$), the $k^{-3}$ asymptotics sets in at earlier at larger scales. As time progresses, larger and larger scales enter the $k^{-3}$ asymptotics.}
\label{fig:15}
\end{figure}

\subsection{Free density-fluctuation bispectrum}

We now turn to the asymptotic behaviour of the free bispectrum, given in (\ref{eq:118}). For finding an expression valid on small scales, an approach based on Laplace-type integrals is possible. For the lowest order term of an asymptotic series, we require an approximate expansion for the exponent $Q_\mathrm{C}^{(3)}$ around its relevant critical point. While $Q_\mathrm{C}^{(3)}$ has a critical point at zero, its Hessian with respect to $q$ is non-invertible at $q = 0$ \cite{2022Waibel}. The previously applied method based on Morse's lemma thus fails, but it can be suitably adapted using the splitting lemma of functions, which extends the validity of the Morse lemma to cases of Hessians with positive corank (see eg. \cite{poston2014catastrophe}).

The splitting lemma states that, if $f$ is a polynomial of order $\ge2$ with a critical point at $x_0$ and a Hessian at $x_0$ with rank $k$, then
\begin{equation}
  f \sim \sum_{j=1}^kx_j^2+g(x_{k+1},\ldots,x_n)
\label{eq:166}
\end{equation}
with either $g = 0$ or $g$ a polynomial of order $\ge3$ which is uniquely determined up to an equivalence transformation, provided the critical point is isolated. Under this assumption, the proof of the splitting lemma can be adapted to the specific form of $Q_\mathrm{C}^{(3)}$.

Since we are studying a statistically isotropic situation, we can without loss of generality orient the coordinate system such that $k_2$ points into $\vec e_z$ direction and $k_3$ falls into the $x$-$z$ plane. Then, the only non-vanishing components of the wave vectors $k_{2,3}$ are $k_{2z}$, $k_{3x}$ and $k_{3z}$, in terms of which the asymptotic behaviour of the bispectrum is given by
\begin{equation}
  \mathcal{B}\left(k_{2z},k_{3x},k_{3z}\right) \sim
  \frac{c_0\,\E^{-3\Sigma/2}}
    {\tau_2^5\tau_3^{1/2}c_4^{1/4}(k)k_{3x}^2k_{2z}^{5/2}}
\label{eq:167}
\end{equation}
for $k_1,k_2,k_3\to\infty$ with the coefficients
\begin{align}
  c_0 &= 9000\sqrt[4]{6.3}\,\pi^{5/2}\Gamma(5/4)\;,\nonumber\\
  c_4(k) &= k_{3x}^4+6k_{3x}^2k_{3z}^2+5k_{3z}^4+k_{2z}^2\left(
    k_{3x}^2+5k_{3z}^2
  \right)+2k_{2z}k_{3z}\left(
    3k_{3x}^2+5k_{3z}^2
  \right)\;,
\label{eq:168}
\end{align}
as shown in \cite{2022Waibel}.

In the triangle formed by $k_{1,2,3}$, let $\alpha$ and $\beta$ be the angles between $k_2$ and $k_3$ and between $k_1$ and $k_3$, respectively. Then, by the sine theorem, the norm of $k_3$ is
\begin{equation}
  \vert k_3\vert = Ak\quad\mbox{with}\quad
  A = \frac{\sin(\alpha+\beta)}{\sin\beta}\;,
\label{eq:169}
\end{equation} 
and its components are $k_{3x} = -Ak\sin\alpha$ and $k_{3z} = -Ak\cos\alpha$. We can then bring the function $c_4(k)$ into the form
\begin{equation}
  c_4(k) = f(\alpha,\beta)\left(\frac{k}{\sin^2\alpha}\right)^4
\label{eq:170}
\end{equation}
with
\begin{equation}
  f(\alpha,\beta) = A^2\sin^8\alpha\left[
    \left(1+A^2\right)\left(1+4\cos^2\alpha\right)-
    2A\cos\alpha\left(3+2\cos^2\alpha\right)
  \right]\;,
\label{eq:171}
\end{equation}
allowing us to write the asymptotic expression for the bispectrum as
\begin{equation}
  \mathcal{B}\left(k,\alpha,\beta\right) \sim
  \frac{c_0}{f^{1/4}(\alpha,\beta)}\,
  \frac{\E^{-3\Sigma/2}}{\tau_2^5\tau_3^{1/2}}\,k^{-11/2}\;.
\label{eq:172}
\end{equation}
Values of the function $f^{1/4}(\alpha,\beta)$ are tabulated for some special cases in Tab.~1.

\begin{figure}[t]
\includegraphics[width=\hsize]{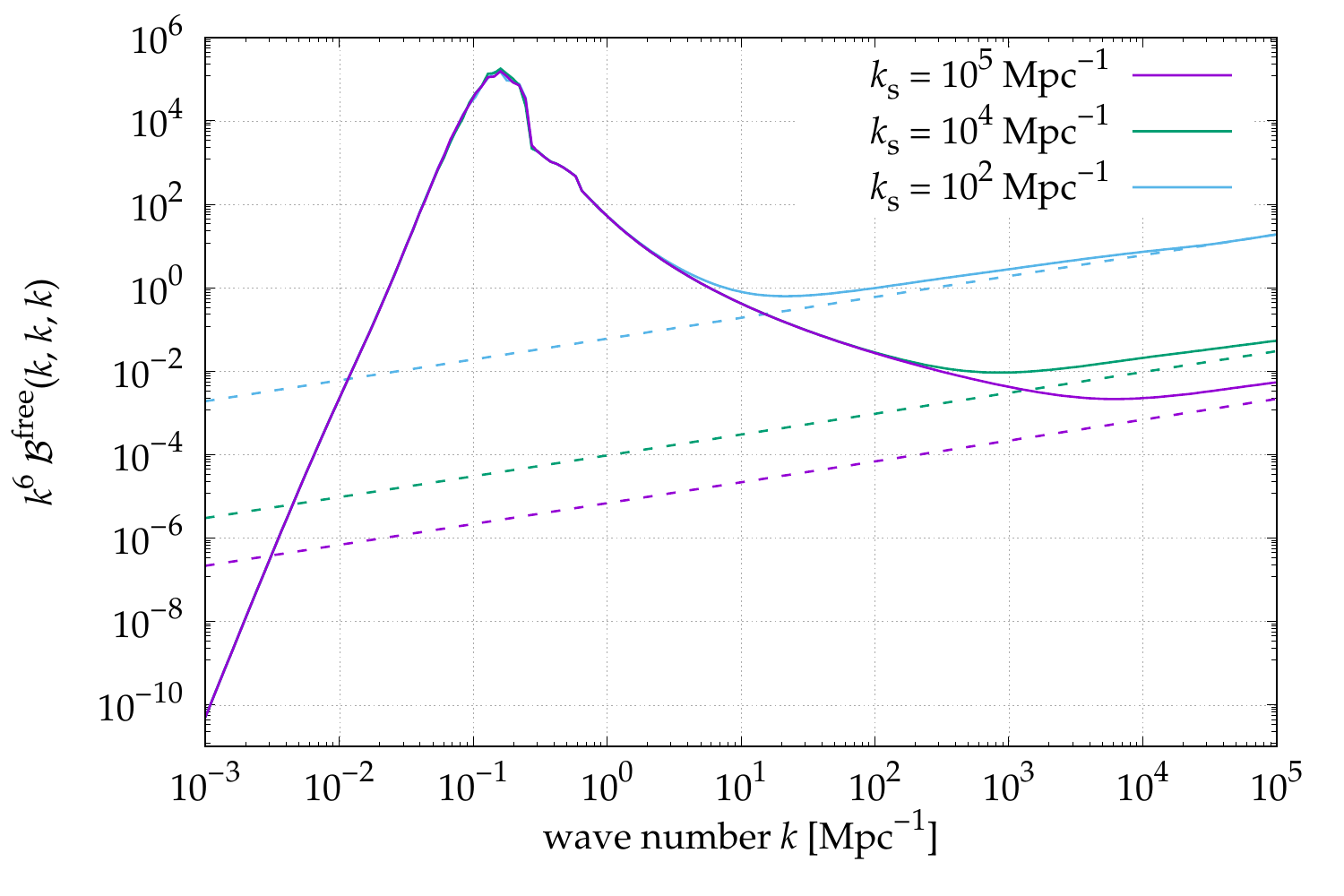}
\caption{The dimensionless free bispectrum $k^6 \mathcal{B}$ at redshift $z=0$ for the equilateral configuration of $k$ vectors for three values of the initial smoothing wave number $k_s$ are shown (solid lines) from (\ref{eq:118}) together with their corresponidng asymptotics (dashed lines) from (\ref{eq:172}).}
\label{fig:16}
\end{figure}

\begin{figure}[t]
  \centering
  \includegraphics[width=0.4\hsize]{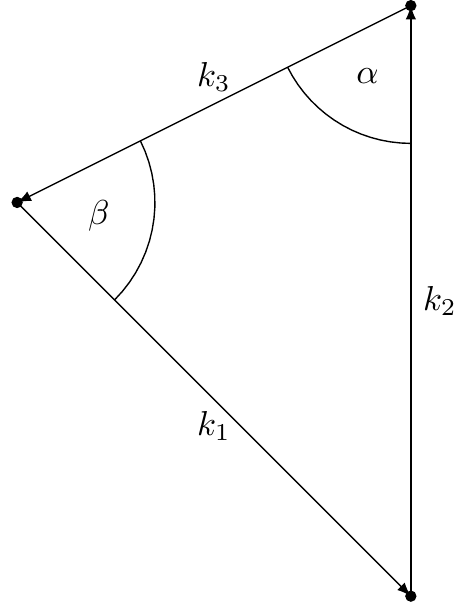}
\caption{Arrangement of the wave vectors $k_{1,2,3}$}
\label{fig:17}
\end{figure}

\begin{table}[ht]
  \caption{Values of the function $f^{1/4}(\alpha,\beta)$ for some special configurations of the triangle $k_{1,2,3}$.}
  \begin{center}
  \begin{tabular}{|l|c|c|c|c|}
    \hline
    case & $\vert k_{1,2,3}\vert$ & $\alpha$ & $\beta$ & $f(\alpha,\beta)$ \\
    \hline
    isosceles & $k$ &
      $\pi/3$ & $\pi/3$ & $(3/4)^4/2$ \\
    right isosceles & $\vert k_2\vert = k = \vert k_3\vert$ &
      $\pi/2$ & $\pi/4$ & $2$ \\
    right isosceles & $\vert k_1\vert = k = \vert k_2\vert$ &
      $\pi/4$ & $\pi/4$ & $1/8$ \\
    acute & $\vert k_2\vert = k = \vert k_3\vert$ &
      $\ll1$ & $\approx\pi/2$ & $\approx\alpha^{10}$ \\
    \hline
  \end{tabular}
  \end{center}
\end{table}

\subsection{Free velocity power spectrum}

The asymptotic expansions (\ref{eq:156}) of the functions $a_{1,2}(q)$ show that
\begin{equation}
  C_{p_1p_2} \sim \frac{\sigma_1^2}{3}\id{3}+\mathcal{O}\left(q^2\right)
  \;,\quad
  D_jQ \sim \mathcal{O}\left(q^2\right)\;,
\label{eq:173}
\end{equation}
thus terms proportional to $k^2$ show up only at $\mathcal{O}(q^4)$ in the terms in parentheses in (\ref{eq:131}). The dependence on the wave number $k$ of the leading-order asymptotic term of the velocity power spectrum $\mathcal{P}_\Pi(k)$ for $k\to\infty$ thus remains unchanged compared to that of the density-fluctuation power spectrum $\mathcal{P}(k)$. Only its amplitude changes because we need to replace the coefficient $b_0$ in the asymptotic expansion (\ref{eq:159}), and thus also the coefficient function $\mathcal{P}^{(0)}$ in (\ref{eq:160}), as
\begin{equation}
  b_0\to-\frac{\sigma_1^2}{3}b_0\;,\quad
  \mathcal{P}^{(0)}\to-\frac{\sigma_1^2}{3}\mathcal{P}^{(0)}\;.
\label{eq:174}
\end{equation}
This takes us to the leading-order asymptotic expression
\begin{equation}
  \mathcal{P}_\Pi(k) \sim \frac{\mathcal{P}^{(0)}_\Pi}{k^3}\,\id{3}
\label{eq:175}
\end{equation}
for $k\to\infty$, with
\begin{equation}
  \mathcal{P}^{(0)}_\Pi = -\frac{\sigma_1^2}{3}\mathcal{P}^{(0)} =
  -(10\pi)^{3/2}\frac{\sigma_1^2}{\tau_2^3}\E^{-\Sigma(t)}\;.
\label{eq:176}
\end{equation} 
The time dependences of the leading-order asymptotic terms in the density-fluctuation and the velocity power spectra are thus the same, their maximum amplitudes are reached at $\Sigma_\mathrm{max} = 3/2$, but the maximum amplitude of the leading-order asymptotic term in the velocity power spectrum is
\begin{equation}
  \mathcal{P}^{(0)}_{\Pi,\mathrm{max}} =
  -\sigma_1^2\left(\frac{6\pi}{\E}\right)^{3/2}\;;
\label{eq:177}
\end{equation} 
compare (\ref{eq:163}). The velocity correlation on small scales is negative and has the same amplitude as the density-fluctuation power spectrum multiplied by the initial velocity dispersion $\sigma_1^2$. This explains the origin of the $k^{-3}$ asymptotics, i.e.\ structures on small scales: initially convergent particle streams cross which leads to caustics.

\subsection{Density-fluctuation power spectrum for interacting particles}

So far, we have neglected particle interactions altogether. This does however not mean that no gravity was included because the Zel'dovich inertial trajectories are subject to the large-scale, linear part of the gravitational interaction, as specified in Sect.~\ref{sec:2}. We shall now proceed to include particle interactions in the mean-field approximation \cite{2021ScPP...10..153B}. 

\subsubsection{Forces in the mean-field approximation}

So far, we have neglected particle interactions beyond those that are already contained in the reference trajectories modelled with the Zel'dovich approximation. The density-fluctuation two-point function written down in (\ref{eq:95}) is still exact, however. The shift tensor $\tens L$, multiplied with the actual particle trajectories, is
\begin{equation}
  \left(\tens L,\bar{\tens x}(t)\right) = -k_1\cdot\left(
    \bar q_1(t)-\bar q_2(t)
  \right)
\label{eq:178}
\end{equation}
according to (\ref{eq:97}), taking into account that $k_1+k_2=0$ due to statistical homogeneity. The spatial trajectories $\bar q_j(t)$ can be split into the free part $\bar q_j^{(0)}(t)$ described by the Zel'dovich approximation, and a part $\bar y(t)$ describing the deviations from the Zel'dovich reference trajectories caused by the particle interactions,
\begin{equation}
  \bar q_j(t) = \underbrace{q_j+tp_j}_{\bar q_j^{(0)}}-
  \underbrace{\int_0^t\D t'g_\mathrm{H}(t,t')m\nabla_j\phi}_{-\bar y_j}\;.
\label{eq:179}
\end{equation}
We are quoting the result (\ref{eq:54}) here which contains the Hamiltonian propagator $g_\mathrm{H}$ given in (\ref{eq:48}), and the gradient of the potential $\phi$ which satisfies the Poisson equation (\ref{eq:55}).

In (\ref{eq:98}), we continued with $\bar q_{1,2}^{(0)}$, neglecting $\bar y_{1,2}$. We shall now include $\bar y_{1,2}$ in a mean-field approximation. For completing the generating functional $Z[\tens L]$, we require the scalar product
\begin{equation}
  \left(\tens L,\bar{\tens y}\right) =
  k_1\cdot\int_0^t\D t'g_\mathrm{H}(t,t')m\left(
    \nabla_1\phi-\nabla_2\phi
  \right)\;.
\label{eq:180}
\end{equation}

We now approximate the potential gradients $\nabla_j\phi$ by a mean-field average,
\begin{equation}
  \nabla_j\phi\to\left\langle\nabla_j\phi\right\rangle\;,
\label{eq:181}
\end{equation}
which we construct in the following way. Let $V(q)$ be an interaction potential linearly superposed by contributions $v(q)$ due to individual particles at positions $q_i$, then
\begin{equation}
  V(q) = \sum_{i=1}^Nv(q-q_i) = \int_yv(q-y)
  \sum_{i=1}^N\delta_\mathrm{D}(y-q_i) = \int_yv(q-y)\rho(y)\;,
\label{eq:182}
\end{equation}
because the sum of delta distributions is the (number) density $\rho$ of the particles. The potential gradient at the position $q_j$ of another particle is
\begin{equation}
  \nabla_{q_j}V(q) = \int_q\delta_\mathrm{D}(q-q_j)\nabla V(q) =
  \int_q\rho_j(q)\nabla V(q)\;,
\label{eq:183}
\end{equation}
where $\rho_j(q)$ is the contribution of particle $j$ to the particle number density. Thus, the potential gradient acting on particle $j$ is
\begin{equation}
  \nabla\phi = \int_q\int_y\rho_j(q)\nabla v(q-y)\rho(y)\;,
\label{eq:184}
\end{equation}
where $v$ is the interaction potential between a pair of individual particles separated by $q-y$. The potential gradient contributed by particles at a distance $q-y$ from particle $j$ is
\begin{equation}
  \nabla\phi(q-y) = \int_q\rho_j(q)\nabla v(q-y)\rho(y)\;.
\label{eq:185}
\end{equation}

In the cosmological situation, we imagine a test particle moving through a collection of other particles, exposed to their collective gravitational field. The number density of these particles can be considered to be arbitrarily high. The force exerted by any individual particle on the test particle is then an arbitrarily small contribution to the total force. Fluctuations of this force caused by individual particles can then be ignored. This situations is typically well described by a mean-field approximation. Averaging the potential gradient $\nabla\phi$ leads to
\begin{equation}
  \left\langle\nabla\phi\right\rangle(q-y) = \int_q\nabla v(q-y)
  \left\langle\rho_j(q)\rho(y)\right\rangle\;,
\label{eq:186}
\end{equation}
containing the two-point function of the (number) density field. By definition of the two-point density auto-correlation function $\xi$,
\begin{equation}
  \left\langle\rho_j(q)\rho(y)\right\rangle = \frac{\bar n^2}{N}\left[
    1+\xi(q-y)
  \right]\;,
\label{eq:187}
\end{equation}
where $\bar n$ is the mean number density. The division by $N$ takes into account that we cross-correlate the one-particle density contribution $\rho_j$ with the density $\rho$.

Inserting (\ref{eq:187}) into (\ref{eq:186}) leaves
\begin{equation}
  \left\langle\nabla\phi\right\rangle(q-y) = \frac{\bar n^2}{N}
  \int_q\nabla v(q-y)\xi(q-y)
\label{eq:188}
\end{equation}
because the contribution by uncorrelated particles averages to zero in a statistically homogeneous and isotropic random field. Now, both factors in the integrand of (\ref{eq:188}) depend on the fixed separation $q-y$ only, but not on the point $q$ any more. This is a consequence of statistical homogeneity. The integral over $q$ thus only results in a volume factor which, together with the prefactor $N^{-1}$, cancels one of the $\bar n$ factors. Thus, the mean potential gradient is
\begin{equation}
  \left\langle\nabla\phi\right\rangle(q-y) = \bar n\nabla v(q-y)\xi(q-y)\;.
\label{eq:189}
\end{equation}

By the Fourier convolution theorem, the Fourier transform of the average potential gradient is the convolution of the Fourier transforms of the individual factors. The Fourier transform between particles is given by (\ref{eq:64}), approximated by (\ref{eq:65}). Thus, the Fourier transform of its gradient is
\begin{equation}
  \widetilde{\nabla v} =
  -\frac{\I A_\varphi k}{\bar n\left(k_0^2+k^2\right)}\;.
\label{eq:190}
\end{equation}
The Fourier transform of the two-point density auto-correlation function $\xi$ is the density-fluctuation power spectrum $P_\delta(k)$. Since we select the motion of particles along Zel'dovich inertial trajectories (\ref{eq:56}) as a reference, the most appropriate power spectrum to insert here would be the free power spectrum $\mathcal{P}(k)$. We shall further simplify this choice below and keep the symbol $P_\delta(k)$ for now. We thus find the Fourier transform of the mean potential gradient to be
\begin{equation}
  \widetilde{\left\langle\nabla\phi\right\rangle}(k) = -\I A_\varphi
  \left(\frac{k}{k_0^2+k^2}\right)\ast P_\delta(k)\;.
\label{eq:191}
\end{equation}

We thus replace the expressions $k_1\cdot(\nabla_1\phi-\nabla_2\phi)$ appearing in the scalar product (\ref{eq:180}) by the averaged expression
\begin{align}
  F(k_1) = k_1\cdot\left(
    \widetilde{\left\langle\nabla\phi\right\rangle}(k_1)-
    \widetilde{\left\langle\nabla\phi\right\rangle}(k_2)
  \right) &= 2k_1\widetilde{\left\langle\nabla\phi\right\rangle}(k_1) 
  \nonumber\\ &=
  -2\I A_\varphi\int_{k'}
  \frac{k_1\cdot(k_1-k')}{k_0^2+(k_1-k')^2}\,P_\delta(k')\;,
\label{eq:192}
\end{align}
where we have used that $\widetilde{\langle\nabla\phi\rangle}(-k_1) = -\widetilde{\langle\nabla\phi\rangle}(k_1)$. For carrying out the convolution, we introduce the cosine $\mu$ of the angle between $k_1$ and $k'$ and define $\kappa = \vert k'\vert/\vert k_1\vert$ as well as $\kappa_0 = \vert k_0\vert/\vert k_1\vert$. This enables us to write
\begin{equation}
  F(k_1) = -2\I A_\varphi\sigma_J^2(k_1)
\label{eq:193}
\end{equation}
where $\sigma_J^2$ is the moment
\begin{equation}
  \sigma_J^2(k_1) = \frac{\vert k_1\vert^3}{(2\pi)^2}
  \int_0^\infty\kappa^2\D\kappa\,P_\delta(\vert k_1\vert\kappa)\,
  J(\kappa,\kappa_0)
\label{eq:194}
\end{equation}
of the density-fluctuation power spectrum with the filter function
\begin{align}
  J(\kappa,\kappa_0) &= \int_{-1}^1\D\mu\,
  \frac{1-\kappa\mu}{1+\kappa_0^2+\kappa^2-2\kappa\mu} \nonumber\\ &=
  1+\frac{1-\kappa^2-\kappa_0^2}{4\kappa}
  \ln\frac{\kappa_0^2+(1+\kappa)^2}{\kappa_0^2+(1-\kappa)^2}\;.
\label{eq:195}
\end{align}

\begin{figure}[t]
  \centering
  \includegraphics[width=\hsize]{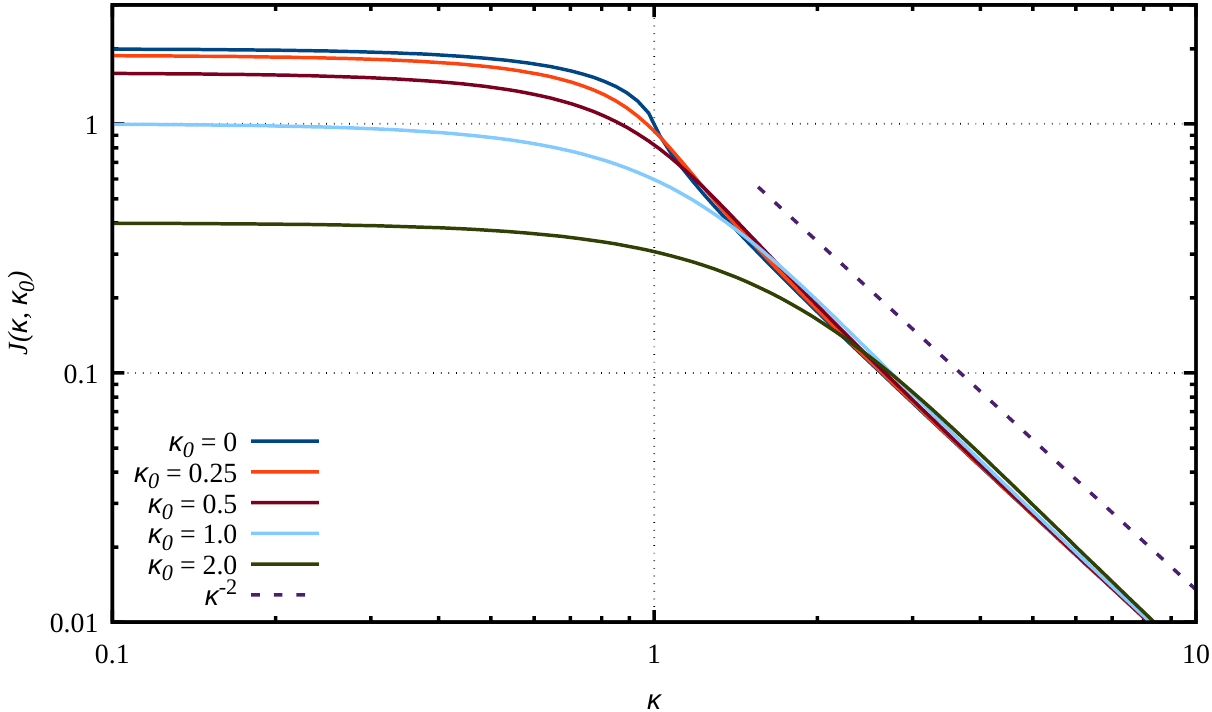}
\caption{Filter function $J$ as defined in (\ref{eq:195}) for different values of $\kappa_0$.}
\label{fig:18}
\end{figure}

\subsubsection{Density-fluctuation power spectrum in the mean-field approximation}

We can now write the mean-field averaged scalar product between $\tens L$ and $\bar{\tens y}$ as
\begin{equation}
  \left\langle\left(\tens L,\bar{\tens y}\right)\right\rangle(k_1) =
  -2\I\int_0^t\D t'g_\mathrm{H}(t,t')mA_\varphi\sigma_J^2(k_1)\;.
\label{eq:196}
\end{equation} 
Since this expression does not depend on the initial phase-space coordinates of the particle ensemble any more, it can be pulled in front of the integral over $\D\Gamma$ in the generating functional. This leads to
\begin{equation}
  Z[\tens L] = \E^{\langle S_\mathrm{I}\rangle}Z_0[\tens L]\;,
\label{eq:197}
\end{equation}
containing the mean-field interaction term
\begin{equation}
  \left\langle S_\mathrm{I}\right\rangle(k_1) =
  \I\left\langle\left(\tens L,\bar{\tens y}\right)\right\rangle(k_1) =
  2\int_0^t\D t'g_\mathrm{H}(t,t')mA_\varphi\sigma_J^2(k_1)\;.
\label{eq:198}
\end{equation} 
Invoking (\ref{eq:109}) and (\ref{eq:110}) once more, the non-linear, density-fluctuation power spectrum, including particle interactions in the mean-field approximation, can be written as
\begin{equation}
  P_\delta^\mathrm{(nl)}(k) = \E^{\langle S_\mathrm{I}\rangle(k)}
  \mathcal{P}(k)\;.
\label{eq:199}
\end{equation}

\begin{figure}[t]
  \centering
  \includegraphics[width=\hsize]{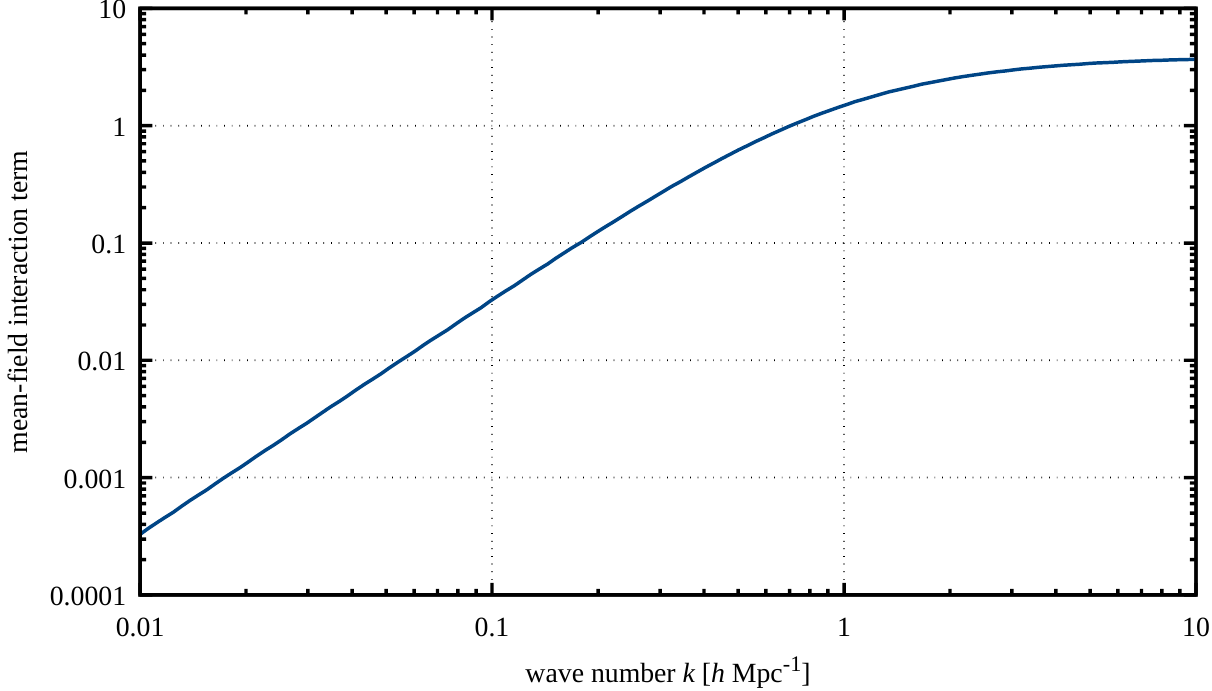}
\caption{The mean-field interaction term $\langle S_\mathrm{I}\rangle$ as defined in (\ref{eq:198}) as a function of wave number $k$.}
\label{fig:19}
\end{figure}

This is a convenient and simple expression whose merits need to be assessed by comparing it with, e.g., density-fluctuation power spectra derived from numerical simulations. In \cite{2021ScPP...10..153B}, we have further simplified it by replacing the free, non-linear power spectrum $\mathcal{P}$ by the linearly evolved power spectrum $P_\delta^\mathrm{(lin)}$, and the power spectrum $P_\delta(k)$ to be inserted into the moment $\sigma_J^2$ given in (\ref{eq:194}) by a suitably damped version of $P_\delta^\mathrm{(lin)}$ as well. The result is the approximate expression
\begin{equation}
  P_\delta^\mathrm{(nl)}(k) \approx \E^{\langle S_\mathrm{I}\rangle(k)}
  P_\delta^\mathrm{(lin)}(k)
\label{eq:200}
\end{equation}
for the mean-field averaged, non-linear power density-fluctuation power spectrum, which reproduces the results from numerical simulations remarkably well. This result may give sufficient credit to the mean-field approximation for the particle interactions.

\begin{figure}[t]
  \centering
  \includegraphics[width=\hsize]{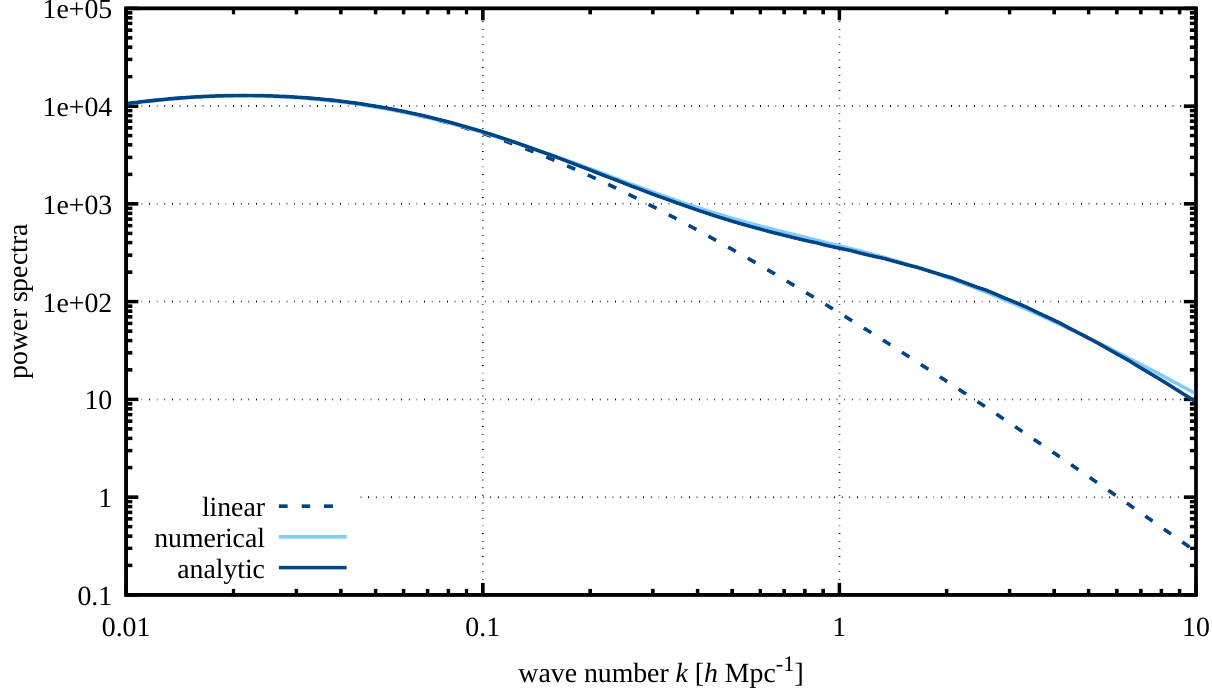}
\caption{The analytic mean-field approximated (dark blue line, (\ref{eq:200})), the non-linear power spectrum obtained numerically by \cite{2016MNRAS.459.1468M} (dark blue line) and the linear power spectrum (dashed line) are shown at redshift $z=0$. }
\label{fig:20}
\end{figure}

In our present context, we are aiming at a different conclusion, however. As (\ref{eq:198}) shows, the scale dependence of the mean-field interaction term is determined by the moment $\sigma_J^2$, which we now repeat in a different way,
\begin{equation}
  \sigma_J^2(k_1) = \int_kP_\delta(k)\,J\left(
    k/k_1,k_0/k_1
  \right)\;.
\label{eq:201}
\end{equation} 
The filter function $J(k/k_1,k_0/k_1)$ flattens for $k<k_1$ at a level decreasing with increasing $k_0/k_1$, and falls off $\propto k^{-2}$ for $k>k_1$. For $k_0/k_1\to0$, the filter function approaches $J\to2$ for $k\ll k_1$. In the small-scale limit, $k_1\to\infty$, the filter function is well approximated by $J\approx 2$ almost everywhere in the integration range in (\ref{eq:201}). Then, the integration covers the entire power spectrum, and the result becomes scale-independent. While the power spectrum in (\ref{eq:201}) is growing linearly, comparing with (\ref{eq:87}) shows that
\begin{equation}
  \sigma_J^2(k_1)\to 2t^2\sigma_2^2
\label{eq:202}
\end{equation}
for $k_1\to\infty$. The mean-field interaction term thus becomes independent of scale in the small-scale limit, which implies that the asymptotic behaviour of the non-linear power spectrum $P_\delta^\mathrm{(nl)}(k)$ for $k\to\infty$ will be the same as that of $\mathcal{P}$,
\begin{equation}
  P_\delta^\mathrm{(nl)}(k) \propto k^{-3}\;,
\label{eq:203}
\end{equation}
while its amplitude will be enhanced compared to (\ref{eq:148}) or the identical expression in (\ref{eq:161}) by the exponentiated mean-field interaction term.

\subsection{Lifting limitations}

Up to this point, we have made several simplifying assumptions. In particular, we have assumed that the moments $\sigma_n^2$ of the initial density-fluctuation power spectrum exist up to the order necessary, and we have neglected density-density and density-momentum correlations in the initial state. We shall now show how these assumptions can be lifted, and that they do not change our conclusions about the asymptotic behaviour of the density-fluctuation and velocity power spectra.

\subsubsection{Asymptotic behaviour for strictly Cold Dark Matter}

In deriving the leading-order asymptotic behaviour of the free power spectrum, we assumed that the second moment $\sigma_2$ of the initial velocity potential power spectrum exists as defined in (\ref{eq:87}). This is a save assumption also for cold dark matter since measurements of the spectral index $n_\mathrm{s}$ return values smaller than unity. For this reason, the initial cold dark matter power spectrum, which is assumed to be accurately described by linear theory, has a tail at large wave numbers falling off like $k^{n_\mathrm{s}-4}\log^2k$. The asymptotic series that we derived above from Erdélyi's theorem, however, requires that the moments $\sigma_n^2$ of arbitrary high order should exist. This is indeed only the case for power spectra that are exponentially cut off above some possibly very large wave number $k_\mathrm{s}$, i.e.\ if structures smaller than a typical length scale of $k_\mathrm{s}^{-1}$ do not exist at the initial time.

Should this not be the case, and it is unknown whether arbitrarily small dark-matter structures can be formed initially, we need stronger methods than previously applied to study the asymptotic behaviour of the free power spectrum. Such methods are provided by analytic continuations of inverse Mellin transforms. Moreover, it turns out that the asymptotics of a power spectrum for strictly cold dark matter correctly describes the intermediate regime of dark matter with large $k_\mathrm{s}$, which is not accessible from our previously derived asymptotic series due to its divergent nature. In this section, we begin with an assumed wide class for the asymptotic behaviour of the initial density-fluctuation power spectrum and use the Mellin-transform technique to derive the asymptotic behaviour of the momentum correlation function $a_\parallel(q)$ for $q\to0$. We then insert the resulting expression into the free, non-linear power spectrum $\mathcal{P}$ from (\ref{eq:111}) and obtain its asymptotic behaviour in the small-scale limit.

We thus begin by assuming that the initial density-fluctuation power spectrum admits an asymptotic expansion of the form
\begin{equation}
  P_\delta^\mathrm{(i)}(k) \sim k^{n_\mathrm{s}-4}
  \sum_{m=0}^\infty k^{-m}\sum_{n=0}^2c_{mn}\log^n k
\label{eq:204}
\end{equation}
for $k\to\infty$ with real coefficients $c_{mn}$. This represents a very wide class of asymptotic behaviour (see eg. \cite{1986ApJ...304...15B, weinberg2008cosmology}).

Next, we need the Mellin transform of a function $f$, which is defined by
\begin{equation}
  \mathcal{M}\left[f;z\right] := \int_0^\infty\D k\,k^{z-1}f(k)\;,
\label{eq:205}
\end{equation}
where the integral typically converges on a strip of the complex plane with $a < \Re\,z < b$. Equipped with the Mellin transform, we study integral transforms of functions $f$ with a kernel $h$, called H-transforms and defined by
\begin{equation}
  I(q) = \int_0^\infty\D k\,h(k)f\left(kq^{-1}\right)
\label{eq:206}
\end{equation} 
Under sufficently general conditions, allowing $I$ to be not absolutely convergent, integral transforms of this type can be expressed by Mellin transforms of the functions involved,
\begin{equation}
  I(q) = \frac{1}{2\pi\I}\int_{c-\I\infty}^{c+\I\infty}\D z\,
  q^z\mathcal{M}\left[f;1-z\right]\mathcal{M}\left[h;z\right]\;,
\label{eq:207}
\end{equation}
where $c$ is a real number falling into the common strip of analyticity of both Mellin transforms. Defining the function
\begin{equation}
  G(z) := \mathcal{M}\left[f;1-z\right]\mathcal{M}\left[h;z\right]
\label{eq:208}
\end{equation}
and assuming $G$ can be analytically continued to the right half-plane, the integral $I(q)$ can be expanded asymptotically as
\begin{equation}
  I(q) = -\sum_{c<\Re\,z<R}\mathrm{Res}\left\{
    q^zG(z)
  \right\}+
  \frac{1}{2\pi\I}\int_{R-\I\infty}^{R+\I\infty}\D z\,q^zG(z)
\label{eq:209}
\end{equation}
for $q\to0^+$, where $G(z)$ has no pole for $\Re\,R$ such the remaining integral is $o(q^R)$ (see eg. \cite{bleistein1975asymptotic}).

In order to derive the asymptotic behaviour of $a_\parallel(q)$ on small scales, we apply this method to the integrals (\ref{eq:84}) and obtain the results
\begin{align}
  a_1(q) & \sim -\frac{1}{6\pi^2}\mathcal{M}[P_\delta^{(i)};1]+
  \frac{q^2}{60\pi^2}\mathcal{M}[P_\delta^{(i)};3] \nonumber \\ &-
  \frac{q^{3-n_\mathrm{s}}}{2\pi^2}
  \sum_{n=0}^2c_{0n}\sum_{j=0}^n\binom{n}{j}\left(
    -\ln q
  \right)^j\mathcal{M}^{(n-j)}[j_1;n_s-4]+
  \mathcal{O}(q^{3-n_s})\;, \nonumber \\ 
  a_2(q) & \sim
  \frac{q^2}{30\pi^2}\mathcal{M}[P_\delta^{(i)};3] \nonumber \\ &+
  \frac{q^{3-n_\mathrm{s}}}{2\pi^2}
  \sum_{n=0}^2c_{0n}\sum_{j=0}^n\binom{n}{j}\left(
    -\ln q
  \right)^j\mathcal{M}^{(n-j)}[j_2;n_s-3]+\mathcal{O}(q^{3-n_s})
\label{eq:210}
\end{align}
for $q\to0$, where $\mathcal{M}^{(n-j)}$ denotes the $(n-j)$-th derivative of the Mellin transform with respect to the function argument. The detailed derivation can be found in \sk{2022Konrad}. The moments $\sigma_n^2$ of the initial velocity potential can be expressed by the Mellin transform as
\begin{equation}
  \sigma_n^2 = \frac{1}{2\pi^2}\mathcal{M}[P_\delta^{(i)};1+2(n-1)]\;,
\label{eq:211}
\end{equation}
which also allows us to identify the analytic continuation of these moments for non-converging integrals. Since $a_{\parallel}(q) = a_1(q) + \mu^2 a_1(q)$, we conclude from (\ref{eq:210}) that we can write
\begin{equation}
  a_\parallel(q) \sim
  -\frac{\sigma_1^2}{3}+\frac{\sigma_2^2}{30}(2\mu^2 +1)q^2-
  \xi(\mu^2,\log q)q^{3-n_\mathrm{s}}+\mathcal{O}(q^{4-n_s})\;,
\label{eq:212}
\end{equation}
as $q\to0$, where the function $\xi(\mu^2,\log q)$ is implicitly given by (\ref{eq:210}).

We can now insert this asymptotic, small-scale expression for $a_\parallel(q)$ into the free power spectrum (\ref{eq:111}) and use a suitable Taylor expansion of the exponential to arrive at the asymptotic expansion
\begin{equation}
  \mathcal{P}(k,t) \sim \frac{1}{(k\tau_2)^3}\sum_{m=0}^M
  \frac{1}{(k\tau_2)^{m(1-n_\mathrm{s})}}
  \sum_{n=0}^{2m}\mathcal{P}_{mn}(\tau_2)\log^n(k\tau_2)\;,
\label{eq:213}
\end{equation}
with time-dependent coefficient functions $\mathcal{P}_{nm}(\tau_2)$ to be specified, turning into constants at late times, $\tau_2\gg1$ \cite{2022Konrad}.

\begin{figure}[t]
\includegraphics[width=\hsize]{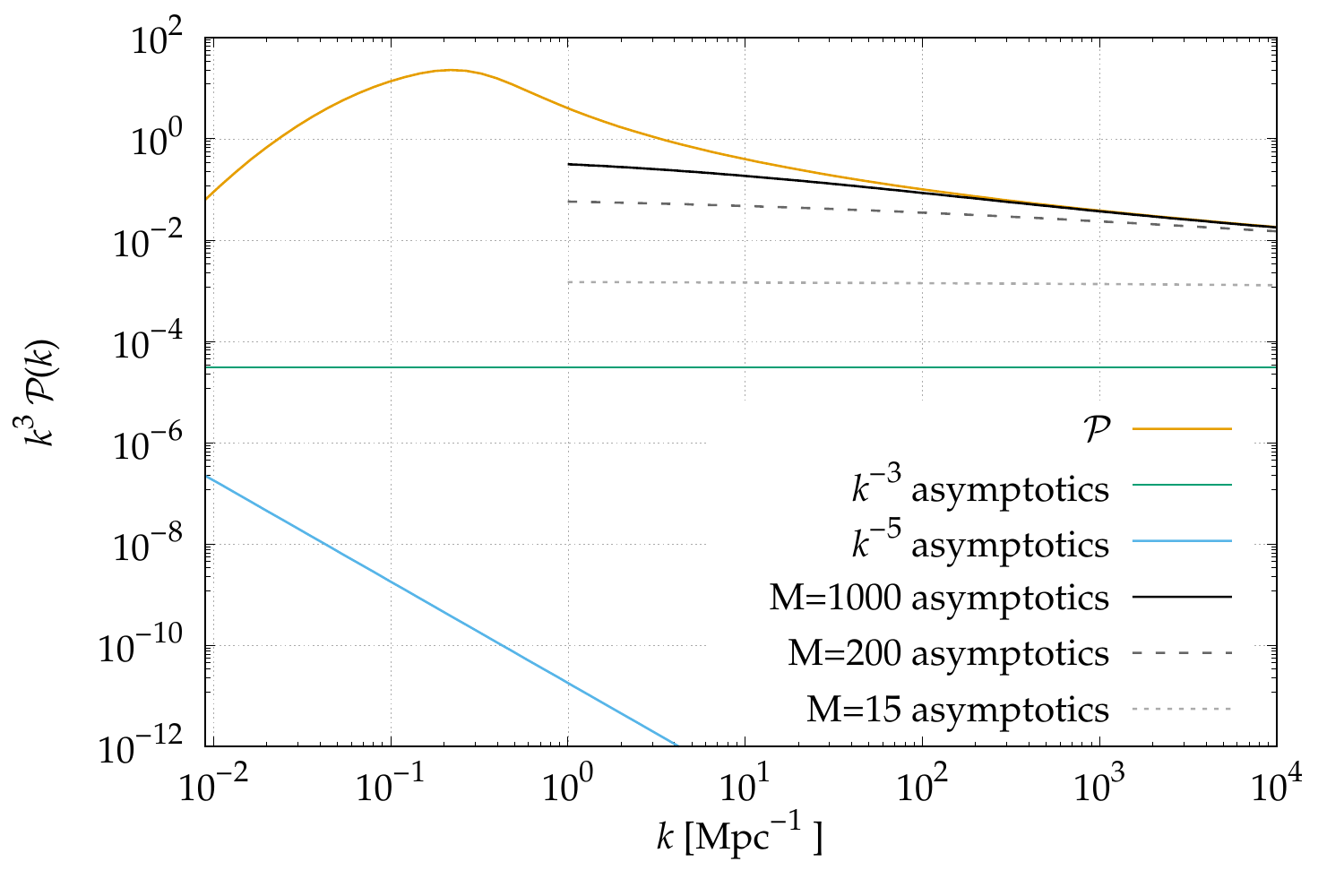}
\caption{The dimensionless free power spectrum $k^3 \mathcal{P}$ for strictly cold dark matter (golden line) at redshift $z=0$ is shown together with the $k^{-3}$ asymptotics (green line) and the asymptotics up to order $k^{-3-M(1-n_s)}$ as defined in (\ref{eq:213}) for three different values of $M$ (solid, dashed and double dashed gray lines). The $k^{-5}$ asymptotics is negligible.}
\label{fig:21}
\end{figure}

\begin{figure}[t]
\includegraphics[width=\hsize]{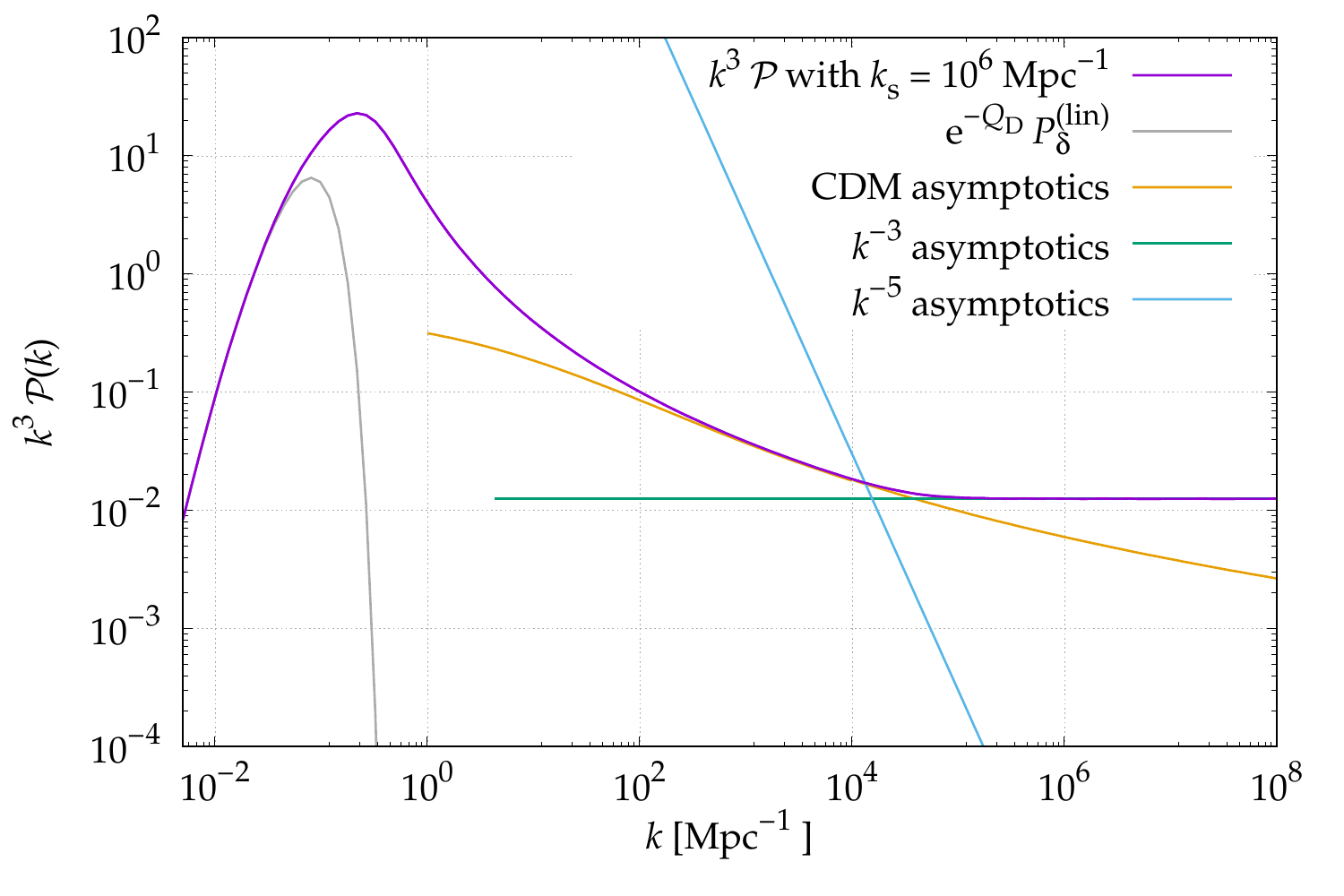}
\caption{The dimensionless free power spectrum $k^3 \mathcal{P}$ for WIMP dark matter with small scale smoothing wave number $k_s = 10^6$ Mpc$^{-1}$ (purple line) at redshift $z=0$ is shown together with the various asymptotics that we derived. At large scales, i.e. up to wave numbers $k \approx 0.02$ Mpc$^{-1}$, $k^3 \mathcal{P}$ is well described by the damped linearly evolved power spectrum (gray line). At small scales, above $k \approx 2 \cdot 10^4$ Mpc$^{-1}$, $k^3 \mathcal{P}$ becomes constant, accurately described by the $k^{-3}$ asymptotics (green line). Below $k \lesssim 1.2 \cdot 10^4$, the asymptotics of the free power spectrum of cold dark matter (golden line) aligns with the initially smoothed spectrum, where we chose $M=1000$. Note, that this is remarkable, as this line contains only information of the unsmoothed spectrum up to $q^{3-n_s}$ order of the initial momentum correlation functions $a_1$ and $a_2$. The crossing point of the $k^{-5}$ asymptotics (blue line) with the $k^{-3}$ asymptotics marks the transition from $k^{-3}$ asymptotics to the CDM asymptotics. To summarize, strictly CDM Zel'dovich power spectra approximate the intermediate regime of WIMP Zeldovich power spectra.}
\label{fig:21b}
\end{figure}

\begin{figure}[t]
\includegraphics[width=0.5\hsize]{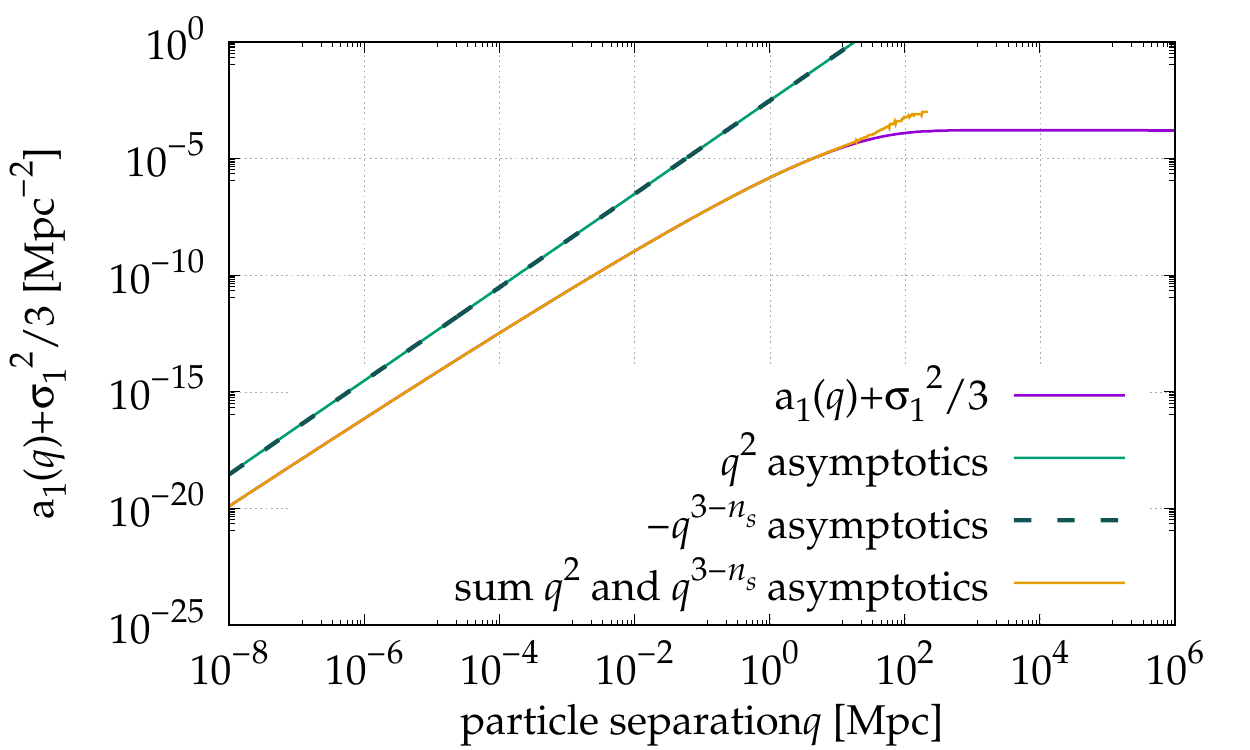}
\includegraphics[width=0.5\hsize]{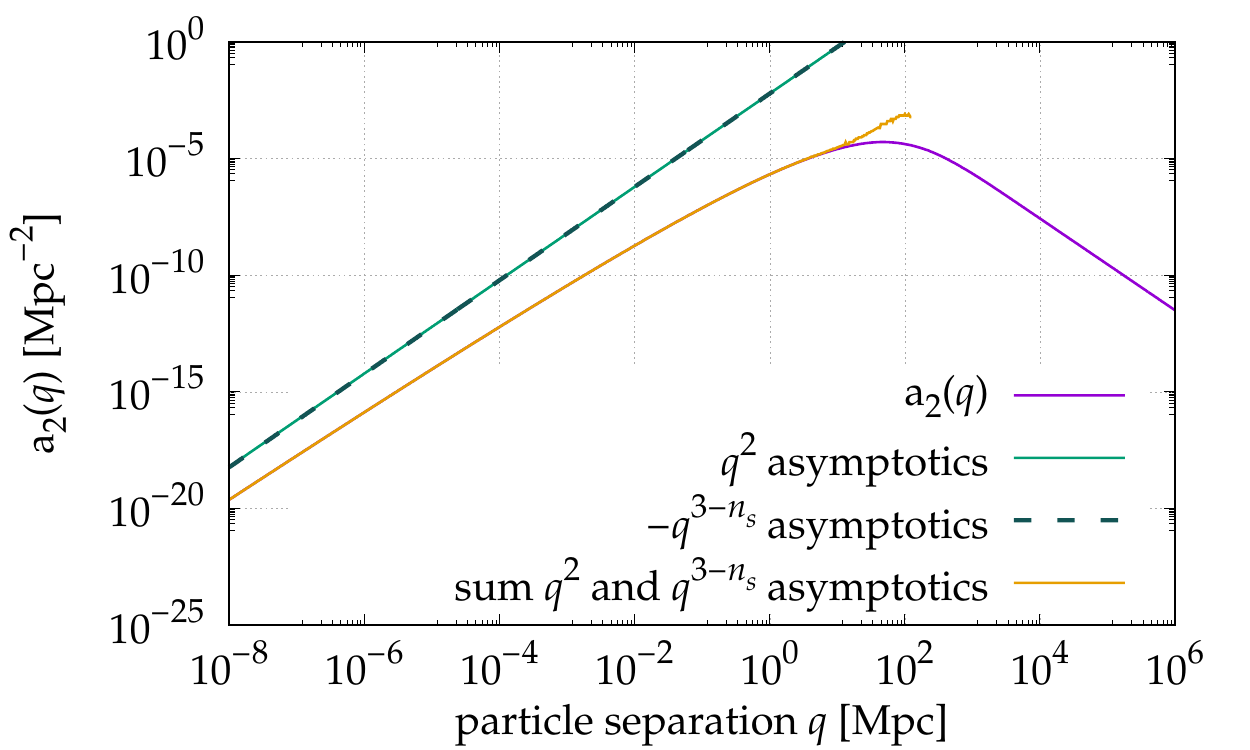}
\caption{The functions $a_1+\frac{\sigma_1^2}{3}$ (left, purple line) and $a_2$ (right, purple line) with their $q^2$ (green lines) and $q^{3-n_s}$ (blue dashed lines) asymptotics are shown. The sum of the two asymptotic orders (golden lines) shows that we indeed need the $q^{3-n_\mathrm{s}}$ order to accurately describe those functions on relevant scales.}
\label{fig:21a}
\end{figure}

\subsubsection{Including all initial correlations in the density-fluctuation power spectrum}

We have so far simplified the probability distribution $P(\tens q,\tens p)$ from (\ref{eq:75}) by approximating the differential operator $\hat D$ by (\ref{eq:77}), leading to the Gaussian (\ref{eq:78}) in the momenta $\tens p$. We shall now demonstrate that this approximation is very well justified at late times \cite{2022Seute}.

With (\ref{eq:75}), the free generating functional (\ref{eq:23}) reads
\begin{equation}
  Z[\tens J] = \hat D\int\D\tens q\D\tens p\int_{\tens s}
  \Phi(\tens R,\tens s)\E^{\I\tens s\cdot\tens p}
  \E^{\I(\tens J,\bar{\tens x}_0)}
\label{eq:214}
\end{equation}
For calculating a power spectrum, we apply two density operators directly and integrate over $q_1$ as before. This results in
\begin{align}
  \left\langle\rho(1)\rho(2)\right\rangle &= N^2Z_0[\tens L] \nonumber\\ &=
  N^2(2\pi)^3\delta_\mathrm{D}(k_1+k_2)\hat D
  \int\D\tens q_{\hat 1}\D\tens p\int_{\tens s}\Phi(\tens R,\tens s)
  \E^{\I(\tens s+\tens L_p)\cdot\tens p+\I k_1\cdot q_2}
\label{eq:215}
\end{align}
with the shift tensor $\tens L_p$ from (\ref{eq:102}). The integral over $\tens p$ results in a delta distribution $\delta_\mathrm{D}(\tens s+\tens L_p)$ which allows carrying out the $\tens s$ integration immediately. This leads to
\begin{equation}
  \left\langle\rho(1)\rho(2)\right\rangle =
  N^2(2\pi)^3\delta_\mathrm{D}(k_1+k_2)
  \hat D\int\D q_{\hat 1}\Phi(\tens R,\tens L_p)
  \E^{\I k_1\cdot q_2}\;.
\label{eq:216}
\end{equation} 
In view of our further calculation, it is important to note that $\tens L_p$ contains entries for two particles only, conveniently labelled as particles $1$ and $2$. If we set $\hat D=V^{-N}\hat 1$ and $\tens R=0$ here and inserted the characteristic function $\Phi$ from (\ref{eq:69}), we reassuringly returned to (\ref{eq:103}).

The characteristic function, evaluated at $(\tens R,\tens L_p)$, reads
\begin{equation}
  \Phi(\tens R,\tens L_p) = \exp\left(
    -\frac{1}{2}\tens L_p^\top C_{pp}\tens L_p
  \right)\E^M\quad\mbox{with}\quad
  M =
    -\frac{1}{2}\tens R^\top C_{\delta\delta}\tens R-
    \tens R^\top C_{\delta p}\tens L_p\;.
\label{eq:217}
\end{equation}
The density-density auto-correlation and density-momentum cross-correlation functions are determined by the initial density-fluctuation power spectrum,
\begin{align}
  C_{\delta\delta}(q) &= \int_lP_\delta^\mathrm{(i)}(l)\,\E^{\I l\cdot q} =
  \frac{1}{2\pi^2}\int_0^\infty l^2\D l\,P_\delta^\mathrm{(i)}(l)\,j_0(lq)
  \;,\nonumber\\
  C_{\delta p}(q) &= \I\int_l
  \frac{l}{l^2}P_\delta^\mathrm{(i)}(l)\,\E^{\I l\cdot q} =
  -\frac{q}{2\pi^2}\int_0^\infty\D l\,P_\delta^\mathrm{(i)}(l)\,j_1(lq)\;.
\label{eq:218}
\end{align} 
Since the differential operator acts on $\tens R$ only, we can continue writing (\ref{eq:216}) in the form
\begin{equation}
  \left\langle\rho(1)\rho(2)\right\rangle =
  N^2(2\pi)^3\delta_\mathrm{D}(k_1+k_2)
  \int\D q_{\hat 1}\exp\left(
    -\frac{1}{2}\tens L_p^\top C_{pp}\tens L_p
  \right)\E^{\I k_1\cdot q_2}\,\hat D\E^M\;.
\label{eq:219}
\end{equation} 

The differential operator $\hat D$, given by (\ref{eq:66}) and (\ref{eq:76}), is
\begin{equation}
  \hat D = \frac{1}{V^N}\prod_{j=1}^N\left.
    \left(1+\I\partial_{R_j}\right)
  \right\vert_{\tens R=0}
\label{eq:220}
\end{equation}
Since $M$ is quadratic in $\tens R$, at most second-order derivatives of $M$ with respect to $\tens R$ can appear. The differential operator $\hat D$ applied to $M$ will thus contain the two types of term
\begin{equation}
  \I\partial_{R_j}M\Big\vert_{\tens R=0} = -\I C_{\delta_jp}\tens L_p
  \quad\mbox{and}\quad
  -\partial_{R_j}\partial_{R_k}M\Big\vert_{\tens R=0} =
  C_{\delta_j\delta_k}
\label{eq:221}
\end{equation}
only. If $j\ne1,2$ in the first type of term, the correlation matrix $C_{\delta_jp}$ depends only on relative coordinates that appear nowhere else in the integrand. Integrating $-\I C_{\delta_jp}\tens L_p$ over the spatial coordinates then results in zero. Likewise, if $j,k\ne1,2$ in the second type of term, the subsequent integration over $C_{\delta_j\delta_k}$ vanishes. Therefore, only those contributions to $\hat DM$ remain which contain $\partial_{R_1}$ or $\partial_{R_2}$. We can thus replace $\hat D\E^M$ in (\ref{eq:219}) by
\begin{align}
  \hat D\E^M &= V^{-N}\left(1+\I\partial_{R_1}\right)
  \left(1+\I\partial_{R_2}\right)\E^M\Big\vert_{R=0}\nonumber\\ &=
  V^{-N}\left[
    1+C_{\delta_1\delta_2}-
    \I C_{\delta_1p}\tens L_p-\I C_{\delta_2p}\tens L_p-
    \left(C_{\delta_1p}\tens L_p\right)
    \left(C_{\delta_2p}\tens L_p\right)
  \right]\;.
\label{eq:222}
\end{align}
Furthermore, since $C_{\delta_jp_j} = 0$,
\begin{equation}
  C_{\delta_1p}\tens L_p = C_{\delta_1p_2}L_{p_2}
\label{eq:223}
\end{equation}
and likewise for $C_{\delta_2p}\tens L_p$. Therefore, we have
\begin{equation}
  \hat D\E^M = V^{-N}\left[
    1+F(k_1,q_2,t)
  \right]
\label{eq:224}
\end{equation}
with
\begin{equation}
  F(k,q,t) = C_{\delta\delta}(q)-
  \I C_{\delta_1p_2}L_{p_2}-\I C_{\delta_2p_1}L_{p_1}-
  \left(C_{\delta_1p_2}L_{p_2}\right)\left(C_{\delta_2p_1}L_{p_1}\right)\;.
\label{eq:225}
\end{equation}

\begin{figure}[t]
\includegraphics[width=\hsize]{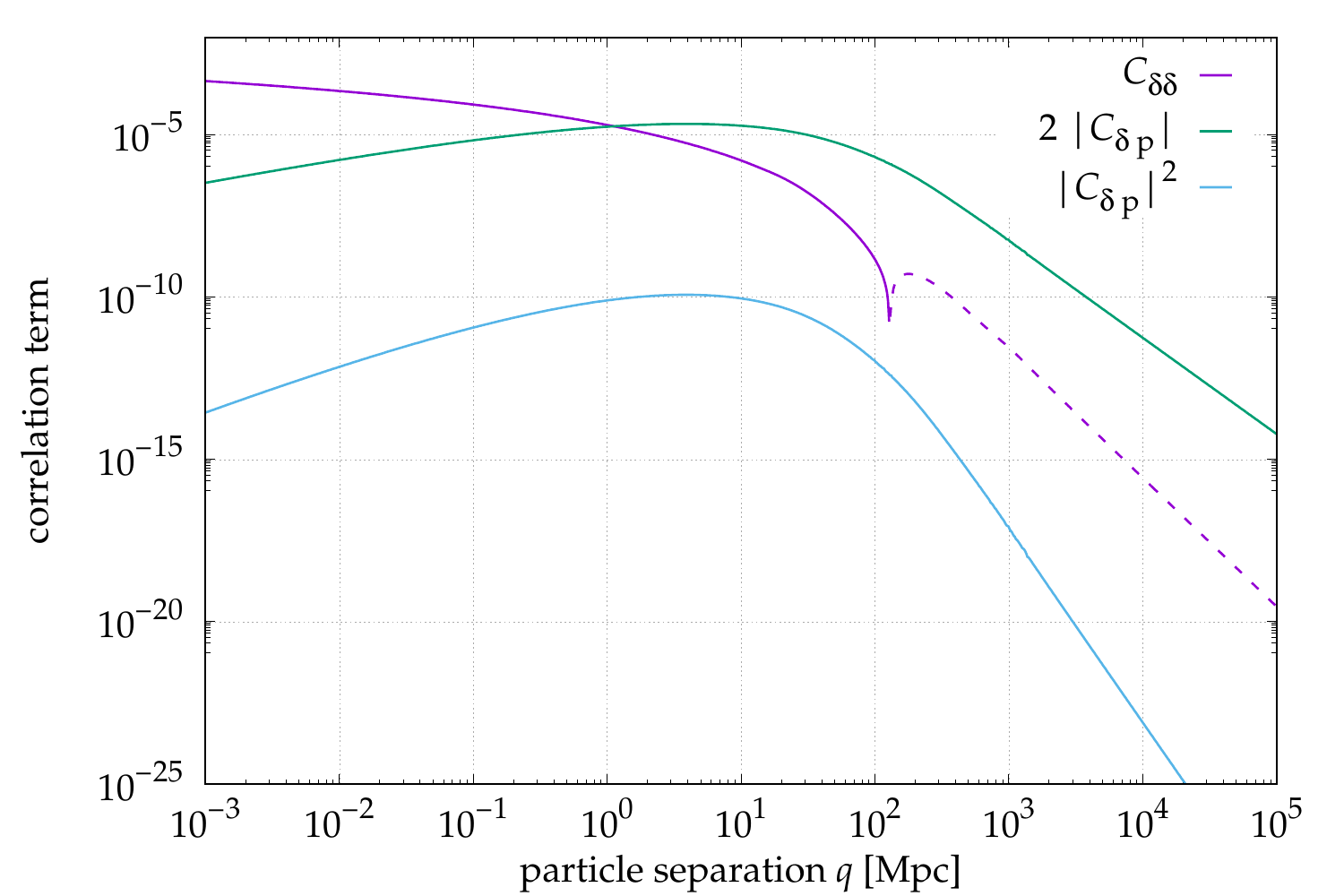}
\caption{The terms that enter the function $F$ as defined in (\ref{eq:226}). The initial 
density-density correlation term ($C_{\delta \delta}$, purple line) enters linearly. The 
linear density-momentum correlation term ($2\vline C_{\delta p}\vline$, green line) enters linearly in $kt$, while the quadratic density-momentum correlation term ($\vline C_{dp}\vline^2$, blue line) enters quadratically in $kt$.}
\label{fig:22}
\end{figure}

\begin{figure}[t]
\includegraphics[width=\hsize]{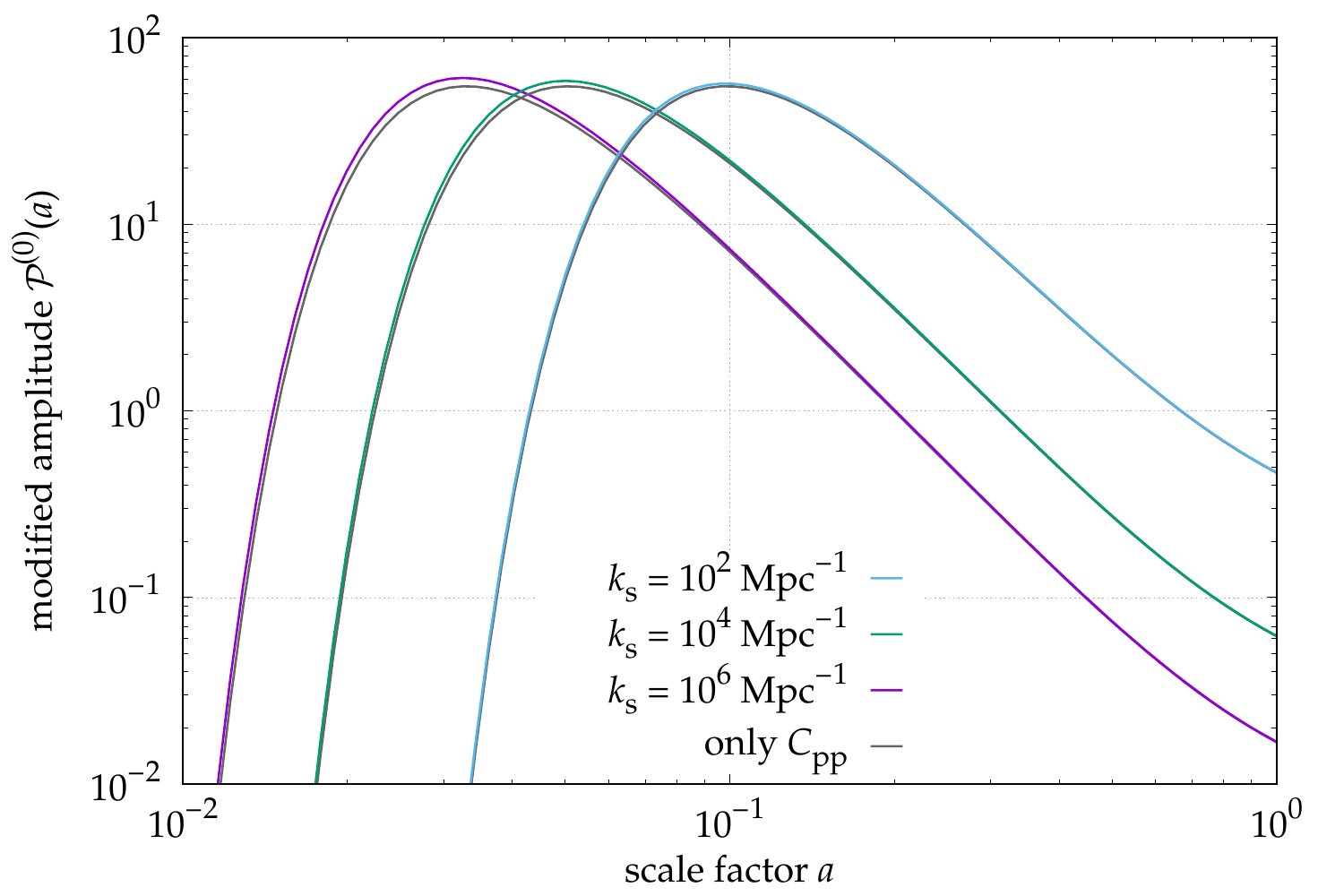}
\caption{The modified asymptotic amplitude $\mathcal{P}^{(0)}$, where initial density-density, density-momentum and momentum-momentum correlations are considered (see (\ref{eq:227})), are shown as a function of scale factor $a$ in colored lines for three different initial small-scale smoothing wave numbers $k_s$. The comparison to the asymptotic amplitudes where only initial momentum-momentum correlations $C_{pp}$ are considered shows that the other correlations have a barely noticeable impact on the small scale structure evolution.}
\label{fig:22a}
\end{figure}

The expression for $C_{\delta p}$ in (\ref{eq:218}) shows that $C_{\delta_jp_k} = -C_{\delta_kp_j}$ because the exchange of indices implies $q\to-q$. Moreover, we know from (\ref{eq:102}) that $L_{p_2} = -L_{p_1} = k_1t$. Thus, we can somewhat simplify $F(k,q,t)$ to read
\begin{equation}
  F(k,q,t) = C_{\delta\delta}(q)-2\I C_{\delta_1p_2}kt-\left(
    C_{\delta_1p_2}kt
  \right)^2\;.
\label{eq:226}
\end{equation} 

Integrating finally over the free particle positions $q_3\ldots q_N$ and comparing (\ref{eq:219}) to (\ref{eq:110}), we can identify the expression
\begin{equation}
  \mathcal{P}(k) = \E^{-Q_\mathrm{D}}
  \int_q\left[1+F(k,q,t)\right]\E^{-t^2k^2a_\parallel(q,\mu)}
  \E^{\I k\cdot q}
\label{eq:227}
\end{equation}
for the free power spectrum including not only momentum auto-correlations, but all correlations between density and momentum fluctuations \cite{2022Seute}.

The expressions (\ref{eq:218}) for $C_{\delta\delta}$ and $C_{\delta p}$ and the series expansions (\ref{eq:155}) for the spherical Bessel functions lead to the asymptotic expansions
\begin{align}
  C_{\delta\delta}(q) &\sim \sum_{n=0}^\infty
  \frac{\left(-q^2\right)^n\sigma_{n+2}^2}{(2n+1)!} \sim
  \sigma_2^2-\frac{q^2}{6}\sigma_3^2 \;,\nonumber\\
  C_{\delta p}(q) &\sim -q\sum_{n=0}^\infty
  \frac{\left(-q^2\right)^n\sigma_{n+2}^2}{(2n+3)(2n+1)!} \sim
  -\frac{q}{3}\left(\sigma_2^2-\frac{q^2}{10}\sigma_3^2\right)
\label{eq:228}
\end{align}
for $q\to0$. Therefore, the factor $(1+F)$ has the asymptotic expansion
\begin{equation}
  1+F \sim 1+\sigma_2^2+\frac{2\I}{3}\sigma_2^2t\,k\cdot q-
  \frac{\sigma_2^4}{9}t^2\left(k\cdot q\right)^2\;.
\label{eq:229}
\end{equation}

Searching for the effect of the factor $(1+F)$ on the leading-order asymptotic behaviour of the free density-fluctuation power spectrum, we thus need to multiply the asymptotic expansion for the function $g$ in (\ref{eq:159}) by each term in the asymptotic expression (\ref{eq:229}) and change $m$ to $m+1$ and $m+2$, respectively. The dependence on the wave number of the leading order terms remain unchanged. The leading-order asymptotic behaviour of the free power spectrum is thus given by
\begin{equation}
  \mathcal{P}(k) \sim \left(1+\frac{4}{9}\sigma_2^2 - \frac{10}{3 t} + \frac{25}{9t^2}\right)
  \frac{\mathcal{P}^{(0)}}{k^3}\;,
\label{eq:230}
\end{equation}
with the coefficient function $\mathcal{P}^{(0)}$ from (\ref{eq:161}), and the $k^{-3}$ behaviour remains \cite{2022Seute}.

\subsubsection{Including all initial correlations in the velocity power spectrum}

As shown before, taking the initial density-density and density-momentum correlations into account in addition to the momentum auto-correlations inserts a factor $(1+F)$ into the integrand of the free power spectrum, with the function $F$ given by (\ref{eq:225}). Therefore, the velocity power spectrum taking the complete set of initial correlations into account is obtained by inserting $(1+F)$ into (\ref{eq:127}),
\begin{equation}
  \mathcal{P}_\Pi(k) = -D_1\otimes D_2\int\D q\,\left[
    1+F(k,q,t)
  \right]\E^{-Q}\,\E^{\I k\cdot q} =
  \int\D q\,\mathcal{F}\,\E^{-Q}\,\E^{\I k\cdot q}
\label{eq:231}
\end{equation}
with the matrix
\begin{equation}
  \mathcal{F} =
  D_1\otimes D_2F-D_1F\otimes D_2Q-D_2F\otimes D_1Q+
  (1+F)\mathcal{Q}\;.
\label{eq:232}
\end{equation} 

As before, we need the leading asymptotic order in $q$ of the matrix $\mathcal{F}$, which is
\begin{equation}
  \mathcal{F} \sim  \left(1+\sigma_2^2+\frac{2\I}{3}\sigma_2^2t\,k\cdot q-
  \frac{\sigma_2^4}{9}t^2\left(k\cdot q\right)^2\right)\frac{\sigma_1^2}{3}\,\id{3}\;,
\label{eq:233}
\end{equation}
implying that the velocity power spectrum including all initial correlations must have the leading-order asymptotic behaviour
\begin{equation}
  \mathcal{P}_\Pi(k) \sim  \left(1+\frac{4}{9}\sigma_2^2 - \frac{10}{3 t} + \frac{25}{9t^2}\right)
  \frac{\mathcal{P}^{(0)}_\Pi}{k^3}\,\id{3}
\label{eq:234}
\end{equation}
with $\mathcal{P}^{(0)}_\Pi$ from (\ref{eq:176}).

\section{Conclusions}
\label{sec:5}

We have reviewed kinetic field theory for classical particle ensembles, putting particular emphasis on the evolution of cosmic structures in collision-less dark matter. Kinetic field theory dissolves the cosmic density field into particles subject to Hamiltonian dynamics and studies the evolution of an initial phase-space probability distribution under the Hamiltonian flow. Compared to other analytic approaches to cosmic structure formation, the essential advantage of kinetic field theory is that trajectories in phase space do not cross. The notorious shell-crossing problem occurring inevitably in methods building upon uniquely valued density and velocity fields in configuration space is thus avoided by construction.

The central mathematical object of kinetic field theory is a generating functional encapsulating the statistically defined initial state of the particle ensemble together with the dynamics of phase-space trajectories. Kinetic field theory does not set up and solve a dynamical equation for smooth density or velocity fields or a phase-space distribution function. Rather, it evolves this generating functional in time and allows extracting statistical information on the evolved particle ensemble by functional derivatives. Formally, it resembles a statistical quantum field theory, however some aspects of it are considerably simpler.

After the introduction in Sect.~\ref{sec:1}, we set up in Sect.~\ref{sec:2} the generating functional of kinetic field theory in an expanding cosmic space-time and chose a suitable Green's function to solve the Hamiltonian equations of motion. We introduced the growth rate of linear density fluctuations as a time coordinate and argued that inertial motion with respect to this time, the so-called Zel'dovich approximation, is particularly appropriate for describing cosmic structure formation. We emphasized that the gravitational potential between particles on Zel'dovich inertial trajectories is sourced only by non-linear fluctuations of the cosmic matter density, which implies that the effective gravitational potential is short-ranged and is approximately of Yukawa form.

We defined density operators in Sect.~\ref{sec:3}, showed how they can be used to extract statistical information from the generating functional, and derived low-order statistical measures for distributions of particles freely streaming along Zel'dovich reference trajectories. We emphasize once more that this kind of reference motion, even though being referred to as free, does include gravitational interaction at early times, and long-range gravitational interaction at all times. In particular, we derived equations for the non-linear, free power spectrum of cosmic density fluctuations and their bispectrum as well as the free velocity power spectrum.

Aiming at rigorous statements on the small-scale behaviour of low-order statistical measures, Sect.~\ref{sec:4} is the core of this paper. There, we used extensions of Laplace's method to derive the small-scale asymptotics of the free density-fluctuation power spectrum, the free bispectrum, and the free  velocity power spectrum. Our central results obtained there are that
\begin{enumerate}
  \item the free density-fluctuation power spectrum and the free velocity power spectrum asymptotically fall off proportional to $k^{-3}$ for wave number $k\to\infty$, and that
  \item the free bispectrum asymptotically falls off like $k^{-11/2}$.
\end{enumerate}
These results assume only that the initial particle momenta are drawn from a Gaussian random field whose power spectrum has finite low-order moments. The nature of the dark matter and the cosmological model are irrelevant. The $k^{-3}$ tail of the free power spectra and the $k^{-11/2}$ tail of the bispectrum evolve by collective streaming of the matter particles, and the exponents are set by the number of spatial dimensions only.

These results persist if some simplifying limitations are given up. We have shown further that gravitational interaction between particles in the mean-field approximation and strictly cold dark matter with an initial power spectrum without small-scale cut-off do not affect the asymptotic $k^{-3}$ dependence of the matter-fluctuation power spectrum. In addition, including initial density-density and density-momentum correlations together with momentum auto-correlations do not change the asymptotic behaviour of the density-fluctuation and the velocity power spectra either.

It would thus appear that the asymptotic behaviour of low-order statistical measures of the cosmic matter and velocity distribution is a consequence of the initial particle momenta being drawn from a Gaussian random field, without further assumptions entering. Late-time, non-linear density-fluctuation power spectra would therefore necessarily develop a $k^{-3}$ tail on small scales, irrespective of the cosmological model and the nature of the dark matter.

Characteristic and universal time scales for structure formation can be derived from the leading-order asymptotic term reaching its maximum amplitude, and from the next-to-leading order term falling below the leading order term. Possible observational consequences and the potential significance of these time scales need to be worked out.

The $k^{-3}$ behaviour of power spectra implies that the power, i.e.\ the product of the power spectra times the number of Fourier modes, flattens off and becomes constant on small scales. Each fixed scale interval will then contribute the same amount of power. This result should provide clues for the universal density profile of gravitationally bound structures. Of course, many questions remain to be addressed and answered, but the conclusions on the small-scale behaviour of non-linear cosmic structures obtained from kinetic field theory here appear promising.

\backmatter

% \bmhead{Supplementary information}

\bmhead{Acknowledgments}
First and foremost, we would like to thank numerous students that helped to develop and to clear the path towards kinetic field theory for cosmic structure formation in the form presented in this review. We are especially thankful for valuable input by Yonadav Barry Ginat, Leif Seute and Ricardo Waibel, whoem we additionally thank for hithero unpublished results on the bispectrum. We thank Lukas Bauer, Robin Bühler, Lukas Heizmann, Patrick Jentsch, Christophe Pixius, Hannes Riechert, Christian Schmidt, Johannes Schwinn, Christian Sorgenfrei, and Stefan Zentarra for helpful input and discussions.
For claryfing discussions and advice, we thank Robert Lilow, Carsten Littek, and Elena Kozlikin.

We gratefully acknowledge fruitful discussions with many colleagues, most notably the always very helpful discussions with Manfred Salmhofer and Jan Pawlowski.

This work is funded by the Deutsche Forschungsgemeinschaft (DFG, German Research Foundation) under Germany's Excellence Strategy EXC 2181/1 - 390900948 (the Heidelberg STRUCTURES Excellence Cluster).

% \section*{Declarations}

% \begin{appendices}

% \section{Section title of first appendix}
% \label{secA1}

% \end{appendices}

\bibliography{main}

%% BioMed_Central_Bib_Style_v1.01

\begin{thebibliography}{75}
% BibTex style file: bmc-mathphys.bst (version 2.1), 2014-07-24
\ifx \bisbn   \undefined \def \bisbn  #1{ISBN #1}\fi
\ifx \binits  \undefined \def \binits#1{#1}\fi
\ifx \bauthor  \undefined \def \bauthor#1{#1}\fi
\ifx \batitle  \undefined \def \batitle#1{#1}\fi
\ifx \bjtitle  \undefined \def \bjtitle#1{#1}\fi
\ifx \bvolume  \undefined \def \bvolume#1{\textbf{#1}}\fi
\ifx \byear  \undefined \def \byear#1{#1}\fi
\ifx \bissue  \undefined \def \bissue#1{#1}\fi
\ifx \bfpage  \undefined \def \bfpage#1{#1}\fi
\ifx \blpage  \undefined \def \blpage #1{#1}\fi
\ifx \burl  \undefined \def \burl#1{\textsf{#1}}\fi
\ifx \doiurl  \undefined \def \doiurl#1{\url{https://doi.org/#1}}\fi
\ifx \betal  \undefined \def \betal{\textit{et al.}}\fi
\ifx \binstitute  \undefined \def \binstitute#1{#1}\fi
\ifx \binstitutionaled  \undefined \def \binstitutionaled#1{#1}\fi
\ifx \bctitle  \undefined \def \bctitle#1{#1}\fi
\ifx \beditor  \undefined \def \beditor#1{#1}\fi
\ifx \bpublisher  \undefined \def \bpublisher#1{#1}\fi
\ifx \bbtitle  \undefined \def \bbtitle#1{#1}\fi
\ifx \bedition  \undefined \def \bedition#1{#1}\fi
\ifx \bseriesno  \undefined \def \bseriesno#1{#1}\fi
\ifx \blocation  \undefined \def \blocation#1{#1}\fi
\ifx \bsertitle  \undefined \def \bsertitle#1{#1}\fi
\ifx \bsnm \undefined \def \bsnm#1{#1}\fi
\ifx \bsuffix \undefined \def \bsuffix#1{#1}\fi
\ifx \bparticle \undefined \def \bparticle#1{#1}\fi
\ifx \barticle \undefined \def \barticle#1{#1}\fi
\bibcommenthead
\ifx \bconfdate \undefined \def \bconfdate #1{#1}\fi
\ifx \botherref \undefined \def \botherref #1{#1}\fi
\ifx \url \undefined \def \url#1{\textsf{#1}}\fi
\ifx \bchapter \undefined \def \bchapter#1{#1}\fi
\ifx \bbook \undefined \def \bbook#1{#1}\fi
\ifx \bcomment \undefined \def \bcomment#1{#1}\fi
\ifx \oauthor \undefined \def \oauthor#1{#1}\fi
\ifx \citeauthoryear \undefined \def \citeauthoryear#1{#1}\fi
\ifx \endbibitem  \undefined \def \endbibitem {}\fi
\ifx \bconflocation  \undefined \def \bconflocation#1{#1}\fi
\ifx \arxivurl  \undefined \def \arxivurl#1{\textsf{#1}}\fi
\csname PreBibitemsHook\endcsname

%%% 1
\bibitem{2006AJ....131.1163S}
\begin{barticle}
\bauthor{\bsnm{{Skrutskie}}, \binits{M.F.}},
\bauthor{\bsnm{{Cutri}}, \binits{R.M.}},
\bauthor{\bsnm{{Stiening}}, \binits{R.}},
\bauthor{\bsnm{{Weinberg}}, \binits{M.D.}},
\bauthor{\bsnm{{Schneider}}, \binits{S.}},
\bauthor{\bsnm{{Carpenter}}, \binits{J.M.}},
\bauthor{\bsnm{{Beichman}}, \binits{C.}},
\bauthor{\bsnm{{Capps}}, \binits{R.}},
\bauthor{\bsnm{{Chester}}, \binits{T.}},
\bauthor{\bsnm{{Elias}}, \binits{J.}},
\bauthor{\bsnm{{Huchra}}, \binits{J.}},
\bauthor{\bsnm{{Liebert}}, \binits{J.}},
\bauthor{\bsnm{{Lonsdale}}, \binits{C.}},
\bauthor{\bsnm{{Monet}}, \binits{D.G.}},
\bauthor{\bsnm{{Price}}, \binits{S.}},
\bauthor{\bsnm{{Seitzer}}, \binits{P.}},
\bauthor{\bsnm{{Jarrett}}, \binits{T.}},
\bauthor{\bsnm{{Kirkpatrick}}, \binits{J.D.}},
\bauthor{\bsnm{{Gizis}}, \binits{J.E.}},
\bauthor{\bsnm{{Howard}}, \binits{E.}},
\bauthor{\bsnm{{Evans}}, \binits{T.}},
\bauthor{\bsnm{{Fowler}}, \binits{J.}},
\bauthor{\bsnm{{Fullmer}}, \binits{L.}},
\bauthor{\bsnm{{Hurt}}, \binits{R.}},
\bauthor{\bsnm{{Light}}, \binits{R.}},
\bauthor{\bsnm{{Kopan}}, \binits{E.L.}},
\bauthor{\bsnm{{Marsh}}, \binits{K.A.}},
\bauthor{\bsnm{{McCallon}}, \binits{H.L.}},
\bauthor{\bsnm{{Tam}}, \binits{R.}},
\bauthor{\bsnm{{Van Dyk}}, \binits{S.}},
\bauthor{\bsnm{{Wheelock}}, \binits{S.}}:
\batitle{{The Two Micron All Sky Survey (2MASS)}}.
\bjtitle{\aj}
\bvolume{131}(\bissue{2}),
\bfpage{1163}--\blpage{1183}
(\byear{2006}).
\doiurl{10.1086/498708}
\end{barticle}
\endbibitem

%%% 2
\bibitem{2020A&A...641A...6P}
\begin{barticle}
\bauthor{\bsnm{{Planck Collaboration}}},
\bauthor{\bsnm{{Aghanim}}, \binits{N.}},
\bauthor{\bsnm{{Akrami}}, \binits{Y.}},
\bauthor{\bsnm{{Ashdown}}, \binits{M.}},
\bauthor{\bsnm{{Aumont}}, \binits{J.}},
\bauthor{\bsnm{{Baccigalupi}}, \binits{C.}},
\bauthor{\bsnm{{Ballardini}}, \binits{M.}},
\bauthor{\bsnm{{Banday}}, \binits{A.J.}},
\bauthor{\bsnm{{Barreiro}}, \binits{R.B.}},
\bauthor{\bsnm{{Bartolo}}, \binits{N.}},
\bauthor{\bsnm{{Basak}}, \binits{S.}},
\bauthor{\bsnm{{Battye}}, \binits{R.}},
\bauthor{\bsnm{{Benabed}}, \binits{K.}},
\bauthor{\bsnm{{Bernard}}, \binits{J.-P.}},
\bauthor{\bsnm{{Bersanelli}}, \binits{M.}},
\bauthor{\bsnm{{Bielewicz}}, \binits{P.}},
\bauthor{\bsnm{{Bock}}, \binits{J.J.}},
\bauthor{\bsnm{{Bond}}, \binits{J.R.}},
\bauthor{\bsnm{{Borrill}}, \binits{J.}},
\bauthor{\bsnm{{Bouchet}}, \binits{F.R.}},
\bauthor{\bsnm{{Boulanger}}, \binits{F.}},
\bauthor{\bsnm{{Bucher}}, \binits{M.}},
\bauthor{\bsnm{{Burigana}}, \binits{C.}},
\bauthor{\bsnm{{Butler}}, \binits{R.C.}},
\bauthor{\bsnm{{Calabrese}}, \binits{E.}},
\bauthor{\bsnm{{Cardoso}}, \binits{J.-F.}},
\bauthor{\bsnm{{Carron}}, \binits{J.}},
\bauthor{\bsnm{{Challinor}}, \binits{A.}},
\bauthor{\bsnm{{Chiang}}, \binits{H.C.}},
\bauthor{\bsnm{{Chluba}}, \binits{J.}},
\bauthor{\bsnm{{Colombo}}, \binits{L.P.L.}},
\bauthor{\bsnm{{Combet}}, \binits{C.}},
\bauthor{\bsnm{{Contreras}}, \binits{D.}},
\bauthor{\bsnm{{Crill}}, \binits{B.P.}},
\bauthor{\bsnm{{Cuttaia}}, \binits{F.}},
\bauthor{\bsnm{{de Bernardis}}, \binits{P.}},
\bauthor{\bsnm{{de Zotti}}, \binits{G.}},
\bauthor{\bsnm{{Delabrouille}}, \binits{J.}},
\bauthor{\bsnm{{Delouis}}, \binits{J.-M.}},
\bauthor{\bsnm{{Di Valentino}}, \binits{E.}},
\bauthor{\bsnm{{Diego}}, \binits{J.M.}},
\bauthor{\bsnm{{Dor{\'e}}}, \binits{O.}},
\bauthor{\bsnm{{Douspis}}, \binits{M.}},
\bauthor{\bsnm{{Ducout}}, \binits{A.}},
\bauthor{\bsnm{{Dupac}}, \binits{X.}},
\bauthor{\bsnm{{Dusini}}, \binits{S.}},
\bauthor{\bsnm{{Efstathiou}}, \binits{G.}},
\bauthor{\bsnm{{Elsner}}, \binits{F.}},
\bauthor{\bsnm{{En{\ss}lin}}, \binits{T.A.}},
\bauthor{\bsnm{{Eriksen}}, \binits{H.K.}},
\bauthor{\bsnm{{Fantaye}}, \binits{Y.}},
\bauthor{\bsnm{{Farhang}}, \binits{M.}},
\bauthor{\bsnm{{Fergusson}}, \binits{J.}},
\bauthor{\bsnm{{Fernandez-Cobos}}, \binits{R.}},
\bauthor{\bsnm{{Finelli}}, \binits{F.}},
\bauthor{\bsnm{{Forastieri}}, \binits{F.}},
\bauthor{\bsnm{{Frailis}}, \binits{M.}},
\bauthor{\bsnm{{Fraisse}}, \binits{A.A.}},
\bauthor{\bsnm{{Franceschi}}, \binits{E.}},
\bauthor{\bsnm{{Frolov}}, \binits{A.}},
\bauthor{\bsnm{{Galeotta}}, \binits{S.}},
\bauthor{\bsnm{{Galli}}, \binits{S.}},
\bauthor{\bsnm{{Ganga}}, \binits{K.}},
\bauthor{\bsnm{{G{\'e}nova-Santos}}, \binits{R.T.}},
\bauthor{\bsnm{{Gerbino}}, \binits{M.}},
\bauthor{\bsnm{{Ghosh}}, \binits{T.}},
\bauthor{\bsnm{{Gonz{\'a}lez-Nuevo}}, \binits{J.}},
\bauthor{\bsnm{{G{\'o}rski}}, \binits{K.M.}},
\bauthor{\bsnm{{Gratton}}, \binits{S.}},
\bauthor{\bsnm{{Gruppuso}}, \binits{A.}},
\bauthor{\bsnm{{Gudmundsson}}, \binits{J.E.}},
\bauthor{\bsnm{{Hamann}}, \binits{J.}},
\bauthor{\bsnm{{Handley}}, \binits{W.}},
\bauthor{\bsnm{{Hansen}}, \binits{F.K.}},
\bauthor{\bsnm{{Herranz}}, \binits{D.}},
\bauthor{\bsnm{{Hildebrandt}}, \binits{S.R.}},
\bauthor{\bsnm{{Hivon}}, \binits{E.}},
\bauthor{\bsnm{{Huang}}, \binits{Z.}},
\bauthor{\bsnm{{Jaffe}}, \binits{A.H.}},
\bauthor{\bsnm{{Jones}}, \binits{W.C.}},
\bauthor{\bsnm{{Karakci}}, \binits{A.}},
\bauthor{\bsnm{{Keih{\"a}nen}}, \binits{E.}},
\bauthor{\bsnm{{Keskitalo}}, \binits{R.}},
\bauthor{\bsnm{{Kiiveri}}, \binits{K.}},
\bauthor{\bsnm{{Kim}}, \binits{J.}},
\bauthor{\bsnm{{Kisner}}, \binits{T.S.}},
\bauthor{\bsnm{{Knox}}, \binits{L.}},
\bauthor{\bsnm{{Krachmalnicoff}}, \binits{N.}},
\bauthor{\bsnm{{Kunz}}, \binits{M.}},
\bauthor{\bsnm{{Kurki-Suonio}}, \binits{H.}},
\bauthor{\bsnm{{Lagache}}, \binits{G.}},
\bauthor{\bsnm{{Lamarre}}, \binits{J.-M.}},
\bauthor{\bsnm{{Lasenby}}, \binits{A.}},
\bauthor{\bsnm{{Lattanzi}}, \binits{M.}},
\bauthor{\bsnm{{Lawrence}}, \binits{C.R.}},
\bauthor{\bsnm{{Le Jeune}}, \binits{M.}},
\bauthor{\bsnm{{Lemos}}, \binits{P.}},
\bauthor{\bsnm{{Lesgourgues}}, \binits{J.}},
\bauthor{\bsnm{{Levrier}}, \binits{F.}},
\bauthor{\bsnm{{Lewis}}, \binits{A.}},
\bauthor{\bsnm{{Liguori}}, \binits{M.}},
\bauthor{\bsnm{{Lilje}}, \binits{P.B.}},
\bauthor{\bsnm{{Lilley}}, \binits{M.}},
\bauthor{\bsnm{{Lindholm}}, \binits{V.}},
\bauthor{\bsnm{{L{\'o}pez-Caniego}}, \binits{M.}},
\bauthor{\bsnm{{Lubin}}, \binits{P.M.}},
\bauthor{\bsnm{{Ma}}, \binits{Y.-Z.}},
\bauthor{\bsnm{{Mac{\'\i}as-P{\'e}rez}}, \binits{J.F.}},
\bauthor{\bsnm{{Maggio}}, \binits{G.}},
\bauthor{\bsnm{{Maino}}, \binits{D.}},
\bauthor{\bsnm{{Mandolesi}}, \binits{N.}},
\bauthor{\bsnm{{Mangilli}}, \binits{A.}},
\bauthor{\bsnm{{Marcos-Caballero}}, \binits{A.}},
\bauthor{\bsnm{{Maris}}, \binits{M.}},
\bauthor{\bsnm{{Martin}}, \binits{P.G.}},
\bauthor{\bsnm{{Martinelli}}, \binits{M.}},
\bauthor{\bsnm{{Mart{\'\i}nez-Gonz{\'a}lez}}, \binits{E.}},
\bauthor{\bsnm{{Matarrese}}, \binits{S.}},
\bauthor{\bsnm{{Mauri}}, \binits{N.}},
\bauthor{\bsnm{{McEwen}}, \binits{J.D.}},
\bauthor{\bsnm{{Meinhold}}, \binits{P.R.}},
\bauthor{\bsnm{{Melchiorri}}, \binits{A.}},
\bauthor{\bsnm{{Mennella}}, \binits{A.}},
\bauthor{\bsnm{{Migliaccio}}, \binits{M.}},
\bauthor{\bsnm{{Millea}}, \binits{M.}},
\bauthor{\bsnm{{Mitra}}, \binits{S.}},
\bauthor{\bsnm{{Miville-Desch{\^e}nes}}, \binits{M.-A.}},
\bauthor{\bsnm{{Molinari}}, \binits{D.}},
\bauthor{\bsnm{{Montier}}, \binits{L.}},
\bauthor{\bsnm{{Morgante}}, \binits{G.}},
\bauthor{\bsnm{{Moss}}, \binits{A.}},
\bauthor{\bsnm{{Natoli}}, \binits{P.}},
\bauthor{\bsnm{{N{\o}rgaard-Nielsen}}, \binits{H.U.}},
\bauthor{\bsnm{{Pagano}}, \binits{L.}},
\bauthor{\bsnm{{Paoletti}}, \binits{D.}},
\bauthor{\bsnm{{Partridge}}, \binits{B.}},
\bauthor{\bsnm{{Patanchon}}, \binits{G.}},
\bauthor{\bsnm{{Peiris}}, \binits{H.V.}},
\bauthor{\bsnm{{Perrotta}}, \binits{F.}},
\bauthor{\bsnm{{Pettorino}}, \binits{V.}},
\bauthor{\bsnm{{Piacentini}}, \binits{F.}},
\bauthor{\bsnm{{Polastri}}, \binits{L.}},
\bauthor{\bsnm{{Polenta}}, \binits{G.}},
\bauthor{\bsnm{{Puget}}, \binits{J.-L.}},
\bauthor{\bsnm{{Rachen}}, \binits{J.P.}},
\bauthor{\bsnm{{Reinecke}}, \binits{M.}},
\bauthor{\bsnm{{Remazeilles}}, \binits{M.}},
\bauthor{\bsnm{{Renzi}}, \binits{A.}},
\bauthor{\bsnm{{Rocha}}, \binits{G.}},
\bauthor{\bsnm{{Rosset}}, \binits{C.}},
\bauthor{\bsnm{{Roudier}}, \binits{G.}},
\bauthor{\bsnm{{Rubi{\~n}o-Mart{\'\i}n}}, \binits{J.A.}},
\bauthor{\bsnm{{Ruiz-Granados}}, \binits{B.}},
\bauthor{\bsnm{{Salvati}}, \binits{L.}},
\bauthor{\bsnm{{Sandri}}, \binits{M.}},
\bauthor{\bsnm{{Savelainen}}, \binits{M.}},
\bauthor{\bsnm{{Scott}}, \binits{D.}},
\bauthor{\bsnm{{Shellard}}, \binits{E.P.S.}},
\bauthor{\bsnm{{Sirignano}}, \binits{C.}},
\bauthor{\bsnm{{Sirri}}, \binits{G.}},
\bauthor{\bsnm{{Spencer}}, \binits{L.D.}},
\bauthor{\bsnm{{Sunyaev}}, \binits{R.}},
\bauthor{\bsnm{{Suur-Uski}}, \binits{A.-S.}},
\bauthor{\bsnm{{Tauber}}, \binits{J.A.}},
\bauthor{\bsnm{{Tavagnacco}}, \binits{D.}},
\bauthor{\bsnm{{Tenti}}, \binits{M.}},
\bauthor{\bsnm{{Toffolatti}}, \binits{L.}},
\bauthor{\bsnm{{Tomasi}}, \binits{M.}},
\bauthor{\bsnm{{Trombetti}}, \binits{T.}},
\bauthor{\bsnm{{Valenziano}}, \binits{L.}},
\bauthor{\bsnm{{Valiviita}}, \binits{J.}},
\bauthor{\bsnm{{Van Tent}}, \binits{B.}},
\bauthor{\bsnm{{Vibert}}, \binits{L.}},
\bauthor{\bsnm{{Vielva}}, \binits{P.}},
\bauthor{\bsnm{{Villa}}, \binits{F.}},
\bauthor{\bsnm{{Vittorio}}, \binits{N.}},
\bauthor{\bsnm{{Wandelt}}, \binits{B.D.}},
\bauthor{\bsnm{{Wehus}}, \binits{I.K.}},
\bauthor{\bsnm{{White}}, \binits{M.}},
\bauthor{\bsnm{{White}}, \binits{S.D.M.}},
\bauthor{\bsnm{{Zacchei}}, \binits{A.}},
\bauthor{\bsnm{{Zonca}}, \binits{A.}}:
\batitle{{Planck 2018 results. VI. Cosmological parameters}}.
\bjtitle{\aap}
\bvolume{641},
\bfpage{6}
(\byear{2020})
{\href{https://arxiv.org/abs/1807.06209}{{arXiv:1807.06209}}}
{[astro-ph.CO]}.
\doiurl{10.1051/0004-6361/201833910}
\end{barticle}
\endbibitem

%%% 3
\bibitem{2012ApJ...757...22C}
\begin{barticle}
\bauthor{\bsnm{{Coe}}, \binits{D.}},
\bauthor{\bsnm{{Umetsu}}, \binits{K.}},
\bauthor{\bsnm{{Zitrin}}, \binits{A.}},
\bauthor{\bsnm{{Donahue}}, \binits{M.}},
\bauthor{\bsnm{{Medezinski}}, \binits{E.}},
\bauthor{\bsnm{{Postman}}, \binits{M.}},
\bauthor{\bsnm{{Carrasco}}, \binits{M.}},
\bauthor{\bsnm{{Anguita}}, \binits{T.}},
\bauthor{\bsnm{{Geller}}, \binits{M.J.}},
\bauthor{\bsnm{{Rines}}, \binits{K.J.}},
\bauthor{\bsnm{{Diaferio}}, \binits{A.}},
\bauthor{\bsnm{{Kurtz}}, \binits{M.J.}},
\bauthor{\bsnm{{Bradley}}, \binits{L.}},
\bauthor{\bsnm{{Koekemoer}}, \binits{A.}},
\bauthor{\bsnm{{Zheng}}, \binits{W.}},
\bauthor{\bsnm{{Nonino}}, \binits{M.}},
\bauthor{\bsnm{{Molino}}, \binits{A.}},
\bauthor{\bsnm{{Mahdavi}}, \binits{A.}},
\bauthor{\bsnm{{Lemze}}, \binits{D.}},
\bauthor{\bsnm{{Infante}}, \binits{L.}},
\bauthor{\bsnm{{Ogaz}}, \binits{S.}},
\bauthor{\bsnm{{Melchior}}, \binits{P.}},
\bauthor{\bsnm{{Host}}, \binits{O.}},
\bauthor{\bsnm{{Ford}}, \binits{H.}},
\bauthor{\bsnm{{Grillo}}, \binits{C.}},
\bauthor{\bsnm{{Rosati}}, \binits{P.}},
\bauthor{\bsnm{{Jim{\'e}nez-Teja}}, \binits{Y.}},
\bauthor{\bsnm{{Moustakas}}, \binits{J.}},
\bauthor{\bsnm{{Broadhurst}}, \binits{T.}},
\bauthor{\bsnm{{Ascaso}}, \binits{B.}},
\bauthor{\bsnm{{Lahav}}, \binits{O.}},
\bauthor{\bsnm{{Bartelmann}}, \binits{M.}},
\bauthor{\bsnm{{Ben\'{\i}tez}}, \binits{N.}},
\bauthor{\bsnm{{Bouwens}}, \binits{R.}},
\bauthor{\bsnm{{Graur}}, \binits{O.}},
\bauthor{\bsnm{{Graves}}, \binits{G.}},
\bauthor{\bsnm{{Jha}}, \binits{S.}},
\bauthor{\bsnm{{Jouvel}}, \binits{S.}},
\bauthor{\bsnm{{Kelson}}, \binits{D.}},
\bauthor{\bsnm{{Moustakas}}, \binits{L.}},
\bauthor{\bsnm{{Maoz}}, \binits{D.}},
\bauthor{\bsnm{{Meneghetti}}, \binits{M.}},
\bauthor{\bsnm{{Merten}}, \binits{J.}},
\bauthor{\bsnm{{Riess}}, \binits{A.}},
\bauthor{\bsnm{{Rodney}}, \binits{S.}},
\bauthor{\bsnm{{Seitz}}, \binits{S.}}:
\batitle{{CLASH: Precise New Constraints on the Mass Profile of the Galaxy
  Cluster A2261}}.
\bjtitle{\apj}
\bvolume{757}(\bissue{1}),
\bfpage{22}
(\byear{2012})
{\href{https://arxiv.org/abs/1201.1616}{{arXiv:1201.1616}}}
{[astro-ph.CO]}.
\doiurl{10.1088/0004-637X/757/1/22}
\end{barticle}
\endbibitem

%%% 4
\bibitem{2013SSRv..177....3B}
\begin{barticle}
\bauthor{\bsnm{{Bartelmann}}, \binits{M.}},
\bauthor{\bsnm{{Limousin}}, \binits{M.}},
\bauthor{\bsnm{{Meneghetti}}, \binits{M.}},
\bauthor{\bsnm{{Schmidt}}, \binits{R.}}:
\batitle{{Internal Cluster Structure}}.
\bjtitle{\ssr}
\bvolume{177}(\bissue{1-4}),
\bfpage{3}--\blpage{29}
(\byear{2013})
{\href{https://arxiv.org/abs/1303.3285}{{arXiv:1303.3285}}}
{[astro-ph.CO]}.
\doiurl{10.1007/s11214-013-9977-6}
\end{barticle}
\endbibitem

%%% 5
\bibitem{1982ApJ...263L...1P}
\begin{barticle}
\bauthor{\bsnm{{Peebles}}, \binits{P.J.E.}}:
\batitle{{Large-scale background temperature and mass fluctuations due to
  scale-invariant primeval perturbations}}.
\bjtitle{\apjl}
\bvolume{263},
\bfpage{1}--\blpage{5}
(\byear{1982}).
\doiurl{10.1086/183911}
\end{barticle}
\endbibitem

%%% 6
\bibitem{dodelson2020modern}
\begin{bbook}
\bauthor{\bsnm{Dodelson}, \binits{S.}},
\bauthor{\bsnm{Schmidt}, \binits{F.}}:
\bbtitle{Modern Cosmology}.
\bpublisher{Academic Press}, \blocation{???}
(\byear{2020})
\end{bbook}
\endbibitem

%%% 7
\bibitem{1993ApJ...416....1K}
\begin{barticle}
\bauthor{\bsnm{{Klypin}}, \binits{A.}},
\bauthor{\bsnm{{Holtzman}}, \binits{J.}},
\bauthor{\bsnm{{Primack}}, \binits{J.}},
\bauthor{\bsnm{{Regos}}, \binits{E.}}:
\batitle{{Structure Formation with Cold plus Hot Dark Matter}}.
\bjtitle{\apj}
\bvolume{416},
\bfpage{1}
(\byear{1993})
{\href{https://arxiv.org/abs/astro-ph/9305011}{{arXiv:astro-ph/9305011}}}
{[astro-ph]}.
\doiurl{10.1086/173210}
\end{barticle}
\endbibitem

%%% 8
\bibitem{2005Natur.435..629S}
\begin{barticle}
\bauthor{\bsnm{{Springel}}, \binits{V.}},
\bauthor{\bsnm{{White}}, \binits{S.D.M.}},
\bauthor{\bsnm{{Jenkins}}, \binits{A.}},
\bauthor{\bsnm{{Frenk}}, \binits{C.S.}},
\bauthor{\bsnm{{Yoshida}}, \binits{N.}},
\bauthor{\bsnm{{Gao}}, \binits{L.}},
\bauthor{\bsnm{{Navarro}}, \binits{J.}},
\bauthor{\bsnm{{Thacker}}, \binits{R.}},
\bauthor{\bsnm{{Croton}}, \binits{D.}},
\bauthor{\bsnm{{Helly}}, \binits{J.}},
\bauthor{\bsnm{{Peacock}}, \binits{J.A.}},
\bauthor{\bsnm{{Cole}}, \binits{S.}},
\bauthor{\bsnm{{Thomas}}, \binits{P.}},
\bauthor{\bsnm{{Couchman}}, \binits{H.}},
\bauthor{\bsnm{{Evrard}}, \binits{A.}},
\bauthor{\bsnm{{Colberg}}, \binits{J.}},
\bauthor{\bsnm{{Pearce}}, \binits{F.}}:
\batitle{{Simulations of the formation, evolution and clustering of galaxies
  and quasars}}.
\bjtitle{\nat}
\bvolume{435}(\bissue{7042}),
\bfpage{629}--\blpage{636}
(\byear{2005})
{\href{https://arxiv.org/abs/astro-ph/0504097}{{arXiv:astro-ph/0504097}}}
{[astro-ph]}.
\doiurl{10.1038/nature03597}
\end{barticle}
\endbibitem

%%% 9
\bibitem{2018MNRAS.475..676S}
\begin{barticle}
\bauthor{\bsnm{{Springel}}, \binits{V.}},
\bauthor{\bsnm{{Pakmor}}, \binits{R.}},
\bauthor{\bsnm{{Pillepich}}, \binits{A.}},
\bauthor{\bsnm{{Weinberger}}, \binits{R.}},
\bauthor{\bsnm{{Nelson}}, \binits{D.}},
\bauthor{\bsnm{{Hernquist}}, \binits{L.}},
\bauthor{\bsnm{{Vogelsberger}}, \binits{M.}},
\bauthor{\bsnm{{Genel}}, \binits{S.}},
\bauthor{\bsnm{{Torrey}}, \binits{P.}},
\bauthor{\bsnm{{Marinacci}}, \binits{F.}},
\bauthor{\bsnm{{Naiman}}, \binits{J.}}:
\batitle{{First results from the IllustrisTNG simulations: matter and galaxy
  clustering}}.
\bjtitle{\mnras}
\bvolume{475}(\bissue{1}),
\bfpage{676}--\blpage{698}
(\byear{2018})
{\href{https://arxiv.org/abs/1707.03397}{{arXiv:1707.03397}}}
{[astro-ph.GA]}.
\doiurl{10.1093/mnras/stx3304}
\end{barticle}
\endbibitem

%%% 10
\bibitem{2017MNRAS.469.1824X}
\begin{barticle}
\bauthor{\bsnm{{Xu}}, \binits{D.}},
\bauthor{\bsnm{{Springel}}, \binits{V.}},
\bauthor{\bsnm{{Sluse}}, \binits{D.}},
\bauthor{\bsnm{{Schneider}}, \binits{P.}},
\bauthor{\bsnm{{Sonnenfeld}}, \binits{A.}},
\bauthor{\bsnm{{Nelson}}, \binits{D.}},
\bauthor{\bsnm{{Vogelsberger}}, \binits{M.}},
\bauthor{\bsnm{{Hernquist}}, \binits{L.}}:
\batitle{{The inner structure of early-type galaxies in the Illustris
  simulation}}.
\bjtitle{\mnras}
\bvolume{469}(\bissue{2}),
\bfpage{1824}--\blpage{1848}
(\byear{2017})
{\href{https://arxiv.org/abs/1610.07605}{{arXiv:1610.07605}}}
{[astro-ph.GA]}.
\doiurl{10.1093/mnras/stx899}
\end{barticle}
\endbibitem

%%% 11
\bibitem{2019MNRAS.484..476C}
\begin{barticle}
\bauthor{\bsnm{{Chua}}, \binits{K.T.E.}},
\bauthor{\bsnm{{Pillepich}}, \binits{A.}},
\bauthor{\bsnm{{Vogelsberger}}, \binits{M.}},
\bauthor{\bsnm{{Hernquist}}, \binits{L.}}:
\batitle{{Shape of dark matter haloes in the Illustris simulation: effects of
  baryons}}.
\bjtitle{\mnras}
\bvolume{484}(\bissue{1}),
\bfpage{476}--\blpage{493}
(\byear{2019})
{\href{https://arxiv.org/abs/1809.07255}{{arXiv:1809.07255}}}
{[astro-ph.GA]}.
\doiurl{10.1093/mnras/sty3531}
\end{barticle}
\endbibitem

%%% 12
\bibitem{2020NatRP...2...42V}
\begin{barticle}
\bauthor{\bsnm{{Vogelsberger}}, \binits{M.}},
\bauthor{\bsnm{{Marinacci}}, \binits{F.}},
\bauthor{\bsnm{{Torrey}}, \binits{P.}},
\bauthor{\bsnm{{Puchwein}}, \binits{E.}}:
\batitle{{Cosmological simulations of galaxy formation}}.
\bjtitle{Nature Reviews Physics}
\bvolume{2}(\bissue{1}),
\bfpage{42}--\blpage{66}
(\byear{2020})
{\href{https://arxiv.org/abs/1909.07976}{{arXiv:1909.07976}}}
{[astro-ph.GA]}.
\doiurl{10.1038/s42254-019-0127-2}
\end{barticle}
\endbibitem

%%% 13
\bibitem{2020Natur.585...39W}
\begin{barticle}
\bauthor{\bsnm{{Wang}}, \binits{J.}},
\bauthor{\bsnm{{Bose}}, \binits{S.}},
\bauthor{\bsnm{{Frenk}}, \binits{C.S.}},
\bauthor{\bsnm{{Gao}}, \binits{L.}},
\bauthor{\bsnm{{Jenkins}}, \binits{A.}},
\bauthor{\bsnm{{Springel}}, \binits{V.}},
\bauthor{\bsnm{{White}}, \binits{S.D.M.}}:
\batitle{{Universal structure of dark matter haloes over a mass range of 20
  orders of magnitude}}.
\bjtitle{\nat}
\bvolume{585}(\bissue{7823}),
\bfpage{39}--\blpage{42}
(\byear{2020})
{\href{https://arxiv.org/abs/1911.09720}{{arXiv:1911.09720}}}
{[astro-ph.CO]}.
\doiurl{10.1038/s41586-020-2642-9}
\end{barticle}
\endbibitem

%%% 14
\bibitem{2002PhR...367....1B}
\begin{barticle}
\bauthor{\bsnm{{Bernardeau}}, \binits{F.}},
\bauthor{\bsnm{{Colombi}}, \binits{S.}},
\bauthor{\bsnm{{Gazta{\~n}aga}}, \binits{E.}},
\bauthor{\bsnm{{Scoccimarro}}, \binits{R.}}:
\batitle{{Large-scale structure of the Universe and cosmological perturbation
  theory}}.
\bjtitle{\physrep}
\bvolume{367}(\bissue{1-3}),
\bfpage{1}--\blpage{248}
(\byear{2002})
{\href{https://arxiv.org/abs/astro-ph/0112551}{{arXiv:astro-ph/0112551}}}
{[astro-ph]}.
\doiurl{10.1016/S0370-1573(02)00135-7}
\end{barticle}
\endbibitem

%%% 15
\bibitem{1970PhRvD...1.2726H}
\begin{barticle}
\bauthor{\bsnm{{Harrison}}, \binits{E.R.}}:
\batitle{{Fluctuations at the Threshold of Classical Cosmology}}.
\bjtitle{\prd}
\bvolume{1}(\bissue{10}),
\bfpage{2726}--\blpage{2730}
(\byear{1970}).
\doiurl{10.1103/PhysRevD.1.2726}
\end{barticle}
\endbibitem

%%% 16
\bibitem{1970ApJ...162..815P}
\begin{barticle}
\bauthor{\bsnm{{Peebles}}, \binits{P.J.E.}},
\bauthor{\bsnm{{Yu}}, \binits{J.T.}}:
\batitle{{Primeval Adiabatic Perturbation in an Expanding Universe}}.
\bjtitle{\apj}
\bvolume{162},
\bfpage{815}
(\byear{1970}).
\doiurl{10.1086/150713}
\end{barticle}
\endbibitem

%%% 17
\bibitem{1972MNRAS.160P...1Z}
\begin{barticle}
\bauthor{\bsnm{{Zeldovich}}, \binits{Y.B.}}:
\batitle{{A hypothesis, unifying the structure and the entropy of the
  Universe}}.
\bjtitle{\mnras}
\bvolume{160},
\bfpage{1}
(\byear{1972}).
\doiurl{10.1093/mnras/160.1.1P}
\end{barticle}
\endbibitem

%%% 18
\bibitem{2020A&A...641A...1P}
\begin{barticle}
\bauthor{\bsnm{{Planck Collaboration}}},
\bauthor{\bsnm{{Aghanim}}, \binits{N.}},
\bauthor{\bsnm{{Akrami}}, \binits{Y.}},
\bauthor{\bsnm{{Arroja}}, \binits{F.}},
\bauthor{\bsnm{{Ashdown}}, \binits{M.}},
\bauthor{\bsnm{{Aumont}}, \binits{J.}},
\bauthor{\bsnm{{Baccigalupi}}, \binits{C.}},
\bauthor{\bsnm{{Ballardini}}, \binits{M.}},
\bauthor{\bsnm{{Banday}}, \binits{A.J.}},
\bauthor{\bsnm{{Barreiro}}, \binits{R.B.}},
\bauthor{\bsnm{{Bartolo}}, \binits{N.}},
\bauthor{\bsnm{{Basak}}, \binits{S.}},
\bauthor{\bsnm{{Battye}}, \binits{R.}},
\bauthor{\bsnm{{Benabed}}, \binits{K.}},
\bauthor{\bsnm{{Bernard}}, \binits{J.-P.}},
\bauthor{\bsnm{{Bersanelli}}, \binits{M.}},
\bauthor{\bsnm{{Bielewicz}}, \binits{P.}},
\bauthor{\bsnm{{Bock}}, \binits{J.J.}},
\bauthor{\bsnm{{Bond}}, \binits{J.R.}},
\bauthor{\bsnm{{Borrill}}, \binits{J.}},
\bauthor{\bsnm{{Bouchet}}, \binits{F.R.}},
\bauthor{\bsnm{{Boulanger}}, \binits{F.}},
\bauthor{\bsnm{{Bucher}}, \binits{M.}},
\bauthor{\bsnm{{Burigana}}, \binits{C.}},
\bauthor{\bsnm{{Butler}}, \binits{R.C.}},
\bauthor{\bsnm{{Calabrese}}, \binits{E.}},
\bauthor{\bsnm{{Cardoso}}, \binits{J.-F.}},
\bauthor{\bsnm{{Carron}}, \binits{J.}},
\bauthor{\bsnm{{Casaponsa}}, \binits{B.}},
\bauthor{\bsnm{{Challinor}}, \binits{A.}},
\bauthor{\bsnm{{Chiang}}, \binits{H.C.}},
\bauthor{\bsnm{{Colombo}}, \binits{L.P.L.}},
\bauthor{\bsnm{{Combet}}, \binits{C.}},
\bauthor{\bsnm{{Contreras}}, \binits{D.}},
\bauthor{\bsnm{{Crill}}, \binits{B.P.}},
\bauthor{\bsnm{{Cuttaia}}, \binits{F.}},
\bauthor{\bsnm{{de Bernardis}}, \binits{P.}},
\bauthor{\bsnm{{de Zotti}}, \binits{G.}},
\bauthor{\bsnm{{Delabrouille}}, \binits{J.}},
\bauthor{\bsnm{{Delouis}}, \binits{J.-M.}},
\bauthor{\bsnm{{D{\'e}sert}}, \binits{F.-X.}},
\bauthor{\bsnm{{Di Valentino}}, \binits{E.}},
\bauthor{\bsnm{{Dickinson}}, \binits{C.}},
\bauthor{\bsnm{{Diego}}, \binits{J.M.}},
\bauthor{\bsnm{{Donzelli}}, \binits{S.}},
\bauthor{\bsnm{{Dor{\'e}}}, \binits{O.}},
\bauthor{\bsnm{{Douspis}}, \binits{M.}},
\bauthor{\bsnm{{Ducout}}, \binits{A.}},
\bauthor{\bsnm{{Dupac}}, \binits{X.}},
\bauthor{\bsnm{{Efstathiou}}, \binits{G.}},
\bauthor{\bsnm{{Elsner}}, \binits{F.}},
\bauthor{\bsnm{{En{\ss}lin}}, \binits{T.A.}},
\bauthor{\bsnm{{Eriksen}}, \binits{H.K.}},
\bauthor{\bsnm{{Falgarone}}, \binits{E.}},
\bauthor{\bsnm{{Fantaye}}, \binits{Y.}},
\bauthor{\bsnm{{Fergusson}}, \binits{J.}},
\bauthor{\bsnm{{Fernandez-Cobos}}, \binits{R.}},
\bauthor{\bsnm{{Finelli}}, \binits{F.}},
\bauthor{\bsnm{{Forastieri}}, \binits{F.}},
\bauthor{\bsnm{{Frailis}}, \binits{M.}},
\bauthor{\bsnm{{Franceschi}}, \binits{E.}},
\bauthor{\bsnm{{Frolov}}, \binits{A.}},
\bauthor{\bsnm{{Galeotta}}, \binits{S.}},
\bauthor{\bsnm{{Galli}}, \binits{S.}},
\bauthor{\bsnm{{Ganga}}, \binits{K.}},
\bauthor{\bsnm{{G{\'e}nova-Santos}}, \binits{R.T.}},
\bauthor{\bsnm{{Gerbino}}, \binits{M.}},
\bauthor{\bsnm{{Ghosh}}, \binits{T.}},
\bauthor{\bsnm{{Gonz{\'a}lez-Nuevo}}, \binits{J.}},
\bauthor{\bsnm{{G{\'o}rski}}, \binits{K.M.}},
\bauthor{\bsnm{{Gratton}}, \binits{S.}},
\bauthor{\bsnm{{Gruppuso}}, \binits{A.}},
\bauthor{\bsnm{{Gudmundsson}}, \binits{J.E.}},
\bauthor{\bsnm{{Hamann}}, \binits{J.}},
\bauthor{\bsnm{{Handley}}, \binits{W.}},
\bauthor{\bsnm{{Hansen}}, \binits{F.K.}},
\bauthor{\bsnm{{Helou}}, \binits{G.}},
\bauthor{\bsnm{{Herranz}}, \binits{D.}},
\bauthor{\bsnm{{Hildebrandt}}, \binits{S.R.}},
\bauthor{\bsnm{{Hivon}}, \binits{E.}},
\bauthor{\bsnm{{Huang}}, \binits{Z.}},
\bauthor{\bsnm{{Jaffe}}, \binits{A.H.}},
\bauthor{\bsnm{{Jones}}, \binits{W.C.}},
\bauthor{\bsnm{{Karakci}}, \binits{A.}},
\bauthor{\bsnm{{Keih{\"a}nen}}, \binits{E.}},
\bauthor{\bsnm{{Keskitalo}}, \binits{R.}},
\bauthor{\bsnm{{Kiiveri}}, \binits{K.}},
\bauthor{\bsnm{{Kim}}, \binits{J.}},
\bauthor{\bsnm{{Kisner}}, \binits{T.S.}},
\bauthor{\bsnm{{Knox}}, \binits{L.}},
\bauthor{\bsnm{{Krachmalnicoff}}, \binits{N.}},
\bauthor{\bsnm{{Kunz}}, \binits{M.}},
\bauthor{\bsnm{{Kurki-Suonio}}, \binits{H.}},
\bauthor{\bsnm{{Lagache}}, \binits{G.}},
\bauthor{\bsnm{{Lamarre}}, \binits{J.-M.}},
\bauthor{\bsnm{{Langer}}, \binits{M.}},
\bauthor{\bsnm{{Lasenby}}, \binits{A.}},
\bauthor{\bsnm{{Lattanzi}}, \binits{M.}},
\bauthor{\bsnm{{Lawrence}}, \binits{C.R.}},
\bauthor{\bsnm{{Le Jeune}}, \binits{M.}},
\bauthor{\bsnm{{Leahy}}, \binits{J.P.}},
\bauthor{\bsnm{{Lesgourgues}}, \binits{J.}},
\bauthor{\bsnm{{Levrier}}, \binits{F.}},
\bauthor{\bsnm{{Lewis}}, \binits{A.}},
\bauthor{\bsnm{{Liguori}}, \binits{M.}},
\bauthor{\bsnm{{Lilje}}, \binits{P.B.}},
\bauthor{\bsnm{{Lilley}}, \binits{M.}},
\bauthor{\bsnm{{Lindholm}}, \binits{V.}},
\bauthor{\bsnm{{L{\'o}pez-Caniego}}, \binits{M.}},
\bauthor{\bsnm{{Lubin}}, \binits{P.M.}},
\bauthor{\bsnm{{Ma}}, \binits{Y.-Z.}},
\bauthor{\bsnm{{Mac\'{\i}as-P{\'e}rez}}, \binits{J.F.}},
\bauthor{\bsnm{{Maggio}}, \binits{G.}},
\bauthor{\bsnm{{Maino}}, \binits{D.}},
\bauthor{\bsnm{{Mandolesi}}, \binits{N.}},
\bauthor{\bsnm{{Mangilli}}, \binits{A.}},
\bauthor{\bsnm{{Marcos-Caballero}}, \binits{A.}},
\bauthor{\bsnm{{Maris}}, \binits{M.}},
\bauthor{\bsnm{{Martin}}, \binits{P.G.}},
\bauthor{\bsnm{{Martinelli}}, \binits{M.}},
\bauthor{\bsnm{{Mart\'{\i}nez-Gonz{\'a}lez}}, \binits{E.}},
\bauthor{\bsnm{{Matarrese}}, \binits{S.}},
\bauthor{\bsnm{{Mauri}}, \binits{N.}},
\bauthor{\bsnm{{McEwen}}, \binits{J.D.}},
\bauthor{\bsnm{{Meerburg}}, \binits{P.D.}},
\bauthor{\bsnm{{Meinhold}}, \binits{P.R.}},
\bauthor{\bsnm{{Melchiorri}}, \binits{A.}},
\bauthor{\bsnm{{Mennella}}, \binits{A.}},
\bauthor{\bsnm{{Migliaccio}}, \binits{M.}},
\bauthor{\bsnm{{Millea}}, \binits{M.}},
\bauthor{\bsnm{{Mitra}}, \binits{S.}},
\bauthor{\bsnm{{Miville-Desch{\^e}nes}}, \binits{M.-A.}},
\bauthor{\bsnm{{Molinari}}, \binits{D.}},
\bauthor{\bsnm{{Moneti}}, \binits{A.}},
\bauthor{\bsnm{{Montier}}, \binits{L.}},
\bauthor{\bsnm{{Morgante}}, \binits{G.}},
\bauthor{\bsnm{{Moss}}, \binits{A.}},
\bauthor{\bsnm{{Mottet}}, \binits{S.}},
\bauthor{\bsnm{{M{\"u}nchmeyer}}, \binits{M.}},
\bauthor{\bsnm{{Natoli}}, \binits{P.}},
\bauthor{\bsnm{{N{\o}rgaard-Nielsen}}, \binits{H.U.}},
\bauthor{\bsnm{{Oxborrow}}, \binits{C.A.}},
\bauthor{\bsnm{{Pagano}}, \binits{L.}},
\bauthor{\bsnm{{Paoletti}}, \binits{D.}},
\bauthor{\bsnm{{Partridge}}, \binits{B.}},
\bauthor{\bsnm{{Patanchon}}, \binits{G.}},
\bauthor{\bsnm{{Pearson}}, \binits{T.J.}},
\bauthor{\bsnm{{Peel}}, \binits{M.}},
\bauthor{\bsnm{{Peiris}}, \binits{H.V.}},
\bauthor{\bsnm{{Perrotta}}, \binits{F.}},
\bauthor{\bsnm{{Pettorino}}, \binits{V.}},
\bauthor{\bsnm{{Piacentini}}, \binits{F.}},
\bauthor{\bsnm{{Polastri}}, \binits{L.}},
\bauthor{\bsnm{{Polenta}}, \binits{G.}},
\bauthor{\bsnm{{Puget}}, \binits{J.-L.}},
\bauthor{\bsnm{{Rachen}}, \binits{J.P.}},
\bauthor{\bsnm{{Reinecke}}, \binits{M.}},
\bauthor{\bsnm{{Remazeilles}}, \binits{M.}},
\bauthor{\bsnm{{Renault}}, \binits{C.}},
\bauthor{\bsnm{{Renzi}}, \binits{A.}},
\bauthor{\bsnm{{Rocha}}, \binits{G.}},
\bauthor{\bsnm{{Rosset}}, \binits{C.}},
\bauthor{\bsnm{{Roudier}}, \binits{G.}},
\bauthor{\bsnm{{Rubi{\~n}o-Mart\'{\i}n}}, \binits{J.A.}},
\bauthor{\bsnm{{Ruiz-Granados}}, \binits{B.}},
\bauthor{\bsnm{{Salvati}}, \binits{L.}},
\bauthor{\bsnm{{Sandri}}, \binits{M.}},
\bauthor{\bsnm{{Savelainen}}, \binits{M.}},
\bauthor{\bsnm{{Scott}}, \binits{D.}},
\bauthor{\bsnm{{Shellard}}, \binits{E.P.S.}},
\bauthor{\bsnm{{Shiraishi}}, \binits{M.}},
\bauthor{\bsnm{{Sirignano}}, \binits{C.}},
\bauthor{\bsnm{{Sirri}}, \binits{G.}},
\bauthor{\bsnm{{Spencer}}, \binits{L.D.}},
\bauthor{\bsnm{{Sunyaev}}, \binits{R.}},
\bauthor{\bsnm{{Suur-Uski}}, \binits{A.-S.}},
\bauthor{\bsnm{{Tauber}}, \binits{J.A.}},
\bauthor{\bsnm{{Tavagnacco}}, \binits{D.}},
\bauthor{\bsnm{{Tenti}}, \binits{M.}},
\bauthor{\bsnm{{Terenzi}}, \binits{L.}},
\bauthor{\bsnm{{Toffolatti}}, \binits{L.}},
\bauthor{\bsnm{{Tomasi}}, \binits{M.}},
\bauthor{\bsnm{{Trombetti}}, \binits{T.}},
\bauthor{\bsnm{{Valiviita}}, \binits{J.}},
\bauthor{\bsnm{{Van Tent}}, \binits{B.}},
\bauthor{\bsnm{{Vibert}}, \binits{L.}},
\bauthor{\bsnm{{Vielva}}, \binits{P.}},
\bauthor{\bsnm{{Villa}}, \binits{F.}},
\bauthor{\bsnm{{Vittorio}}, \binits{N.}},
\bauthor{\bsnm{{Wandelt}}, \binits{B.D.}},
\bauthor{\bsnm{{Wehus}}, \binits{I.K.}},
\bauthor{\bsnm{{White}}, \binits{M.}},
\bauthor{\bsnm{{White}}, \binits{S.D.M.}},
\bauthor{\bsnm{{Zacchei}}, \binits{A.}},
\bauthor{\bsnm{{Zonca}}, \binits{A.}}:
\batitle{{Planck 2018 results. I. Overview and the cosmological legacy of
  Planck}}.
\bjtitle{\aap}
\bvolume{641},
\bfpage{1}
(\byear{2020})
{\href{https://arxiv.org/abs/1807.06205}{{arXiv:1807.06205}}}
{[astro-ph.CO]}.
\doiurl{10.1051/0004-6361/201833880}
\end{barticle}
\endbibitem

%%% 19
\bibitem{2016MNRAS.459.1468M}
\begin{barticle}
\bauthor{\bsnm{{Mead}}, \binits{A.J.}},
\bauthor{\bsnm{{Heymans}}, \binits{C.}},
\bauthor{\bsnm{{Lombriser}}, \binits{L.}},
\bauthor{\bsnm{{Peacock}}, \binits{J.A.}},
\bauthor{\bsnm{{Steele}}, \binits{O.I.}},
\bauthor{\bsnm{{Winther}}, \binits{H.A.}}:
\batitle{{Accurate halo-model matter power spectra with dark energy, massive
  neutrinos and modified gravitational forces}}.
\bjtitle{\mnras}
\bvolume{459}(\bissue{2}),
\bfpage{1468}--\blpage{1488}
(\byear{2016})
{\href{https://arxiv.org/abs/1602.02154}{{arXiv:1602.02154}}}
{[astro-ph.CO]}.
\doiurl{10.1093/mnras/stw681}
\end{barticle}
\endbibitem

%%% 20
\bibitem{2021MNRAS.506.2871S}
\begin{barticle}
\bauthor{\bsnm{{Springel}}, \binits{V.}},
\bauthor{\bsnm{{Pakmor}}, \binits{R.}},
\bauthor{\bsnm{{Zier}}, \binits{O.}},
\bauthor{\bsnm{{Reinecke}}, \binits{M.}}:
\batitle{{Simulating cosmic structure formation with the GADGET-4 code}}.
\bjtitle{\mnras}
\bvolume{506}(\bissue{2}),
\bfpage{2871}--\blpage{2949}
(\byear{2021})
{\href{https://arxiv.org/abs/2010.03567}{{arXiv:2010.03567}}}
{[astro-ph.IM]}.
\doiurl{10.1093/mnras/stab1855}
\end{barticle}
\endbibitem

%%% 21
\bibitem{2020A&A...634A.127A}
\begin{barticle}
\bauthor{\bsnm{{Asgari}}, \binits{M.}},
\bauthor{\bsnm{{Tr{\"o}ster}}, \binits{T.}},
\bauthor{\bsnm{{Heymans}}, \binits{C.}},
\bauthor{\bsnm{{Hildebrandt}}, \binits{H.}},
\bauthor{\bsnm{{van den Busch}}, \binits{J.L.}},
\bauthor{\bsnm{{Wright}}, \binits{A.H.}},
\bauthor{\bsnm{{Choi}}, \binits{A.}},
\bauthor{\bsnm{{Erben}}, \binits{T.}},
\bauthor{\bsnm{{Joachimi}}, \binits{B.}},
\bauthor{\bsnm{{Joudaki}}, \binits{S.}},
\bauthor{\bsnm{{Kannawadi}}, \binits{A.}},
\bauthor{\bsnm{{Kuijken}}, \binits{K.}},
\bauthor{\bsnm{{Lin}}, \binits{C.-A.}},
\bauthor{\bsnm{{Schneider}}, \binits{P.}},
\bauthor{\bsnm{{Zuntz}}, \binits{J.}}:
\batitle{{KiDS+VIKING-450 and DES-Y1 combined: Mitigating baryon feedback
  uncertainty with COSEBIs}}.
\bjtitle{\aap}
\bvolume{634},
\bfpage{127}
(\byear{2020})
{\href{https://arxiv.org/abs/1910.05336}{{arXiv:1910.05336}}}
{[astro-ph.CO]}.
\doiurl{10.1051/0004-6361/201936512}
\end{barticle}
\endbibitem

%%% 22
\bibitem{2017MNRAS.471.4412K}
\begin{barticle}
\bauthor{\bsnm{{K{\"o}hlinger}}, \binits{F.}},
\bauthor{\bsnm{{Viola}}, \binits{M.}},
\bauthor{\bsnm{{Joachimi}}, \binits{B.}},
\bauthor{\bsnm{{Hoekstra}}, \binits{H.}},
\bauthor{\bsnm{{van Uitert}}, \binits{E.}},
\bauthor{\bsnm{{Hildebrandt}}, \binits{H.}},
\bauthor{\bsnm{{Choi}}, \binits{A.}},
\bauthor{\bsnm{{Erben}}, \binits{T.}},
\bauthor{\bsnm{{Heymans}}, \binits{C.}},
\bauthor{\bsnm{{Joudaki}}, \binits{S.}},
\bauthor{\bsnm{{Klaes}}, \binits{D.}},
\bauthor{\bsnm{{Kuijken}}, \binits{K.}},
\bauthor{\bsnm{{Merten}}, \binits{J.}},
\bauthor{\bsnm{{Miller}}, \binits{L.}},
\bauthor{\bsnm{{Schneider}}, \binits{P.}},
\bauthor{\bsnm{{Valentijn}}, \binits{E.A.}}:
\batitle{{KiDS-450: the tomographic weak lensing power spectrum and constraints
  on cosmological parameters}}.
\bjtitle{\mnras}
\bvolume{471}(\bissue{4}),
\bfpage{4412}--\blpage{4435}
(\byear{2017})
{\href{https://arxiv.org/abs/1706.02892}{{arXiv:1706.02892}}}
{[astro-ph.CO]}.
\doiurl{10.1093/mnras/stx1820}
\end{barticle}
\endbibitem

%%% 23
\bibitem{1995ApJ...455....7M}
\begin{barticle}
\bauthor{\bsnm{{Ma}}, \binits{C.-P.}},
\bauthor{\bsnm{{Bertschinger}}, \binits{E.}}:
\batitle{{Cosmological Perturbation Theory in the Synchronous and Conformal
  Newtonian Gauges}}.
\bjtitle{\apj}
\bvolume{455},
\bfpage{7}
(\byear{1995})
{\href{https://arxiv.org/abs/astro-ph/9506072}{{arXiv:astro-ph/9506072}}}
{[astro-ph]}.
\doiurl{10.1086/176550}
\end{barticle}
\endbibitem

%%% 24
\bibitem{2001A&A...379....8V}
\begin{barticle}
\bauthor{\bsnm{{Valageas}}, \binits{P.}}:
\batitle{{Dynamics of gravitational clustering. I. Building perturbative
  expansions}}.
\bjtitle{\aap}
\bvolume{379},
\bfpage{8}--\blpage{20}
(\byear{2001})
{\href{https://arxiv.org/abs/astro-ph/0107015}{{arXiv:astro-ph/0107015}}}
{[astro-ph]}.
\doiurl{10.1051/0004-6361:20011309}
\end{barticle}
\endbibitem

%%% 25
\bibitem{2004ApJ...612...28M}
\begin{barticle}
\bauthor{\bsnm{{Ma}}, \binits{C.-P.}},
\bauthor{\bsnm{{Bertschinger}}, \binits{E.}}:
\batitle{{A Cosmological Kinetic Theory for the Evolution of Cold Dark Matter
  Halos with Substructure: Quasi-Linear Theory}}.
\bjtitle{\apj}
\bvolume{612}(\bissue{1}),
\bfpage{28}--\blpage{49}
(\byear{2004})
{\href{https://arxiv.org/abs/astro-ph/0311049}{{arXiv:astro-ph/0311049}}}
{[astro-ph]}.
\doiurl{10.1086/421766}
\end{barticle}
\endbibitem

%%% 26
\bibitem{2006PhRvD..73f3519C}
\begin{barticle}
\bauthor{\bsnm{{Crocce}}, \binits{M.}},
\bauthor{\bsnm{{Scoccimarro}}, \binits{R.}}:
\batitle{{Renormalized cosmological perturbation theory}}.
\bjtitle{\prd}
\bvolume{73}(\bissue{6}),
\bfpage{063519}
(\byear{2006})
{\href{https://arxiv.org/abs/astro-ph/0509418}{{arXiv:astro-ph/0509418}}}
{[astro-ph]}.
\doiurl{10.1103/PhysRevD.73.063519}
\end{barticle}
\endbibitem

%%% 27
\bibitem{2006PhRvD..73f3520C}
\begin{barticle}
\bauthor{\bsnm{{Crocce}}, \binits{M.}},
\bauthor{\bsnm{{Scoccimarro}}, \binits{R.}}:
\batitle{{Memory of initial conditions in gravitational clustering}}.
\bjtitle{\prd}
\bvolume{73}(\bissue{6}),
\bfpage{063520}
(\byear{2006})
{\href{https://arxiv.org/abs/astro-ph/0509419}{{arXiv:astro-ph/0509419}}}
{[astro-ph]}.
\doiurl{10.1103/PhysRevD.73.063520}
\end{barticle}
\endbibitem

%%% 28
\bibitem{2008JCAP...10..036P}
\begin{barticle}
\bauthor{\bsnm{{Pietroni}}, \binits{M.}}:
\batitle{{Flowing with time: a new approach to non-linear cosmological
  perturbations}}.
\bjtitle{\jcap}
\bvolume{2008}(\bissue{10}),
\bfpage{036}
(\byear{2008})
{\href{https://arxiv.org/abs/0806.0971}{{arXiv:0806.0971}}}
{[astro-ph]}.
\doiurl{10.1088/1475-7516/2008/10/036}
\end{barticle}
\endbibitem

%%% 29
\bibitem{2011JCAP...06..015A}
\begin{barticle}
\bauthor{\bsnm{{Anselmi}}, \binits{S.}},
\bauthor{\bsnm{{Matarrese}}, \binits{S.}},
\bauthor{\bsnm{{Pietroni}}, \binits{M.}}:
\batitle{{Next-to-leading resummations in cosmological perturbation theory}}.
\bjtitle{\jcap}
\bvolume{2011}(\bissue{6}),
\bfpage{015}
(\byear{2011})
{\href{https://arxiv.org/abs/1011.4477}{{arXiv:1011.4477}}}
{[astro-ph.CO]}.
\doiurl{10.1088/1475-7516/2011/06/015}
\end{barticle}
\endbibitem

%%% 30
\bibitem{2012JCAP...12..013A}
\begin{barticle}
\bauthor{\bsnm{{Anselmi}}, \binits{S.}},
\bauthor{\bsnm{{Pietroni}}, \binits{M.}}:
\batitle{{Nonlinear power spectrum from resummed perturbation theory: a leap
  beyond the BAO scale}}.
\bjtitle{\jcap}
\bvolume{2012}(\bissue{12}),
\bfpage{013}
(\byear{2012})
{\href{https://arxiv.org/abs/1205.2235}{{arXiv:1205.2235}}}
{[astro-ph.CO]}.
\doiurl{10.1088/1475-7516/2012/12/013}
\end{barticle}
\endbibitem

%%% 31
\bibitem{2012JCAP...01..019P}
\begin{barticle}
\bauthor{\bsnm{{Pietroni}}, \binits{M.}},
\bauthor{\bsnm{{Mangano}}, \binits{G.}},
\bauthor{\bsnm{{Saviano}}, \binits{N.}},
\bauthor{\bsnm{{Viel}}, \binits{M.}}:
\batitle{{Coarse-grained cosmological perturbation theory}}.
\bjtitle{\jcap}
\bvolume{2012}(\bissue{1}),
\bfpage{019}
(\byear{2012})
{\href{https://arxiv.org/abs/1108.5203}{{arXiv:1108.5203}}}
{[astro-ph.CO]}.
\doiurl{10.1088/1475-7516/2012/01/019}
\end{barticle}
\endbibitem

%%% 32
\bibitem{1992MNRAS.254..729B}
\begin{barticle}
\bauthor{\bsnm{{Buchert}}, \binits{T.}}:
\batitle{{Lagrangian theory of gravitational instability of Friedman-Lemaitre
  cosmologies and the 'Zel'dovich approximation'}}.
\bjtitle{\mnras}
\bvolume{254},
\bfpage{729}--\blpage{737}
(\byear{1992}).
\doiurl{10.1093/mnras/254.4.729}
\end{barticle}
\endbibitem

%%% 33
\bibitem{1993MNRAS.264..375B}
\begin{barticle}
\bauthor{\bsnm{{Buchert}}, \binits{T.}},
\bauthor{\bsnm{{Ehlers}}, \binits{J.}}:
\batitle{{Lagrangian theory of gravitational instability of Friedman-Lemaitre
  cosmologies -- second-order approach: an improved model for non-linear
  clustering}}.
\bjtitle{\mnras}
\bvolume{264},
\bfpage{375}--\blpage{387}
(\byear{1993}).
\doiurl{10.1093/mnras/264.2.375}
\end{barticle}
\endbibitem

%%% 34
\bibitem{1994MNRAS.267..811B}
\begin{barticle}
\bauthor{\bsnm{{Buchert}}, \binits{T.}}:
\batitle{{Lagrangian Theory of Gravitational Instability of Friedman-Lemaitre
  Cosmologies - a Generic Third-Order Model for Nonlinear Clustering}}.
\bjtitle{\mnras}
\bvolume{267},
\bfpage{811}
(\byear{1994})
{\href{https://arxiv.org/abs/astro-ph/9309055}{{arXiv:astro-ph/9309055}}}
{[astro-ph]}.
\doiurl{10.1093/mnras/267.4.811}
\end{barticle}
\endbibitem

%%% 35
\bibitem{1995A&A...296..575B}
\begin{barticle}
\bauthor{\bsnm{{Bouchet}}, \binits{F.R.}},
\bauthor{\bsnm{{Colombi}}, \binits{S.}},
\bauthor{\bsnm{{Hivon}}, \binits{E.}},
\bauthor{\bsnm{{Juszkiewicz}}, \binits{R.}}:
\batitle{{Perturbative Lagrangian approach to gravitational instability.}}
\bjtitle{\aap}
\bvolume{296},
\bfpage{575}
(\byear{1995})
{\href{https://arxiv.org/abs/astro-ph/9406013}{{arXiv:astro-ph/9406013}}}
{[astro-ph]}
\end{barticle}
\endbibitem

%%% 36
\bibitem{1997GReGr..29..733E}
\begin{barticle}
\bauthor{\bsnm{{Ehlers}}, \binits{J.}},
\bauthor{\bsnm{{Buchert}}, \binits{T.}}:
\batitle{{Newtonian Cosmology in Lagrangian Formulation: Foundations and
  Perturbation Theory}}.
\bjtitle{General Relativity and Gravitation}
\bvolume{29}(\bissue{6}),
\bfpage{733}--\blpage{764}
(\byear{1997})
{\href{https://arxiv.org/abs/astro-ph/9609036}{{arXiv:astro-ph/9609036}}}
{[astro-ph]}.
\doiurl{10.1023/A:1018885922682}
\end{barticle}
\endbibitem

%%% 37
\bibitem{2008PhRvD..77f3530M}
\begin{barticle}
\bauthor{\bsnm{{Matsubara}}, \binits{T.}}:
\batitle{{Resumming cosmological perturbations via the Lagrangian picture:
  One-loop results in real space and in redshift space}}.
\bjtitle{\prd}
\bvolume{77}(\bissue{6}),
\bfpage{063530}
(\byear{2008})
{\href{https://arxiv.org/abs/0711.2521}{{arXiv:0711.2521}}}
{[astro-ph]}.
\doiurl{10.1103/PhysRevD.77.063530}
\end{barticle}
\endbibitem

%%% 38
\bibitem{2008PhRvD..78h3503B}
\begin{barticle}
\bauthor{\bsnm{{Bernardeau}}, \binits{F.}},
\bauthor{\bsnm{{Valageas}}, \binits{P.}}:
\batitle{{Propagators in Lagrangian space}}.
\bjtitle{\prd}
\bvolume{78}(\bissue{8}),
\bfpage{083503}
(\byear{2008})
{\href{https://arxiv.org/abs/0805.0805}{{arXiv:0805.0805}}}
{[astro-ph]}.
\doiurl{10.1103/PhysRevD.78.083503}
\end{barticle}
\endbibitem

%%% 39
\bibitem{2012JCAP...06..021R}
\begin{barticle}
\bauthor{\bsnm{{Rampf}}, \binits{C.}},
\bauthor{\bsnm{{Buchert}}, \binits{T.}}:
\batitle{{Lagrangian perturbations and the matter bispectrum I: fourth-order
  model for non-linear clustering}}.
\bjtitle{\jcap}
\bvolume{2012}(\bissue{6}),
\bfpage{021}
(\byear{2012})
{\href{https://arxiv.org/abs/1203.4260}{{arXiv:1203.4260}}}
{[astro-ph.CO]}.
\doiurl{10.1088/1475-7516/2012/06/021}
\end{barticle}
\endbibitem

%%% 40
\bibitem{2013PhRvD..87h3522V}
\begin{barticle}
\bauthor{\bsnm{{Valageas}}, \binits{P.}},
\bauthor{\bsnm{{Nishimichi}}, \binits{T.}},
\bauthor{\bsnm{{Taruya}}, \binits{A.}}:
\batitle{{Matter power spectrum from a Lagrangian-space regularization of
  perturbation theory}}.
\bjtitle{\prd}
\bvolume{87}(\bissue{8}),
\bfpage{083522}
(\byear{2013})
{\href{https://arxiv.org/abs/1302.4533}{{arXiv:1302.4533}}}
{[astro-ph.CO]}.
\doiurl{10.1103/PhysRevD.87.083522}
\end{barticle}
\endbibitem

%%% 41
\bibitem{2012JCAP...07..051B}
\begin{barticle}
\bauthor{\bsnm{{Baumann}}, \binits{D.}},
\bauthor{\bsnm{{Nicolis}}, \binits{A.}},
\bauthor{\bsnm{{Senatore}}, \binits{L.}},
\bauthor{\bsnm{{Zaldarriaga}}, \binits{M.}}:
\batitle{{Cosmological non-linearities as an effective fluid}}.
\bjtitle{\jcap}
\bvolume{2012}(\bissue{7}),
\bfpage{051}
(\byear{2012})
{\href{https://arxiv.org/abs/1004.2488}{{arXiv:1004.2488}}}
{[astro-ph.CO]}.
\doiurl{10.1088/1475-7516/2012/07/051}
\end{barticle}
\endbibitem

%%% 42
\bibitem{2014JCAP...05..022P}
\begin{barticle}
\bauthor{\bsnm{{Porto}}, \binits{R.A.}},
\bauthor{\bsnm{{Senatore}}, \binits{L.}},
\bauthor{\bsnm{{Zaldarriaga}}, \binits{M.}}:
\batitle{{The Lagrangian-space Effective Field Theory of large scale
  structures}}.
\bjtitle{\jcap}
\bvolume{2014}(\bissue{5}),
\bfpage{022}
(\byear{2014})
{\href{https://arxiv.org/abs/1311.2168}{{arXiv:1311.2168}}}
{[astro-ph.CO]}.
\doiurl{10.1088/1475-7516/2014/05/022}
\end{barticle}
\endbibitem

%%% 43
\bibitem{2014JCAP...07..057C}
\begin{barticle}
\bauthor{\bsnm{{Carrasco}}, \binits{J.J.M.}},
\bauthor{\bsnm{{Foreman}}, \binits{S.}},
\bauthor{\bsnm{{Green}}, \binits{D.}},
\bauthor{\bsnm{{Senatore}}, \binits{L.}}:
\batitle{{The Effective Field Theory of Large Scale Structures at two loops}}.
\bjtitle{\jcap}
\bvolume{2014}(\bissue{7}),
\bfpage{057}
(\byear{2014})
{\href{https://arxiv.org/abs/1310.0464}{{arXiv:1310.0464}}}
{[astro-ph.CO]}.
\doiurl{10.1088/1475-7516/2014/07/057}
\end{barticle}
\endbibitem

%%% 44
\bibitem{2014PhRvD..89d3521H}
\begin{barticle}
\bauthor{\bsnm{{Hertzberg}}, \binits{M.P.}}:
\batitle{{Effective field theory of dark matter and structure formation:
  Semianalytical results}}.
\bjtitle{\prd}
\bvolume{89}(\bissue{4}),
\bfpage{043521}
(\byear{2014})
{\href{https://arxiv.org/abs/1208.0839}{{arXiv:1208.0839}}}
{[astro-ph.CO]}.
\doiurl{10.1103/PhysRevD.89.043521}
\end{barticle}
\endbibitem

%%% 45
\bibitem{2015JCAP...05..007B}
\begin{barticle}
\bauthor{\bsnm{{Baldauf}}, \binits{T.}},
\bauthor{\bsnm{{Mercolli}}, \binits{L.}},
\bauthor{\bsnm{{Mirbabayi}}, \binits{M.}},
\bauthor{\bsnm{{Pajer}}, \binits{E.}}:
\batitle{{The bispectrum in the Effective Field Theory of Large Scale
  Structure}}.
\bjtitle{\jcap}
\bvolume{2015}(\bissue{5}),
\bfpage{007}
(\byear{2015})
{\href{https://arxiv.org/abs/1406.4135}{{arXiv:1406.4135}}}
{[astro-ph.CO]}.
\doiurl{10.1088/1475-7516/2015/05/007}
\end{barticle}
\endbibitem

%%% 46
\bibitem{2018PhRvD..97f3526L}
\begin{barticle}
\bauthor{\bsnm{{Lewandowski}}, \binits{M.}},
\bauthor{\bsnm{{Senatore}}, \binits{L.}},
\bauthor{\bsnm{{Prada}}, \binits{F.}},
\bauthor{\bsnm{{Zhao}}, \binits{C.}},
\bauthor{\bsnm{{Chuang}}, \binits{C.-H.}}:
\batitle{{EFT of large scale structures in redshift space}}.
\bjtitle{\prd}
\bvolume{97}(\bissue{6}),
\bfpage{063526}
(\byear{2018})
{\href{https://arxiv.org/abs/1512.06831}{{arXiv:1512.06831}}}
{[astro-ph.CO]}.
\doiurl{10.1103/PhysRevD.97.063526}
\end{barticle}
\endbibitem

%%% 47
\bibitem{2019JCAP...11..027K}
\begin{barticle}
\bauthor{\bsnm{{Konstandin}}, \binits{T.}},
\bauthor{\bsnm{{Porto}}, \binits{R.A.}},
\bauthor{\bsnm{{Rubira}}, \binits{H.}}:
\batitle{{The effective field theory of large scale structure at three loops}}.
\bjtitle{\jcap}
\bvolume{2019}(\bissue{11}),
\bfpage{027}
(\byear{2019})
{\href{https://arxiv.org/abs/1906.00997}{{arXiv:1906.00997}}}
{[astro-ph.CO]}.
\doiurl{10.1088/1475-7516/2019/11/027}
\end{barticle}
\endbibitem

%%% 48
\bibitem{2020JCAP...07..011F}
\begin{barticle}
\bauthor{\bsnm{{Fonseca de la Bella}}, \binits{L.}},
\bauthor{\bsnm{{Regan}}, \binits{D.}},
\bauthor{\bsnm{{Seery}}, \binits{D.}},
\bauthor{\bsnm{{Parkinson}}, \binits{D.}}:
\batitle{{Impact of bias and redshift-space modelling for the halo power
  spectrum: testing the effective field theory of large-scale structure}}.
\bjtitle{\jcap}
\bvolume{2020}(\bissue{7}),
\bfpage{011}
(\byear{2020})
{\href{https://arxiv.org/abs/1805.12394}{{arXiv:1805.12394}}}
{[astro-ph.CO]}.
\doiurl{10.1088/1475-7516/2020/07/011}
\end{barticle}
\endbibitem

%%% 49
\bibitem{2016NJPh...18d3020B}
\begin{barticle}
\bauthor{\bsnm{{Bartelmann}}, \binits{M.}},
\bauthor{\bsnm{{Fabis}}, \binits{F.}},
\bauthor{\bsnm{{Berg}}, \binits{D.}},
\bauthor{\bsnm{{Kozlikin}}, \binits{E.}},
\bauthor{\bsnm{{Lilow}}, \binits{R.}},
\bauthor{\bsnm{{Viermann}}, \binits{C.}}:
\batitle{{A microscopic, non-equilibrium, statistical field theory for cosmic
  structure formation}}.
\bjtitle{New Journal of Physics}
\bvolume{18}(\bissue{4}),
\bfpage{043020}
(\byear{2016})
{\href{https://arxiv.org/abs/1411.0806}{{arXiv:1411.0806}}}
{[cond-mat.stat-mech]}.
\doiurl{10.1088/1367-2630/18/4/043020}
\end{barticle}
\endbibitem

%%% 50
\bibitem{2017NJPh...19h3001B}
\begin{barticle}
\bauthor{\bsnm{{Bartelmann}}, \binits{M.}},
\bauthor{\bsnm{{Fabis}}, \binits{F.}},
\bauthor{\bsnm{{Kozlikin}}, \binits{E.}},
\bauthor{\bsnm{{Lilow}}, \binits{R.}},
\bauthor{\bsnm{{Dombrowski}}, \binits{J.}},
\bauthor{\bsnm{{Mildenberger}}, \binits{J.}}:
\batitle{{Kinetic field theory: effects of momentum correlations on the cosmic
  density-fluctuation power spectrum}}.
\bjtitle{New Journal of Physics}
\bvolume{19}(\bissue{8}),
\bfpage{083001}
(\byear{2017})
{\href{https://arxiv.org/abs/1611.09503}{{arXiv:1611.09503}}}
{[astro-ph.CO]}.
\doiurl{10.1088/1367-2630/aa7e6f}
\end{barticle}
\endbibitem

%%% 51
\bibitem{2018JSMTE..04.3214F}
\begin{barticle}
\bauthor{\bsnm{{Fabis}}, \binits{F.}},
\bauthor{\bsnm{{Kozlikin}}, \binits{E.}},
\bauthor{\bsnm{{Lilow}}, \binits{R.}},
\bauthor{\bsnm{{Bartelmann}}, \binits{M.}}:
\batitle{{Kinetic field theory: exact free evolution of Gaussian phase-space
  correlations}}.
\bjtitle{Journal of Statistical Mechanics: Theory and Experiment}
\bvolume{4}(\bissue{4}),
\bfpage{043214}
(\byear{2018})
{\href{https://arxiv.org/abs/1710.01611}{{arXiv:1710.01611}}}
{[cond-mat.stat-mech]}.
\doiurl{10.1088/1742-5468/aab850}
\end{barticle}
\endbibitem

%%% 52
\bibitem{2019AnP...53100446B}
\begin{barticle}
\bauthor{\bsnm{{Bartelmann}}, \binits{M.}},
\bauthor{\bsnm{{Kozlikin}}, \binits{E.}},
\bauthor{\bsnm{{Lilow}}, \binits{R.}},
\bauthor{\bsnm{{Littek}}, \binits{C.}},
\bauthor{\bsnm{{Fabis}}, \binits{F.}},
\bauthor{\bsnm{{Kostyuk}}, \binits{I.}},
\bauthor{\bsnm{{Viermann}}, \binits{C.}},
\bauthor{\bsnm{{Heisenberg}}, \binits{L.}},
\bauthor{\bsnm{{Konrad}}, \binits{S.}},
\bauthor{\bsnm{{Geiss}}, \binits{D.}}:
\batitle{{Cosmic Structure Formation with Kinetic Field Theory}}.
\bjtitle{Annalen der Physik}
\bvolume{531}(\bissue{11}),
\bfpage{1800446}
(\byear{2019})
{\href{https://arxiv.org/abs/1905.01179}{{arXiv:1905.01179}}}
{[astro-ph.CO]}.
\doiurl{10.1002/andp.201800446}
\end{barticle}
\endbibitem

%%% 53
\bibitem{2021JCAP...06..035K}
\begin{barticle}
\bauthor{\bsnm{{Kozlikin}}, \binits{E.}},
\bauthor{\bsnm{{Lilow}}, \binits{R.}},
\bauthor{\bsnm{{Fabis}}, \binits{F.}},
\bauthor{\bsnm{{Bartelmann}}, \binits{M.}}:
\batitle{{A first comparison of Kinetic Field Theory with Eulerian Standard
  Perturbation Theory}}.
\bjtitle{\jcap}
\bvolume{2021}(\bissue{6}),
\bfpage{035}
(\byear{2021})
{\href{https://arxiv.org/abs/2012.05812}{{arXiv:2012.05812}}}
{[astro-ph.CO]}.
\doiurl{10.1088/1475-7516/2021/06/035}
\end{barticle}
\endbibitem

%%% 54
\bibitem{2021ScPP...10..153B}
\begin{barticle}
\bauthor{\bsnm{{Bartelmann}}, \binits{M.}},
\bauthor{\bsnm{{Dombrowski}}, \binits{J.}},
\bauthor{\bsnm{{Konrad}}, \binits{S.}},
\bauthor{\bsnm{{Kozlikin}}, \binits{E.}},
\bauthor{\bsnm{{Lilow}}, \binits{R.}},
\bauthor{\bsnm{{Littek}}, \binits{C.}},
\bauthor{\bsnm{{Pixius}}, \binits{C.}},
\bauthor{\bsnm{{Fabis}}, \binits{F.}}:
\batitle{{Kinetic field theory: Non-linear cosmic power spectra in the
  mean-field approximation}}.
\bjtitle{SciPost Physics}
\bvolume{10}(\bissue{6}),
\bfpage{153}
(\byear{2021})
{\href{https://arxiv.org/abs/2011.04979}{{arXiv:2011.04979}}}
{[astro-ph.CO]}.
\doiurl{10.21468/SciPostPhys.10.6.153}
\end{barticle}
\endbibitem

%%% 55
\bibitem{2021arXiv211007427K}
\begin{botherref}
\oauthor{\bsnm{{Konrad}}, \binits{S.}},
\oauthor{\bsnm{{Bartelmann}}, \binits{M.}}:
{On the asymptotic behaviour of cosmic density-fluctuation power spectra}.
arXiv e-prints,
2110--07427
(2021)
{\href{https://arxiv.org/abs/2110.07427}{{arXiv:2110.07427}}}
{[astro-ph.CO]}
\end{botherref}
\endbibitem

%%% 56
\bibitem{2019JCAP...04..001L}
\begin{barticle}
\bauthor{\bsnm{{Lilow}}, \binits{R.}},
\bauthor{\bsnm{{Fabis}}, \binits{F.}},
\bauthor{\bsnm{{Kozlikin}}, \binits{E.}},
\bauthor{\bsnm{{Viermann}}, \binits{C.}},
\bauthor{\bsnm{{Bartelmann}}, \binits{M.}}:
\batitle{{Resummed Kinetic Field Theory: general formalism and linear structure
  growth from Newtonian particle dynamics}}.
\bjtitle{\jcap}
\bvolume{2019}(\bissue{4}),
\bfpage{001}
(\byear{2019})
{\href{https://arxiv.org/abs/1809.06942}{{arXiv:1809.06942}}}
{[astro-ph.CO]}.
\doiurl{10.1088/1475-7516/2019/04/001}
\end{barticle}
\endbibitem

%%% 57
\bibitem{2019JCAP...05..017G}
\begin{barticle}
\bauthor{\bsnm{{Geiss}}, \binits{D.}},
\bauthor{\bsnm{{Lilow}}, \binits{R.}},
\bauthor{\bsnm{{Fabis}}, \binits{F.}},
\bauthor{\bsnm{{Bartelmann}}, \binits{M.}}:
\batitle{{Resummed Kinetic Field Theory: using Mesoscopic Particle
  Hydrodynamics to describe baryonic matter in a cosmological framework}}.
\bjtitle{\jcap}
\bvolume{2019}(\bissue{5}),
\bfpage{017}
(\byear{2019})
{\href{https://arxiv.org/abs/1811.07741}{{arXiv:1811.07741}}}
{[astro-ph.CO]}.
\doiurl{10.1088/1475-7516/2019/05/017}
\end{barticle}
\endbibitem

%%% 58
\bibitem{2021JCAP...01..046G}
\begin{barticle}
\bauthor{\bsnm{{Geiss}}, \binits{D.}},
\bauthor{\bsnm{{Kostyuk}}, \binits{I.}},
\bauthor{\bsnm{{Lilow}}, \binits{R.}},
\bauthor{\bsnm{{Bartelmann}}, \binits{M.}}:
\batitle{{Resummed kinetic field theory: a model of coupled baryonic and dark
  matter}}.
\bjtitle{\jcap}
\bvolume{2021}(\bissue{1}),
\bfpage{046}
(\byear{2021})
{\href{https://arxiv.org/abs/2007.09484}{{arXiv:2007.09484}}}
{[astro-ph.CO]}.
\doiurl{10.1088/1475-7516/2021/01/046}
\end{barticle}
\endbibitem

%%% 59
\bibitem{2015PhRvD..91h3524B}
\begin{barticle}
\bauthor{\bsnm{{Bartelmann}}, \binits{M.}}:
\batitle{{Trajectories of point particles in cosmology and the Zel'dovich
  approximation}}.
\bjtitle{\prd}
\bvolume{91}(\bissue{8}),
\bfpage{083524}
(\byear{2015})
{\href{https://arxiv.org/abs/1411.0805}{{arXiv:1411.0805}}}
{[gr-qc]}.
\doiurl{10.1103/PhysRevD.91.083524}
\end{barticle}
\endbibitem

%%% 60
\bibitem{peebles1980large}
\begin{botherref}
\oauthor{\bsnm{Peebles}, \binits{P.}}:
The large scale structure of the universe.
Princeton Univ
(1980)
\end{botherref}
\endbibitem

%%% 61
\bibitem{1970A&A.....5...84Z}
\begin{barticle}
\bauthor{\bsnm{{Zel'Dovich}}, \binits{Y.B.}}:
\batitle{{Reprint of 1970A\&A.....5...84Z. Gravitational instability: an
  approximate theory for large density perturbations.}}
\bjtitle{\aap}
\bvolume{500},
\bfpage{13}--\blpage{18}
(\byear{1970})
\end{barticle}
\endbibitem

%%% 62
\bibitem{2020A&A...641A...9P}
\begin{barticle}
\bauthor{\bsnm{{Planck Collaboration}}},
\bauthor{\bsnm{{Akrami}}, \binits{Y.}},
\bauthor{\bsnm{{Arroja}}, \binits{F.}},
\bauthor{\bsnm{{Ashdown}}, \binits{M.}},
\bauthor{\bsnm{{Aumont}}, \binits{J.}},
\bauthor{\bsnm{{Baccigalupi}}, \binits{C.}},
\bauthor{\bsnm{{Ballardini}}, \binits{M.}},
\bauthor{\bsnm{{Banday}}, \binits{A.J.}},
\bauthor{\bsnm{{Barreiro}}, \binits{R.B.}},
\bauthor{\bsnm{{Bartolo}}, \binits{N.}},
\bauthor{\bsnm{{Basak}}, \binits{S.}},
\bauthor{\bsnm{{Benabed}}, \binits{K.}},
\bauthor{\bsnm{{Bernard}}, \binits{J.-P.}},
\bauthor{\bsnm{{Bersanelli}}, \binits{M.}},
\bauthor{\bsnm{{Bielewicz}}, \binits{P.}},
\bauthor{\bsnm{{Bond}}, \binits{J.R.}},
\bauthor{\bsnm{{Borrill}}, \binits{J.}},
\bauthor{\bsnm{{Bouchet}}, \binits{F.R.}},
\bauthor{\bsnm{{Bucher}}, \binits{M.}},
\bauthor{\bsnm{{Burigana}}, \binits{C.}},
\bauthor{\bsnm{{Butler}}, \binits{R.C.}},
\bauthor{\bsnm{{Calabrese}}, \binits{E.}},
\bauthor{\bsnm{{Cardoso}}, \binits{J.-F.}},
\bauthor{\bsnm{{Casaponsa}}, \binits{B.}},
\bauthor{\bsnm{{Challinor}}, \binits{A.}},
\bauthor{\bsnm{{Chiang}}, \binits{H.C.}},
\bauthor{\bsnm{{Colombo}}, \binits{L.P.L.}},
\bauthor{\bsnm{{Combet}}, \binits{C.}},
\bauthor{\bsnm{{Crill}}, \binits{B.P.}},
\bauthor{\bsnm{{Cuttaia}}, \binits{F.}},
\bauthor{\bsnm{{de Bernardis}}, \binits{P.}},
\bauthor{\bsnm{{de Rosa}}, \binits{A.}},
\bauthor{\bsnm{{de Zotti}}, \binits{G.}},
\bauthor{\bsnm{{Delabrouille}}, \binits{J.}},
\bauthor{\bsnm{{Delouis}}, \binits{J.-M.}},
\bauthor{\bsnm{{Di Valentino}}, \binits{E.}},
\bauthor{\bsnm{{Diego}}, \binits{J.M.}},
\bauthor{\bsnm{{Dor{\'e}}}, \binits{O.}},
\bauthor{\bsnm{{Douspis}}, \binits{M.}},
\bauthor{\bsnm{{Ducout}}, \binits{A.}},
\bauthor{\bsnm{{Dupac}}, \binits{X.}},
\bauthor{\bsnm{{Dusini}}, \binits{S.}},
\bauthor{\bsnm{{Efstathiou}}, \binits{G.}},
\bauthor{\bsnm{{Elsner}}, \binits{F.}},
\bauthor{\bsnm{{En{\ss}lin}}, \binits{T.A.}},
\bauthor{\bsnm{{Eriksen}}, \binits{H.K.}},
\bauthor{\bsnm{{Fantaye}}, \binits{Y.}},
\bauthor{\bsnm{{Fergusson}}, \binits{J.}},
\bauthor{\bsnm{{Fernandez-Cobos}}, \binits{R.}},
\bauthor{\bsnm{{Finelli}}, \binits{F.}},
\bauthor{\bsnm{{Frailis}}, \binits{M.}},
\bauthor{\bsnm{{Fraisse}}, \binits{A.A.}},
\bauthor{\bsnm{{Franceschi}}, \binits{E.}},
\bauthor{\bsnm{{Frolov}}, \binits{A.}},
\bauthor{\bsnm{{Galeotta}}, \binits{S.}},
\bauthor{\bsnm{{Galli}}, \binits{S.}},
\bauthor{\bsnm{{Ganga}}, \binits{K.}},
\bauthor{\bsnm{{G{\'e}nova-Santos}}, \binits{R.T.}},
\bauthor{\bsnm{{Gerbino}}, \binits{M.}},
\bauthor{\bsnm{{Gonz{\'a}lez-Nuevo}}, \binits{J.}},
\bauthor{\bsnm{{G{\'o}rski}}, \binits{K.M.}},
\bauthor{\bsnm{{Gratton}}, \binits{S.}},
\bauthor{\bsnm{{Gruppuso}}, \binits{A.}},
\bauthor{\bsnm{{Gudmundsson}}, \binits{J.E.}},
\bauthor{\bsnm{{Hamann}}, \binits{J.}},
\bauthor{\bsnm{{Handley}}, \binits{W.}},
\bauthor{\bsnm{{Hansen}}, \binits{F.K.}},
\bauthor{\bsnm{{Herranz}}, \binits{D.}},
\bauthor{\bsnm{{Hivon}}, \binits{E.}},
\bauthor{\bsnm{{Huang}}, \binits{Z.}},
\bauthor{\bsnm{{Jaffe}}, \binits{A.H.}},
\bauthor{\bsnm{{Jones}}, \binits{W.C.}},
\bauthor{\bsnm{{Jung}}, \binits{G.}},
\bauthor{\bsnm{{Keih{\"a}nen}}, \binits{E.}},
\bauthor{\bsnm{{Keskitalo}}, \binits{R.}},
\bauthor{\bsnm{{Kiiveri}}, \binits{K.}},
\bauthor{\bsnm{{Kim}}, \binits{J.}},
\bauthor{\bsnm{{Krachmalnicoff}}, \binits{N.}},
\bauthor{\bsnm{{Kunz}}, \binits{M.}},
\bauthor{\bsnm{{Kurki-Suonio}}, \binits{H.}},
\bauthor{\bsnm{{Lamarre}}, \binits{J.-M.}},
\bauthor{\bsnm{{Lasenby}}, \binits{A.}},
\bauthor{\bsnm{{Lattanzi}}, \binits{M.}},
\bauthor{\bsnm{{Lawrence}}, \binits{C.R.}},
\bauthor{\bsnm{{Le Jeune}}, \binits{M.}},
\bauthor{\bsnm{{Levrier}}, \binits{F.}},
\bauthor{\bsnm{{Lewis}}, \binits{A.}},
\bauthor{\bsnm{{Liguori}}, \binits{M.}},
\bauthor{\bsnm{{Lilje}}, \binits{P.B.}},
\bauthor{\bsnm{{Lindholm}}, \binits{V.}},
\bauthor{\bsnm{{L{\'o}pez-Caniego}}, \binits{M.}},
\bauthor{\bsnm{{Ma}}, \binits{Y.-Z.}},
\bauthor{\bsnm{{Mac\'{\i}as-P{\'e}rez}}, \binits{J.F.}},
\bauthor{\bsnm{{Maggio}}, \binits{G.}},
\bauthor{\bsnm{{Maino}}, \binits{D.}},
\bauthor{\bsnm{{Mandolesi}}, \binits{N.}},
\bauthor{\bsnm{{Marcos-Caballero}}, \binits{A.}},
\bauthor{\bsnm{{Maris}}, \binits{M.}},
\bauthor{\bsnm{{Martin}}, \binits{P.G.}},
\bauthor{\bsnm{{Mart\'{\i}nez-Gonz{\'a}lez}}, \binits{E.}},
\bauthor{\bsnm{{Matarrese}}, \binits{S.}},
\bauthor{\bsnm{{Mauri}}, \binits{N.}},
\bauthor{\bsnm{{McEwen}}, \binits{J.D.}},
\bauthor{\bsnm{{Meerburg}}, \binits{P.D.}},
\bauthor{\bsnm{{Meinhold}}, \binits{P.R.}},
\bauthor{\bsnm{{Melchiorri}}, \binits{A.}},
\bauthor{\bsnm{{Mennella}}, \binits{A.}},
\bauthor{\bsnm{{Migliaccio}}, \binits{M.}},
\bauthor{\bsnm{{Miville-Desch{\^e}nes}}, \binits{M.-A.}},
\bauthor{\bsnm{{Molinari}}, \binits{D.}},
\bauthor{\bsnm{{Moneti}}, \binits{A.}},
\bauthor{\bsnm{{Montier}}, \binits{L.}},
\bauthor{\bsnm{{Morgante}}, \binits{G.}},
\bauthor{\bsnm{{Moss}}, \binits{A.}},
\bauthor{\bsnm{{M{\"u}nchmeyer}}, \binits{M.}},
\bauthor{\bsnm{{Natoli}}, \binits{P.}},
\bauthor{\bsnm{{Oppizzi}}, \binits{F.}},
\bauthor{\bsnm{{Pagano}}, \binits{L.}},
\bauthor{\bsnm{{Paoletti}}, \binits{D.}},
\bauthor{\bsnm{{Partridge}}, \binits{B.}},
\bauthor{\bsnm{{Patanchon}}, \binits{G.}},
\bauthor{\bsnm{{Perrotta}}, \binits{F.}},
\bauthor{\bsnm{{Pettorino}}, \binits{V.}},
\bauthor{\bsnm{{Piacentini}}, \binits{F.}},
\bauthor{\bsnm{{Polenta}}, \binits{G.}},
\bauthor{\bsnm{{Puget}}, \binits{J.-L.}},
\bauthor{\bsnm{{Rachen}}, \binits{J.P.}},
\bauthor{\bsnm{{Racine}}, \binits{B.}},
\bauthor{\bsnm{{Reinecke}}, \binits{M.}},
\bauthor{\bsnm{{Remazeilles}}, \binits{M.}},
\bauthor{\bsnm{{Renzi}}, \binits{A.}},
\bauthor{\bsnm{{Rocha}}, \binits{G.}},
\bauthor{\bsnm{{Rubi{\~n}o-Mart\'{\i}n}}, \binits{J.A.}},
\bauthor{\bsnm{{Ruiz-Granados}}, \binits{B.}},
\bauthor{\bsnm{{Salvati}}, \binits{L.}},
\bauthor{\bsnm{{Savelainen}}, \binits{M.}},
\bauthor{\bsnm{{Scott}}, \binits{D.}},
\bauthor{\bsnm{{Shellard}}, \binits{E.P.S.}},
\bauthor{\bsnm{{Shiraishi}}, \binits{M.}},
\bauthor{\bsnm{{Sirignano}}, \binits{C.}},
\bauthor{\bsnm{{Sirri}}, \binits{G.}},
\bauthor{\bsnm{{Smith}}, \binits{K.}},
\bauthor{\bsnm{{Spencer}}, \binits{L.D.}},
\bauthor{\bsnm{{Stanco}}, \binits{L.}},
\bauthor{\bsnm{{Sunyaev}}, \binits{R.}},
\bauthor{\bsnm{{Suur-Uski}}, \binits{A.-S.}},
\bauthor{\bsnm{{Tauber}}, \binits{J.A.}},
\bauthor{\bsnm{{Tavagnacco}}, \binits{D.}},
\bauthor{\bsnm{{Tenti}}, \binits{M.}},
\bauthor{\bsnm{{Toffolatti}}, \binits{L.}},
\bauthor{\bsnm{{Tomasi}}, \binits{M.}},
\bauthor{\bsnm{{Trombetti}}, \binits{T.}},
\bauthor{\bsnm{{Valiviita}}, \binits{J.}},
\bauthor{\bsnm{{Van Tent}}, \binits{B.}},
\bauthor{\bsnm{{Vielva}}, \binits{P.}},
\bauthor{\bsnm{{Villa}}, \binits{F.}},
\bauthor{\bsnm{{Vittorio}}, \binits{N.}},
\bauthor{\bsnm{{Wandelt}}, \binits{B.D.}},
\bauthor{\bsnm{{Wehus}}, \binits{I.K.}},
\bauthor{\bsnm{{Zacchei}}, \binits{A.}},
\bauthor{\bsnm{{Zonca}}, \binits{A.}}:
\batitle{{Planck 2018 results. IX. Constraints on primordial non-Gaussianity}}.
\bjtitle{\aap}
\bvolume{641},
\bfpage{9}
(\byear{2020})
{\href{https://arxiv.org/abs/1905.05697}{{arXiv:1905.05697}}}
{[astro-ph.CO]}.
\doiurl{10.1051/0004-6361/201935891}
\end{barticle}
\endbibitem

%%% 63
\bibitem{bleistein1975asymptotic}
\begin{bbook}
\bauthor{\bsnm{Bleistein}, \binits{N.}},
\bauthor{\bsnm{Handelsman}, \binits{R.A.}}:
\bbtitle{Asymptotic Expansions of Integrals}.
\bpublisher{Ardent Media}, \blocation{???}
(\byear{1975})
\end{bbook}
\endbibitem

%%% 64
\bibitem{wong2001asymptotic}
\begin{bbook}
\bauthor{\bsnm{Wong}, \binits{R.}}:
\bbtitle{Asymptotic Approximations of Integrals}.
\bpublisher{SIAM}, \blocation{???}
(\byear{2001})
\end{bbook}
\endbibitem

%%% 65
\bibitem{2020JCAP...06..033C}
\begin{barticle}
\bauthor{\bsnm{{Chen}}, \binits{S.-F.}},
\bauthor{\bsnm{{Pietroni}}, \binits{M.}}:
\batitle{{Asymptotic expansions for Large Scale Structure}}.
\bjtitle{\jcap}
\bvolume{2020}(\bissue{6}),
\bfpage{033}
(\byear{2020})
{\href{https://arxiv.org/abs/2002.11357}{{arXiv:2002.11357}}}
{[astro-ph.CO]}.
\doiurl{10.1088/1475-7516/2020/06/033}
\end{barticle}
\endbibitem

%%% 66
\bibitem{fulks1961asymptotics}
\begin{barticle}
\bauthor{\bsnm{Fulks}, \binits{W.}},
\bauthor{\bsnm{Sather}, \binits{J.}}:
\batitle{Asymptotics. ii. laplace's method for multiple integrals.}
\bjtitle{Pacific Journal of Mathematics}
\bvolume{11}(\bissue{1}),
\bfpage{185}--\blpage{192}
(\byear{1961})
\end{barticle}
\endbibitem

%%% 67
\bibitem{doetsch1955anwendungen}
\begin{bbook}
\bauthor{\bsnm{Doetsch}, \binits{G.}}:
\bbtitle{Anwendungen der Laplace-Transformation}.
\bpublisher{Birkh{\"a}user}, \blocation{???}
(\byear{1955})
\end{bbook}
\endbibitem

%%% 68
\bibitem{erdelyi1956asymptotic}
\begin{bbook}
\bauthor{\bsnm{Erd{\'e}lyi}, \binits{A.}}:
\bbtitle{Asymptotic Expansions}
vol. \bseriesno{3}.
\bpublisher{Courier Corporation}, \blocation{???}
(\byear{1956})
\end{bbook}
\endbibitem

%%% 69
\bibitem{erdelyi1961general}
\begin{barticle}
\bauthor{\bsnm{Erd{\'e}lyi}, \binits{A.}}:
\batitle{General asymptotic expansions of laplace integrals}.
\bjtitle{Archive for Rational Mechanics and Analysis}
\bvolume{7}(\bissue{1}),
\bfpage{1}--\blpage{20}
(\byear{1961})
\end{barticle}
\endbibitem

%%% 70
\bibitem{2022Waibel}
\begin{botherref}
\oauthor{\bsnm{{Waibel}}, \binits{R.e.a.}}:
{On the asymptotic behaviour of cosmic density-fluctuation bispectra}.
in preparation
(2022)
\end{botherref}
\endbibitem

%%% 71
\bibitem{poston2014catastrophe}
\begin{bbook}
\bauthor{\bsnm{Poston}, \binits{T.}},
\bauthor{\bsnm{Stewart}, \binits{I.}}:
\bbtitle{Catastrophe Theory and Its Applications}.
\bpublisher{Courier Corporation}, \blocation{???}
(\byear{2014})
\end{bbook}
\endbibitem

%%% 72
\bibitem{1986ApJ...304...15B}
\begin{barticle}
\bauthor{\bsnm{{Bardeen}}, \binits{J.M.}},
\bauthor{\bsnm{{Bond}}, \binits{J.R.}},
\bauthor{\bsnm{{Kaiser}}, \binits{N.}},
\bauthor{\bsnm{{Szalay}}, \binits{A.S.}}:
\batitle{{The Statistics of Peaks of Gaussian Random Fields}}.
\bjtitle{\apj}
\bvolume{304},
\bfpage{15}
(\byear{1986}).
\doiurl{10.1086/164143}
\end{barticle}
\endbibitem

%%% 73
\bibitem{weinberg2008cosmology}
\begin{bbook}
\bauthor{\bsnm{Weinberg}, \binits{S.}}:
\bbtitle{Cosmology}.
\bpublisher{Oxford university press}, \blocation{???}
(\byear{2008})
\end{bbook}
\endbibitem

%%% 74
\bibitem{2022Konrad}
\begin{botherref}
\oauthor{\bsnm{{Konrad}}, \binits{S.}},
\oauthor{\bsnm{{Ginat}}, \binits{B.Y.}},
\oauthor{\bsnm{{Bartelmann}}, \binits{M.}}:
{On the asymptotic behaviour of cosmic density-fluctuation power spectra of
  cold dark matter}.
in preparation
(2022)
\end{botherref}
\endbibitem

%%% 75
\bibitem{2022Seute}
\begin{botherref}
\oauthor{\bsnm{{Seute}}, \binits{L.e.a.}}:
{On the asymptotic behaviour of cosmic density-fluctuation power spectra,
  including full initial phase-space correlations}.
in preparation
(2022)
\end{botherref}
\endbibitem

\end{thebibliography}

\end{document}